\documentclass[12pt, oneside]{article}
\usepackage{graphicx} 
\usepackage{color} 
\usepackage[normalem]{ulem} 
\usepackage{multirow} 
\usepackage{amsfonts} 
\usepackage{amsmath}
\usepackage{hyperref}
\usepackage{extsizes}
\usepackage{physics} 
\usepackage{comment} 
\usepackage{bm} 
\usepackage{xcolor} 
\usepackage{authblk} 
\usepackage[sort&compress,numbers]{natbib}

\hypersetup{
  colorlinks,
  citecolor=blue!50!green!90!black,
  linkcolor=blue!80!green!90!black,
  urlcolor=blue!80!green!90!black}

\newcommand{\blue}[1]{{\color{black}#1}}

\newcommand{\hehe}[1]{{\color{black!65!green!90!yellow}#1}}

\newcommand{\iu}{\mathrm{i}\mkern1mu}
\newcommand{\eu}{\mathrm{e}\mkern1mu}

\usepackage[
    left=30mm, 
    right=30mm, 
    top=23mm,
    bottom=23mm, 
    columnsep=9mm 
]{geometry}

\title{
Achiral nanostructures: perturbative harmonic generation and dichroism under vortex and vector beams illumination}

\author[1]{Anastasia Nikitina\thanks{Email: anastasia.nikitina@metalab.ifmo.ru}}
\author[2]{Kristina Frizyuk\thanks{Email: frzyuk@gmail.com}}

\affil[1]{The School of Physics and Engineering, ITMO University, Saint-Petersburg, Russia}
\affil[2]{Department of Information Engineering, University of Brescia, Brescia, Italy}
\date{} 

\begin{document}

\maketitle
\begin{abstract}
    In this study, we investigate the nonlinear optical phenomena emerging from the interaction of vortex and vector beams with achiral nanoparticles or nanostructures. 
    We reveal the conditions under which linear or nonlinear dichroism can be observed.
    Despite the achiral symmetry of the nanostructure, the interplay between the symmetries of the vortex beam, the nanostructure, and the crystalline lattice of the nanostructure material may result in circular dichroism in the nonlinear regime.
    We derive a formula that describes the conditions for the appearance of circular dichroism across a broad range of scenarios, taking into account all the symmetries. 
    Building on these findings, we have determined the conditions for both linear and nonlinear dichroism when illuminated by vector beams. 
    We believe that this work provides important insights that can enhance the design of chiral sensors and optical traps, making them more versatile and effective.
\end{abstract}

\tableofcontents
\section{Introduction}
\allowdisplaybreaks
\sloppy
Chirality, being a natural characteristic, is crucial in various scientific fields such as physics and biology~\cite{Pasteur, Rosenfeld1929-QuantenmechanischeTh, Cahn1966-SpecificationofMole}.
It's often observed in natural structures like amino acids or proteins.
\begin{figure}[ht!]
    \centering
    \includegraphics[width=0.65\linewidth]{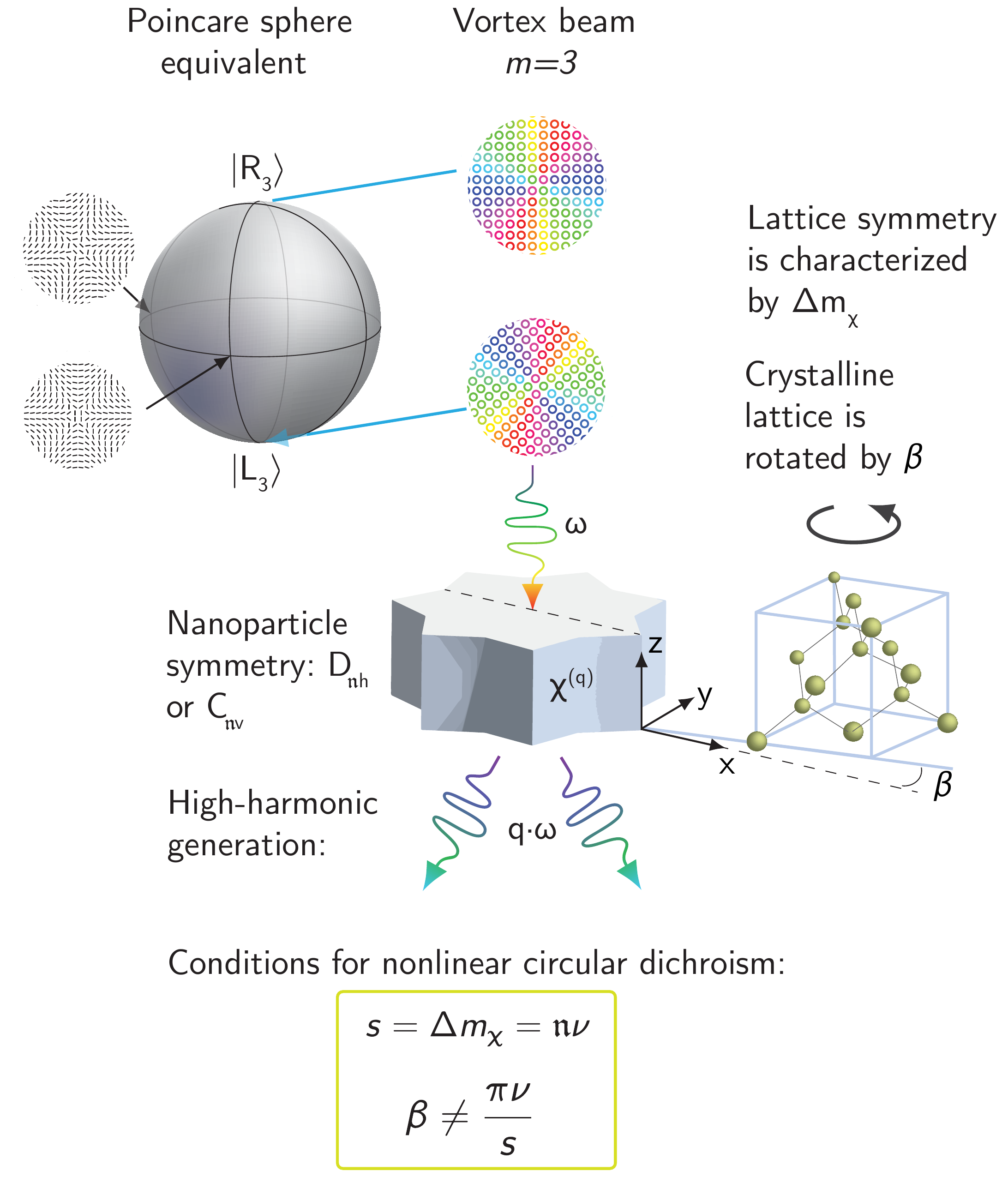}
    \caption{Schematic of the concept. 
    Incident vortex beam, characterized by an angular momentum projection $\pm m$ is depicted using higher order Poincare sphere~\cite{Padgett1999-Poincare-sphereequiv, Shen2021-RayswavesSU(2)sy}. 
    The value $m = 3$ is taken, as an example. 
    The circular dichroism in high-harmonic generation can be obtained when we excite the nanoparticle of an achiral symmetry {$\text{D}_{\mathfrak{n} \text{h}}$ or $\text{C}_{\mathfrak{n}\text{v}}$} by either $\ket{R_m}$ or $\ket{L_m}$ beam. 
    Another two states which are shown with the black lines (polarization ellipses with one zero axis) 
    are vector beams~\cite{Zhang2017-Gouyphaseinducedpo}. 
    The nanoparticle generates high harmonic due to material properties, where $q$ is an order of high harmonic. 
    The crystalline lattice is rotated with respect to the nanoparticle symmetry axis by an angle $\beta$ and characterized by differences $\Delta m_\chi$ determined by $\hat{\chi}^{(q)}$ tensor. 
    The main result is shown in a box, where $\nu \in \mathbb Z$.}
    \label{mainscheme}
\end{figure}
Circular dichroism spectroscopy is a well-known method for investigating the properties of chiral substances.
While it is commonly defined as a difference in the absorption of a left- and right-circularly polarized light~\cite{rodger1997circular, fasman2013circular}, it can be generalized for more complex systems~\cite{Stephens1970-TheoryofMagneticCi, 
Hecht1994-RayleighandRamanop, 
Stephens1985-VibrationalCircular, 
Manakov1996-Anewtechniqueinth},
including nonlinear circular dichroism~\cite{Koshelev2024-Scatteringmatrixfor, 
Byers1994-Second-harmonicgener,
Cai2024-Perfectintrinsicand,
Chacon2020-Circulardichroismin,  
Huttunen2009-Absolutenonlinearop, 
Spreyer2022-SecondHarmonicOptic}
or circular dichroism under excitation by light beams with angular momentum~\cite{Forbes2021-Opticalvortexdichro,  
zambrana2022vortex,
Bibikova2024-Topologicalcircular, 
Zambrana-Puyalto2014-Angularmomentum-indu, 
Andrews2004-Onopticalvortexint}. 
Previously, various types of circular dichroism were investigated for continuous media, and focused studies on individual nanoparticles or metamaterials have only recently begun to emerge~\cite{Bautista2012-Second-HarmonicGener, 
Koshelev2023-Scatteringmatrixfor, 
Govorov2010-TheoryofCircularDi, 
Zhu2018-Giantintrinsicchiro, 
Wang2017-CircularDichroismSt, 
Rebholz2024-SeparatingtheMateri, 
Czajkowski2022-Localversusbulkcir}. 
In recent works, we have demonstrated the existence of nonlinear, particularly second-harmonic circular dichroism (SH-CD) in achiral dielectric nanostructures~\cite{Frizyuk2021-NonlinearCircularDi, 
Nikitina2023-Nonlinearcirculardi} with intrinsic nonlinearity.
{In fact}, it was defined as the difference between values of second-harmonic intensity for right- and left-circularly polarized plane wave excitation, and it was proven that this nonlinear effect appears under certain conditions on the orientation of the nanoparticle relative to its material's lattice.

Here, we show that these results can be generalized to the case of any high-harmonic generation as well as any excitation by a vortex beam with an angular momentum (See Fig.~\ref{mainscheme}) proposing a generalized framework for understanding nonlinear circular dichroism in complex light-matter interactions.
These results can be of particular importance for chiral sensing enhancement by nanoparticles, since we show that in some cases, the dichroic signal of the nanoparticle itself cannot be neglected. 
{We primarily focus on this case, as it provides a general framework for studying many different systems.}

{Another scenario is analogous to linear dichroism in the linear regime, i.e. the dichroism under vector beams illumination~\cite{Zhang2017-Gouyphaseinducedpo, Rosales-Guzman2018-Areviewofcomplexv}. 
Here, the dichroism is possible even in the linear regime, and we provide the conditions, under which it can or can not be observed.}

{This work is organized as follows. 
In Section~\ref{theor framework}, we give a general framework of the problem. 
Section~\ref{sec:mainformula} contains the main result: the formula, which describes the conditions of dichroism occurrence in all cases. 
Section~\ref{sec:cases} provides the strict consideration of the second-harmonic generation for particular materials and shapes of the nanostructures in terms of modal and multipolar analysis. 
The first case~\ref{sec:GaAsC3v} is one in which dichroism is possible, and the second~\ref{sec:GaAsC4v} is one in which it is not.
This approach leads to the derivation of the main formula~\eqref{rule1}, in Section~\ref{universalmaincondition} and also shows the variety of options and interactions that arise in intermediate processes. 
Section~\ref{sec:app:plots} provides the numerical calculations to support the theoretical predictions.
Section~\ref{sec:discussions} gives discussions of some important subtleties.
{In Section~\ref{vectorbeam} we discuss the conditions for observing dichroism under vector beam illumination in linear~\eqref{lincond} and nonlinear regimes~\eqref{vectorcondq}.}
In Section~\ref{sec:conclusions}, we give a brief conclusion.

\section{Theoretical framework}
\label{theor framework}
\subsection{Incident beams}
Before introducing the formula for general notation of the nonlinear circular dichroism, we describe the nature of vortex beams. 
We consider the following description for vortex beams in cylindrical coordinates, for which a higher-order Poincare sphere can be introduced~\cite{Padgett1999-Poincare-sphereequiv, 
Shen2021-RayswavesSU(2)sy, 
Aita2024-Propagationoffocuse, 
Soares2006-Hermite-Besselbeams}. 
{In paraxial regime, they can be described as:
\begin{equation}
    |R_m\rangle {\sim} \exp(+\iu(m-1)\varphi)(\hat{\vb x} + \iu\hat{\vb y})/\sqrt{2},
    \label{eq_rl}
\end{equation}
\begin{equation}
    |L_m\rangle {\sim} \exp(-\iu(m-1)\varphi)(\hat{\vb x} - \iu\hat{\vb y})/\sqrt{2},
    \label{eq_ll}
\end{equation}
where $|R_m\rangle$ and $|L_m\rangle$ are related to the right- and left-handed beam, respectively, {which are mirror images of each other}.
Here $m\in \mathbb Z$, {$\varphi$ is the azimuthal angle}, and $ \hat{\vb{x}}$ and $\hat{\vb y}$ are unit basis vectors. 
A detailed description of the properties of vortex beams and the connection between angular momentum and topological charge can be found in the Appendix~\ref{app:mult}.
}
We omit the dependence of the radial coordinate $r$~\cite{Zhan2006-Propertiesofcircula} because it does not change the general symmetry, in which we are interested.
{We also use 
\begin{equation}
    (\hat{\vb x}\pm \iu \hat{{\vb y}}) = e^{\pm \iu \varphi}(\hat{\bm \rho}\pm \iu \hat{{\bm \varphi}})
\end{equation} and rewrite:
\begin{equation}
    |R_m\rangle  
    = e^{ + \iu m^{}\varphi}\frac{\hat{\bm \rho} + \iu \hat{{\bm \varphi}}}{\sqrt{2}},
    \label{eq_rl2}
\end{equation}
\begin{equation}
    |L_m\rangle = 
    e^{ - \iu m^{}\varphi}\frac{\hat{\bm \rho} - \iu \hat{{\bm \varphi}}}{\sqrt{2}}.
    \label{eq_ll2}
\end{equation}
Note that $\pm m$ refers to the total angular momentum projection on the $z$-axis and plays the important role in our considerations. 
Handedness refers to the helicity of the beam~\cite{Fernandez-Corbaton2012-Helicityandangular}.}
We consider this problem by implementing the multipolar decomposition because, for the multipoles, total angular momentum projection is well-defined~\cite{Akhiezer}.
{We use} vector spherical harmonics~\cite{Stratton, Bohren1998Mar} (spherical multipoles), {although their application may seem} redundant in this context. 
{However, in fact,} employing any kind of cylindrical harmonics~\cite{Frezza2018-Introductiontoelect} does not necessarily simplify or clarify the problem, as we are only interested in the total angular momentum projection $m$ regardless of the basis used.

Another type of beams we are interested in are vector beams, related to the diameter of the generalized Poincaré sphere (see Fig.~\ref{mainscheme}). 
They are introduced as:
\begin{equation}
    |\text{B}_{me}\rangle \propto |L_m\rangle + |R_m\rangle
    \label{vb_e}
\end{equation}
\begin{equation}
    |\text{B}_{mo}\rangle \propto |L_m\rangle - |R_m\rangle.
    \label{vb_o}
\end{equation}
Indexes $e$ and $o$ denote the parity under reflection in $ y=0 $ plane.

\subsection{Multipolar content of the vortex beams}
{For further convenience, we introduce $m^{\text{in}}=|m|$. 
The multipolar content of vortex beams with particular helicity~\cite{Fernandez-Corbaton2012-Helicityandangular} and total angular momentum projection is known from literature} {(Eq.~(4) in~\cite{Molina-Terriza2008-Determinationofthe}} and Eq.~(53) in~\cite{Vavilin2024-ThepolychromaticT-m}). 
Thus, we are interested in the following combinations of vector spherical harmonics defined in~\cite{Bohren1998Mar} (see Appendix~\ref{sec:app:1}):
\begin{align}
    \label{sei}
    (\vb N_{em^{\text{in}}n} \pm \iu\vb N_{om^{\text{in}}n}) + (\pm\vb M_{em^{\text{in}}n} + \iu\vb M_{om^{\text{in}}n}) \sim  e^{\pm \iu m^{\text{in}} \varphi}(\hat{\bm \rho}\pm \iu \hat{{\bm \varphi}}),
\end{align}
which also can be compared with multipolar content of circularly polarized plane wave for $m^{\text{in}}=1$~\cite{Bohren1998Mar}, 
and {also}
\begin{align}
    \label{sette}
    (\vb N_{em^{\text{in}}n} \mp \iu\vb N_{om^{\text{in}}n}) - (\mp\vb M_{em^{\text{in}}n} + \iu\vb M_{om^{\text{in}}n}) \sim  e^{\mp \iu m^{\text{in}} \varphi}(\hat{\bm \rho} \pm \iu \hat{{\bm \varphi}}),
\end{align}
where $\bm \rho$ and $\bm \varphi$ are unit basis vectors in cylindrical coordinates. 
Here we use the sign $\sim$ {which shows that such linear combinations contribute to the multipolar content of the corresponding beam, and also possess the same symmetry behavior under rotations around the $z$-axis}.
{Further analysis is applicable to all potential incident beams that can be expressed via these multipolar combinations. 
We have addressed paraxial approximation as commonly employed and familiar; however, it is the analysis of the multipolar composition that yields the conclusive answer.}

{Importantly, when nanostructures are excited by a vortex beam, combinations as described in equations~\eqref{sei} and~\eqref{sette} do not occur for internal or scattered fields~\cite{Bohren1998Mar} due to the resonant excitation of the eigenmodes. 
Additionally, the resonant frequencies of magnetic and electric multipoles are generally different, except in helicity-preserving systems where $\varepsilon/\mu=\text{const}$~\cite{Zambrana-Puyalto2013-Dualitysymmetryand, 
Lasa-Alonso2023-Ontheoriginofthe, 
Graf2019-AchiralHelicityPre, 
Fernandez-Corbaton2013-Forwardandbackward}.}

Thus, we further prove that we can just consider the total angular momentum projection of a beam $m^{\text{in}}$ (except the case $m^{\text{in}}=0$, which is commented in discussions), which obeys a ``modified by nanostructure'' conservation law, and the problem becomes almost fully analogous to the one considered in~\cite{Nikitina2023-Nonlinearcirculardi}. 
Now we can move to the discussion of the nonlinear response.

\subsection{{Perturbative harmonic generation}}
We consider achiral nanoparticles or nanostructures, which can be of C$_{\mathfrak{n}  \text v}$, D$_{\mathfrak{n} \text h}$ symmetry (Schoenflies notation).
The nanoparticles are made from a material with a noncentrosymmetric crytalline lattice (e.g., GaAs, BaTiO$_3$, or LiNbO$_3$).
In such dielectric nanostructures, second harmonic generation is commonly described by the nonlinear susceptibility tensor $\hat\chi^{(2)}$~\cite{Boyd2003}:
\begin{align}
    P^{2\omega}_i(\vb r) = \varepsilon_0 \chi^{(2)}_{ijk} E_j^{\omega}(\vb r) E_k^{\omega}(\vb r),
    \label{eq_shg_baza}
\end{align}
where $P^{2\omega}_i(\vb r)$ is a nonlinear polarization, and $\vb E^{\omega}(\vb r)$ is a fundamental field inside the nanoparticle or nanostructure. 
The second-harmonic generation is given separately as a most common case and will be further considered in some examples. 
For the {perturbative (See discussion in Appendix~\ref{sec:app:pert}) high-harmonic} generation of order $q$, the nonlinear polarization $P^{q\omega}_i(\vb r)$ can be defined as~\cite{Shcherbakov2021-Generationofevenan}:
\begin{align}
    P^{q\omega}_i(\vb r) = \varepsilon_0 \chi^{(q)}_{ijk\dots h} \underbrace{E^{\omega}_j(\vb r) E^{\omega}_k(\vb r)\dots E^{\omega}_h(\vb r)}_{q},
    \label{eq_shg_baza_anyharmonic}
\end{align}
The far-field outside the nanoparticle can be described with the help of Green's function~\cite{Novotny, Doost2014-Resonant-stateexpans}:
\begin{align}
    \label{efieldsh_anyharmonic} 
    \vb E^{q\omega}(\mathbf{r}) 
    & = (q\omega)^2 \mu_0 \int_{V} \dd V' \hat{\vb G}(q\omega, \mathbf{r}, \mathbf{r}') \vb P^{q\omega}(\mathbf{r}') = 
    \\\nonumber 
    & = (q\omega)^2\mu_0\int\limits_V \dd V' \sum_{j} \frac{\mathbf{E}^{q\omega}_{j}(\mathbf{r}) \otimes \mathbf{E}^{q\omega}_{j}\left(\mathbf{r}^{\prime}\right)}{{2 k}\left(k-k_{j}\right)} \mathbf{P}^{q\omega} (\mathbf{r}')=
    \\
    & =(q\omega)^2\mu_0\sum_{j}\frac{1}{{2 k}\left(k-k_{j}\right)} \mathbf{E}^{q\omega}_{j}(\mathbf{r})
    \nonumber \int\limits_V \dd V'   \mathbf{E}^{q\omega}_{j}\left(\mathbf{r}^{\prime}\right) \mathbf{P}^{q\omega} (\mathbf{r}'),
\end{align}
where {$\omega$ is the fundamental frequency, $\mu_0$ is the vacuum permeability, and $\mathbf{E}^{q\omega}_{j}(\mathbf{r})$ is the field of the system's  quasi-normal mode (eigenfrequency $\omega_j=ck_j$)~\cite{Doost2014-Resonant-stateexpans, Lalanne2018-LightInteractionwit,
Gigli2020-Quasinormal-ModeNon} 
at the appropriate frequency $q\omega=ck$. 
The integration is over the particle's volume.} 
Green's function is defined as
$\nabla \times \nabla \times \hat{\vb G}(\omega, \mathbf{r}, \mathbf{r}') = \left(\frac{\omega}{c}\right)^2 \varepsilon(\mathbf{r}, \omega)\hat{\vb G}(\omega, \mathbf{r}, \mathbf{r}') + \hat{\vb I}\delta(\mathbf{r} - \mathbf{r}')$~\cite{Novotny},
where $ \hat{\vb I}$  is the unit dyadic,
$\varepsilon(\mathbf{r}, \omega) = \varepsilon_2(\omega)$ for nanostructure, and  $\varepsilon(\mathbf{r}) = 1$  for vacuum.
{We assume $\varepsilon_2(\omega)$ to be a scalar quantity, and the structure is not birefringent (which is rigorous for cubic lattices).}
{Note that strictly speaking, the equation~\eqref{efieldsh_anyharmonic} converges only for the region inside the nanoparticle, and $\vb E^{q\omega}(\mathbf{r})$ in the outer region should be replaced by its analytic continuation (See e.g.~\cite{Wu2024-Understandingthephy} or SI in ~\cite{Bogdanov2019-Boundstatesinthec, Poleva2023-Multipolartheoryof}). 
However, this detail is omitted subsequently, as the symmetry of the fields remains unchanged after this procedure.}
In this work, we assume that the incident beam has a certain azimuthal symmetry.

{Circular dichroism in the second-harmonic signal can be defined as:
\begin{equation}
    \text{SH-CD}_{m, \lambda} = \frac{(I_{R_m}^{2\omega}-I_{L_m}^{2\omega})}{(I_{R_m}^{2\omega}+I_{{{L_m}}}^{2\omega})},
    \label{SHCDsecond}
\end{equation}}
where {$I_{R_m}^{2\omega},\  I_{L_m}^{2\omega}$} is the corresponding total intensity of a second-harmonic {described by $\vb E^{q\omega}(\vb r)$, excited by {a right- or left-handed} beam, respectively, and integrated over the sphere in the far-field (the center of the sphere should coincide with the center of the nanostructure)}. 
{This definition is analogous to circular-helical dichroism~\cite{Ye2019-ProbingMolecularChi} or circular-vortex dichroism~\cite{Forbes2021-Orbitalangularmomen}, which can be found in literature.}
We can introduce analogously higher-order nonlinear circular dichroism~\cite{Koshelev2023-ResonantChiralEffec, Chen2016-GiantNonlinearOptic}:
\begin{equation}
    \text{HH-CD}_{m, \lambda} = \frac{(I_{R_m}^{q\omega}-I_{{L}_m}^{q\omega})}{(I_{R_m}^{q\omega}+I_{{L}_m}^{q\omega})}.
    \label{HHCD}
\end{equation}
In the previous work~\cite{Nikitina2023-Nonlinearcirculardi}, the general form of the decomposition of the nonlinear polarization $\mathbf{P}^{2\omega}(\vb r)$~\eqref{eq_shg_baza} has been obtained for an achiral nanostructure with 
$\text{C}_{\mathfrak{n}\text{v}}$ or $\text{D}_{\mathfrak{n}\text{h}}$ symmetry irradiated by a normally incident circularly polarized plane wave.
{To obtain polarization in the general case, two steps are necessary. 
{Firstly, one should describe} the field inside the nanostructure on the fundamental frequency{: its multipolar decomposition contains the functions with $m$ equal to the sum of the total angular momentum projection of the incident field and}
additional momentum projections, $\mathfrak{n}\nu$ 
{(where $\nu \in \mathbb{Z}$)}, 
which are multiples of the rotational symmetry of the nanoparticle {characterized by the number $\mathfrak{n}$}. 
{Secondly, it is required to} rewrite the  
$\hat{\chi}^{(q)}$ tensor in cylindrical coordinates,} as it was done, e.g. in~\cite{Toftul2023-Nonlinearity-Induced, 
Gladyshev2024-FastSimulationofLi} for  $\hat{\chi}^{(2)}$ {or for $\hat{\chi}^{(3)}$ in the Appendix~\ref{sec:app:tens}.}
After that, we get all possible additional momentum projections $m_\chi$ (it is the multiplier which stands in the terms like {$\exp(m_\chi\varphi)$} if one looks at the tensor components written in cylindrical coordinates), as well as differences $\Delta m_\chi=m_\chi-m'_\chi$ between all possible different values.
{Note that the values of the $m_\chi$ are related to the rotational symmetry of the lattice, but in fact, the symmetry of the tensor is in some sense ``higher'' than the symmetry of the lattice itself, being restricted by the tensor transformation properties and its rank. }
Now, due to the fact that the geometry of the problem is the same but nanostructures are irradiated by a vortex beam, we can introduce the same expression as in~\cite{Nikitina2023-Nonlinearcirculardi} with only one generalization: the projection of the momentum of the incident vortex $\pm m^{\text{in}}$ instead of the values $\pm 1$, as it was for the plane wave.
Thus, the decomposition of the nonlinear polarization $\mathbf{P}^{2\omega}(\vb r)$ for the second-harmonic can be written in the cylindrical coordinates $r,z,\varphi$ in the following form:
\begin{align}
    \mathbf{P}^{2\omega}(\mathbf{r}) \propto \!\sum\limits_{{m_\chi},\nu} \eu^{- m_\chi\iu\beta} \bigg[\mathbf{P}^{2\omega}_{(m_{\chi} + 2m^{\text{in}}),\nu}(r,z) \eu^{(m_\chi \pm 2m^{\text{in}} + \mathfrak{n}\nu)\iu\varphi}\bigg], \label{polmitchi2}
\end{align}
{where 
$\mathfrak n$ describes the symmetry group of a nanostructure (C$_{\mathfrak{n}  \text v}$ or D$_{\mathfrak{n} \text h}$), $\nu \in \mathbb Z$, and {$\beta$ is a relative angle between the nanoparticle and its crystalline lattice, as it shown in Fig.~\ref{mainscheme}}. {The values $m_{\chi}$ reflect the lattice symmetry, while the value $m^{\text{in}}$ characterizes the incident vortex}. 
Here we explicitly separate the dependence of the nonlinear polarization $\mathbf{P}^{q\omega}(\mathbf{r})$ on coordinate $\varphi$.
In a similar way, the high-harmonic nonlinear polarization $\mathbf{P}^{q\omega}(\vb r)$ can be written as:
\begin{align}
    \mathbf{P}^{q\omega}(\mathbf{r}) \propto \!\sum\limits_{{m_\chi},\nu} \eu^{- m_\chi\iu\beta} \bigg[\mathbf{P}^{q\omega}_{(m_{\chi} + qm^{\text{in}}),\nu}(r,z) \eu^{(m_\chi \pm qm^{\text{in}} + \mathfrak{n}\nu)\iu\varphi}\bigg]. \label{polmitchiq}
\end{align}
In the equations~\eqref{polmitchi2} and~\eqref{polmitchiq} {we use the proportionality sign to emphasize the symmetrical nature of the phenomenon and the insignificance of constants and other dependencies, such as dependency on wavelength.} 
{Nevertheless, this consideration is justified because}, for determining the multipolar content in an axially symmetric system, we mainly need the dependence on $\varphi$ (See Suppl. Info. in~\cite{Gladyshev2024-FastSimulationofLi}).

\section{Conditions for the high harmonic circular dichroism}
\label{sec:mainformula}
Surprisingly, despite the particular complexity and tremendous construction of intermediate calculations, the final expression for the possibility of circular dichroism can be written in a short and elegant form.
Moreover, the formula applies universally across all cases, regardless of the incident beam, nanostructure symmetry, or high-order of harmonic generation. 
Consequently, it aligns precisely with the one presented in~\cite{Nikitina2023-Nonlinearcirculardi}, and proving {its universality} is the main result of our work. 
The theoretical investigation shows independence of the obtained results on the angular momentum projection absolute value $|m^{\text{in}}|$ of the incident optical vortex. 

Let us here recall that $\Delta m_\chi$ being one of the possible differences between $m_\chi$ of a $\hat\chi^{(q)}$ tensor written in cylindrical coordinates, generally reflects the crystalline lattice symmetry, and $\mathfrak n$ reflects the rotational symmetry of a nanoparticle.
\hehe{Thus, the HH-CD$_{m, \lambda}$ can be obtained if:
\begin{enumerate}
    \item {$\exists \nu \in \mathbb{Z}$ such that $\Delta m_\chi = \nu \mathfrak{n}$, so we introduce the number $s$:
    \begin{equation}
         s = \Delta m_\chi = \nu \mathfrak{n}. \label{rule1}
    \end{equation}}
    \item{The relative angle $\beta$ of a crystalline lattice rotation with respect to nanoparticle is:
    \begin{equation}
        \beta \neq \frac{\pi \nu}{s} \ \ \ \forall \nu \in \mathbb Z. \label{pinu}
    \end{equation}}
    \item {If there are several possible $s$ in~\eqref{rule1}, one should prohibit only those angles  $\beta = \pi \nu/s$ in~\eqref{pinu} that coincide for all $s = \Delta m_\chi$.}
\end{enumerate}}

Note that this formula does not know anything about the value of the total angular momentum of the incident beam, but for the high-harmonic generation, the result can depend on $q$.
The difference will be {in} the nonlinear susceptibility tensor $\hat{\chi}^{(q)}$ and, as a consequence, the values of the difference between the angular momentum projections $\Delta m_\chi$. 
Overall, the introduced rule in~\eqref{rule1} and~\eqref{pinu} can be used for nonlinear circular dichroism in high-harmonic generation in nanostructures with any achiral symmetry and material characterized by the nonlinear susceptibility tensor $\hat{\chi}^{(q)}$ irradiated by any optical vortex. 
We also note that for the second harmonic generation, the maximum value of $\Delta m_\chi$ is equal to 6, thus, according to the rule~\eqref{rule1}, circular dichroism in the second-harmonic signal is possible only for C$_{\mathfrak{n} \text v}$, D$_{\mathfrak{n} \text h}$ symmetries with $\mathfrak{n} \leq 6$. 
In a more general $q$-order harmonic generation case, the maximum value of $\Delta m_\chi$ is equal to $2(q+1)$, thus circular dichroism in the $q$-order harmonic signal is possible only for C$_{\mathfrak{n} \text v}$, D$_{\mathfrak{n} \text h}$ symmetries with
\begin{equation}
    \label{highestn}
    \mathfrak{n} \leq 2(q+1).
\end{equation}
In what follows, we consider some examples to show the variety of cases that may arise when all the intermediate calculations are carefully and rigorously considered and show that they will indeed lead to the same result in the end.

\section{Analysing particular cases}
\label{sec:cases}
To provide a clue as to {how the main formula works} and offer a proof, we investigate specific cases. Through these examinations, we identify patterns that ultimately lead us to derive the general formula.
It is done in the following steps:
\begin{enumerate}
    \item Find the nonlinear polarization induced inside the nanoparticle with a particular symmetry {made of particular material} by a particular vortex. 
    \item Find the symmetry of the eigenmodes of the nanoparticle that can be excited by this nonlinear polarization using the equation~\eqref{efieldsh_anyharmonic}. 
    \item Analyze the phases of the excited modes as well as their dependence on the incident light polarization and angle~$\beta$.
    \item {Calculate the total integral intensity $I^{q\omega}$, where the integral is taken by the sphere in the faf-field. 
    For that, the two-mode approximation when two modes of the same symmetry are excited with a phase $\alpha$ between them can be used.}
    \item Analyze the dependence of the total integral intensity on the sign of the incident beam. 
    If the case of left- and right-handed beams {don't exactly} mirror each other, then the nonlinear circular dichroism exists.
    \item {Find conditions for the appearance of the nonlinear circular dichroism in the particular nanostructure.}
\end{enumerate}
{In what follows, we analyse the second harmonic, but to generalize the results for the higher harmonic, one should substitute $q$ instead of 2, and consider another (in many cases, similar) set of $m_\chi$ values, which could be obtained as in Appendix~\ref{sec:app:tens}.}
\subsection{Second harmonic circular dichroism in GaAs[111]\texorpdfstring{$\parallel$}{Lg}z \texorpdfstring{$\text{C}_{3\text{v}}$}{Lg} nanostructure}
\label{sec:GaAsC3v}
In this section, we show that in the nanostructure GaAs[111]$\parallel$z with symmetry $\text{C}_{3\text{v}}$ nonlinear circular dichroism appears in the second-harmonic signal for angles $\beta \neq \pi\nu/3$, where $\nu \in \mathbb {Z}$, regardless of the absolute value of the angular momentum projection $|m_{\text{in}}|$.

\subsubsection{Nonlinear polarization}
{The second-order} polarization $\mathbf{P}^{2\omega}(\mathbf{r})$ for the nanostructure GaAs[111]$\parallel$z~\eqref{polmitchi2} 
reads as follows~\cite{Nikitina2023-Nonlinearcirculardi}
\begin{align}
    \label{GaAs[111]polnew}
    \mathbf{P}^{2\omega}(r,z,\varphi)\propto
    &\bigg\{\sum\limits_{\nu} \mathbf{P}_{(3+2m^{\text{in}}),\nu}^{2\omega}(r,z)\eu^{\pm (3 + 2m^{\text{in}})\iu\varphi + \mathfrak{n}\nu \iu\varphi}\bigg\}\eu^{\mp 3\iu\beta}+
    \\\nonumber
    + &\bigg\{\sum\limits_{\nu} \mathbf{P}_{(-3+2m^{\text{in}}),\nu}^{2\omega}(r,z)\eu^{\pm (-3+2m^{\text{in}})\iu\varphi + \mathfrak{n}\nu \iu\varphi}\bigg\}\eu^{\pm 3\iu\beta} + 
    \\ \nonumber + 
    &\bigg\{ \sum\limits_{\nu} \mathbf{P}_{2m^{\text{in}},\nu}^{2\omega}(r,z)\eu^{\pm 2m^{\text{in}}\iu\varphi + \mathfrak{n}\nu \iu\varphi}\bigg\},
\end{align}
{because for the $\hat{\chi}^{(2)}$ tensor for GaAs[111]$\parallel$z projections $m_\chi = 0,\pm 3$, as it was shown in the previous work~\cite{Nikitina2023-Nonlinearcirculardi}.} 
We are interested in the group symmetry $\text{C}_{3\text{v}}$, e.g., a trimer of cylinders or a prism with a right triangle at the base, thus the index $\mathfrak{n}$ is equal to $3$.
For convenience, we will call it trimer further. 
{We also note that in $\mathbf{P}_{m\nu}^{2\omega}(r,z)$ 
sign of $\varphi$-component depends on the handedness, but one can check (See Suppl. Info. in~\cite{Frizyuk2021-NonlinearCircularDi}) that it does not alter the general result. }
We omit the notation, which contains this difference.
Nonlinear polarization $\mathbf{P}^{2\omega}(\mathbf{r})$~\eqref{GaAs[111]polnew} excites in the trimer its eigenmodes at the second harmonic frequency $2\omega$.
Depending on the momentum projection $m^{\text{in}}$ of an incident field, the excited modes can be of any possible symmetry, i.e., transformed under all possible irreducible representations $\text{A}_1$, $\text{A}_2$, or $\text{E}$ of a $\text{C}_{3\text{v}}$ group~\cite{Xiong_Xiong_Yang_Yang_Chen_Wang_Xu_Xu_Xu_Liu_2020, Gladyshev_Frizyuk_Bogdanov_2020}.
{Specifically, with the help of coupling integrals 
$\text{D}_j = \int\limits_V \dd V' \mathbf{E}^{2\omega}_{j} \left(\mathbf{r}^{\prime}\right) \mathbf{P}^{2\omega} (\mathbf{r}')$
{(see~\eqref{efieldsh_anyharmonic}),} we will prove that modes transformed under $\text{E}$ can be excited only if $m^{\text{in}}=3s-2$ and $m^{\text{in}}=3s-1$, and modes transformed under $\text{A}_1$ and $\text{A}_2$ can be excited only if $m^{\text{in}}=3s$ where $ s \in \mathbb {Z}_+$. }
\subsubsection{Eigenmode presentations}
For that, let us write down the expansions of 
trimer's eigenmodes transformed under $\text{E}$, and $\text{A}_1$, $\text{A}_2$ in magnetic and electric vector spherical harmonics $\mathbf{M}_{^e_omn}, \mathbf{N}_{^e_omn}$ one by one, {which definition can be found in Appendix~\ref{sec:app:1}}. 
{One should pay attention to the multipolar content of the modes. Importantly, each mode of a particular symmetry consist of the multipoles, which $m$s differ by $\mathfrak n$.}
\paragraph{Eigenmodes, which are transformed under E irrep}
Decomposition of the eigenmodes transformed under $\text{E}$, according to the tables with a given multipolar content of the eigenmodes~\cite{Gladyshev_Frizyuk_Bogdanov_2020, 
Xiong_Xiong_Yang_Yang_Chen_Wang_Xu_Xu_Xu_Liu_2020} consists only of the vector spherical harmonics with the projections of the angular momentum $m=3s-1$ and $m=3s-2$, where $ s \in \mathbb {Z}_+$ with, in general, complex coefficients.
However, we note that $\text{E}$ is a 2-dimensional irreducible representation, thus  there are two orthogonal eigenmodes transformed under this representation. 
{We defer further discussion to Appendix~\ref{sec:app:4:1} and present here the expansions of the two orthogonal modes $\mathbf{E}^{2\omega, x\pm \iu y}_{j} \left(\mathbf{r}^{\prime}\right)$ transformed under $\text{E}$ representation:}
\begin{align}
        \label{modesеtrimernewfinaloutappendix}
        \nonumber
        \mathbf{E}^{2\omega, x\pm \iu y}_{j} \left(\mathbf{r}^{\prime}\right) = \sum_{n}\bigg\{ &\mathbf{E}^{}_{j,1n}(r,z)\eu^{\pm 1\iu\varphi} 
        +  \mathbf{E}^{}_{j,2n}(r,z)\eu^{\mp 2\iu\varphi} + 
        \\ + &\mathbf{E}^{}_{j,4n}(r,z)\eu^{\pm 4\iu\varphi}  + \mathbf{E}^{}_{j,5n}(r,z)\eu^{\mp 5\iu\varphi} 
         +\dots\bigg\},
\end{align}
{where index $j$ denotes a specific eigenmode and coefficients $\mathbf{E}^{}_{j,mn}(r,z)$ contain complex constants used in the expansions of the eigenmodes and all other functions and constants independent on coordinate $\varphi$, originating from expressions for the magnetic $\mathbf{M}_{^e_omn}(r,z)$ and electric $\mathbf{N}_{^e_omn}(r,z)$ vector spherical harmonics. 
Note that detailed derivations can be found in Appendix~\ref{sec:app:4:1}.} 
\paragraph{Eigenmodes, which are transformed under $\text{A}_1$ and $\text{A}_2$ irrep}
{Decompositions of eigenmodes transformed under irreducible representations $\text{A}_1$ and $\text{A}_2$ contain vector spherical harmonics $\mathbf{M}_{^e_o(3s)n}$ and $\mathbf{N}_{^e_o(3s)n}$, where $ s \in \mathbb {Z}_+$.} Eigenmodes transformed under $\text{A}_1$ are even under reflection in the plane XZ (for a particular orientation of a trimer), while eigenmodes transformed under $\text{A}_2$ are odd.  
{Following the same method used for eigenmodes transformed under $\text{E}$ representation, we can derive expressions for eigenmodes transformed under $\text{A}_1$ and $\text{A}_2$ representations. 
They will take the following form:}
    \begin{align}
        \label{modesA1A2finalA1} 
        {\mathbf{E}^{2\omega, \text{A}_1}_{j} \left(\mathbf{r}^{\prime}\right)}= 
         \sum_{n} \bigg\{[&{E}^{r}_{j,e3n} \left(r,z\right)\hat{\mathbf{e}}_r + {E}^z_{j,e3n} \left(r,z\right)\hat{\mathbf{e}}_z]\cos{3\varphi} +\\\nonumber + &{E}^{\varphi}_{j,e3n} \left(r,z\right)\hat{\mathbf{e}}_\varphi\sin{3\varphi} + \dots\bigg\},
    \end{align}
    \begin{align}
        \label{modesA1A2finalA2}
        {\mathbf{E}^{2\omega, \text{A}_2}_{j} \left(\mathbf{r}^{\prime}\right)}= \sum_{n} \bigg\{[&{E}^{r}_{j,o3n} \left(r,z\right)\hat{\mathbf{e}}_r + {E}^z_{j,o3n} \left(r,z\right)\hat{\mathbf{e}}_z]\sin{3\varphi} +  \\\nonumber +
         &{E}^{\varphi}_{j,o3n} \left(r,z\right)\hat{\mathbf{e}}_\varphi\cos{3\varphi} + \dots\bigg\},
    \end{align}
{where coefficients 
${E}^{r}_{j,^e_o3n}\left(r,z\right)$,  
${E}^{z}_{j,^e_o3n}\left(r,z\right)$,  
${E}^{\varphi}_{j,^e_o3n}\left(r,z\right)$ 
contain all functions and constants independent on coordinate $\varphi$. 
{For transparency, we have included all intermediate steps of the calculations relevant to this case in Appendix~\ref{sec:app:4:1}.}}

\subsubsection{{Coupling integrals}}
Next, we calculate coupling integrals 
$\text{D}_j = \int\limits_{\text{C}_{3\text{v}}} \dd V' \mathbf{E}^{2\omega, x\pm\iu y}_{j} \left(\mathbf{r}^{\prime}\right) \mathbf{P}^{2\omega} (\mathbf{r}')$ 
for all eigenmodes to understand which of them are excited.
\paragraph{Eigenmodes, which are transformed under $\text{E}$ irrep}
Let us use expressions~\eqref{modesеtrimernewfinaloutappendix} for eigenmodes 
$\mathbf{E}^{2\omega, x\pm\iu y}_{j} \left(\mathbf{r}^{\prime}\right)$ 
and decomposition of nonlinear polarization 
$\mathbf{P}^{2\omega}(\mathbf{r})$~\eqref{GaAs[111]polnew}
to calculate the coupling integrals 
$\text{D}_j$. 
Overall, we get four distinct coupling integrals $\text{D}_j$. 
In particular, there are two integrals for eigenmode $\mathbf{E}^{2\omega, x+\iu y}_{j} \left(\mathbf{r}^{\prime}\right)$ 
excited by left- and right-handed circularly polarized beam:
\begin{align}
    \label{integralbig1}
    \ket{L_m}\colon \; &\text{D}_j = \int\limits_{\text{C}_{3\text{v}}} \dd V' \mathbf{E}^{2\omega, x+\iu y}_{j} \left(\mathbf{r}^{\prime}\right) \mathbf{P}^{2\omega} (\mathbf{r}^{\prime}) \propto 
    \\\nonumber & \propto \int\limits_{\text{C}_{3\text{v}}} \dd V' \sum_{n,\nu}\bigg(\bigg[\mathbf{E}_{j,1n}\eu^{+1\iu\varphi} + \mathbf{E}_{j,2n}\eu^{- 2\iu\varphi} 
     + \mathbf{E}_{j,4n}\eu^{+4\iu\varphi} + \mathbf{E}_{j,5n}\eu^{- 5\iu\varphi} +\dots\bigg]\cdot
    \\\nonumber & \cdot\bigg[ \bigg\{\mathbf{P}_{{(3+2m^{\text{in}})},\nu}^{2\omega}\eu^{+ {(3+2m^{\text{in}})}\iu\varphi + 3\nu \iu\varphi}\bigg\}\eu^{- 3\iu\beta} + \\\nonumber & + \bigg\{
    \mathbf{P}_{{(-3+2m^{\text{in}})},\nu}^{2\omega}\eu^{+ {(-3+2m^{\text{in}})}\iu\varphi + 3\nu \iu\varphi}\bigg\}\eu^{+ 3\iu\beta} + 
    \\\nonumber & + \bigg\{\mathbf{P}_{{2m^{\text{in}}},\nu}^{2\omega}\eu^{+ {2m^{\text{in}}}\iu\varphi + 3\nu \iu\varphi}\bigg\}\bigg]\bigg),
\end{align}
\begin{align}
    \label{integralbig2}
    \ket{R_m}\colon \; 
    & \text{D}_j = \int\limits_{\text{C}_{3\text{v}}} \dd V' \mathbf{E}^{2\omega, x+\iu y}_{j} \left(\mathbf{r}^{\prime}\right) \mathbf{P}^{2\omega} (\mathbf{r}^{\prime}) \propto 
    \\\nonumber 
    & \propto \int\limits_{\text{C}_{3\text{v}}} \dd V' \sum_{n,\nu}\bigg(\bigg[\mathbf{E}_{j,1n}\eu^{+1\iu\varphi} + \mathbf{E}_{j,2n}\eu^{- 2\iu\varphi} + 
     \mathbf{E}_{j,4n}\eu^{+4\iu\varphi} + \mathbf{E}_{j,5n}\eu^{- 5\iu\varphi} +\dots\bigg]\cdot
    \\\nonumber 
    &\cdot\bigg[ \bigg\{\mathbf{P}_{\blue{(3+2m^{\text{in}})},\nu}^{2\omega}\eu^{- \blue{(3+2m^{\text{in}})}\iu\varphi + 3\nu \iu\varphi}\bigg\}\eu^{+ 3\iu\beta} + 
    \\\nonumber 
    & + \bigg\{
    \mathbf{P}_{{(-3+2m^{\text{in}})},\nu}^{2\omega}\eu^{- \blue{(-3+2m^{\text{in}})}\iu\varphi + 3\nu \iu\varphi}\bigg\}\eu^{- 3\iu\beta} + 
    \\\nonumber 
    & + \bigg\{\mathbf{P}_{\blue{2m^{\text{in}}},\nu}^{2\omega}\eu^{- \blue{2m^{\text{in}}}\iu\varphi + 3\nu \iu\varphi}\bigg\}\bigg]\bigg),
\end{align}
and there are two other integrals for eigenmode $\mathbf{E}^{2\omega, x-\iu y}_{j} \left(\mathbf{r}^{\prime}\right)$ excited by left- and right-handed circularly polarized beams:
\begin{align}
    \label{integralbig3}
    \ket{L_m}\colon \; 
    &\text{D}_j = \int\limits_{\text{C}_{3\text{v}}} \dd V' \mathbf{E}^{2\omega, x-\iu y}_{j} \left(\mathbf{r}^{\prime}\right) \mathbf{P}^{2\omega} (\mathbf{r}^{\prime}) \propto 
    \\\nonumber & \propto \int\limits_{\text{C}_{3\text{v}}} \dd V' \sum_{n,\nu}\bigg(\bigg[\mathbf{E}_{j,1n}\eu^{-1\iu\varphi} + \mathbf{E}_{j,2n}\eu^{+ 2\iu\varphi} + 
     \mathbf{E}_{j,4n}\eu^{-4\iu\varphi} + \mathbf{E}_{j,5n}\eu^{+ 5\iu\varphi} +\dots\bigg]\cdot
    \\\nonumber & \cdot\bigg[ \bigg\{\mathbf{P}_{\blue{(3+2m^{\text{in}})},\nu}^{2\omega}\eu^{+ \blue{(3+2m^{\text{in}})}\iu\varphi + 3\nu \iu\varphi}\bigg\}\eu^{- 3\iu\beta} + 
    \\\nonumber & + \bigg\{
    \mathbf{P}_{{(-3+2m^{\text{in}})},\nu}^{2\omega}\eu^{+ \blue{(-3+2m^{\text{in}})}\iu\varphi + 3\nu \iu\varphi}\bigg\}\eu^{+ 3\iu\beta} + 
    \\\nonumber & + \bigg\{\mathbf{P}_{\blue{2m^{\text{in}}},\nu}^{2\omega}\eu^{+ {2m^{\text{in}}}\iu\varphi + 3\nu \iu\varphi}\bigg\}\bigg]\bigg),
\end{align}
\begin{align}
    \label{integralbig4}
    \ket{R_m}\colon \; 
    &\text{D}_j = \int\limits_{\text{C}_{3\text{v}}} \dd V' \mathbf{E}^{2\omega, x-\iu y}_{j} \left(\mathbf{r}^{\prime}\right) \mathbf{P}^{2\omega} (\mathbf{r}^{\prime}) \propto 
    \\\nonumber & \propto \int\limits_{\text{C}_{3\text{v}}} \dd V' \sum_{n,\nu}\bigg(\bigg[\mathbf{E}_{j,1n}\eu^{-1\iu\varphi} + \mathbf{E}_{j,2n}\eu^{+ 2\iu\varphi} + 
     \mathbf{E}_{j,4n}\eu^{-4\iu\varphi} + \mathbf{E}_{j,5n}\eu^{+ 5\iu\varphi} +\dots\bigg]\cdot
    \\\nonumber & \cdot\bigg[ \bigg\{\mathbf{P}_{{(3+2m^{\text{in}})},\nu}^{2\omega}\eu^{- {(3+2m^{\text{in}})}\iu\varphi + 3\nu \iu\varphi}\bigg\}\eu^{+ 3\iu\beta} + 
    \\\nonumber & + \bigg\{
    \mathbf{P}_{\blue{(-3+2m^{\text{in}})},\nu}^{2\omega}\eu^{- \blue{(-3+2m^{\text{in}})}\iu\varphi + 3\nu \iu\varphi}\bigg\}\eu^{- 3\iu\beta} + 
    \\\nonumber & + \bigg\{\mathbf{P}_{\blue{2m^{\text{in}}},\nu}^{2\omega}\eu^{- \blue{2m^{\text{in}}}\iu\varphi + 3\nu \iu\varphi}\bigg\}\bigg]\bigg).
\end{align}
\paragraph{Eigenmodes, which are transformed under $\text{A}_1$ and $\text{A}_2$ irrep}
{Next, let us use expressions~\eqref{modesA1A2finalA1} and~\eqref{modesA1A2finalA2} for eigenmodes $\mathbf{E}^{2\omega, \text{A}_1}_{j} \left(\mathbf{r}^{\prime}\right)$ and $\mathbf{E}^{2\omega, \text{A}_2}_{j} \left(\mathbf{r}^{\prime}\right)$ and again the decomposition of nonlinear polarization $\mathbf{P}^{2\omega}(\mathbf{r})$~\eqref{GaAs[111]polnew} to calculate the coupling integrals $\text{D}_j$ for this case. 
There are two integrals for eigenmode $\mathbf{E}^{2\omega, \text{A}_1}_{j} \left(\mathbf{r}^{\prime}\right)$ excited by left- and right-handed circularly polarized beams:}
\begin{align}
    \label{integralbig5}
     \ket{L_m}\colon & \; \text{D}_j = \int\limits_{\text{C}_{3\text{v}}} \dd V' \mathbf{E}^{2\omega, \text{A}_1}_{j} \left(\mathbf{r}^{\prime}\right) \mathbf{P}^{2\omega} (\mathbf{r}^{\prime}) \propto
     \\\nonumber \propto   \int\limits_{\text{C}_{3\text{v}}} \dd V' &\sum_{n,\nu}\bigg(\bigg[\left[{E}^{r}_{j,e3n} \left(r,z\right)\hat{\mathbf{e}}_r + {E}^z_{j,e3n} \left(r,z\right)\hat{\mathbf{e}}_z\right]
     \cos{3\varphi} +
      {E}^{\varphi}_{j,e3n} \left(r,z\right)\hat{\mathbf{e}}_\varphi\sin{3\varphi} + \dots\bigg]\cdot
     \\\nonumber & \cdot\bigg[ \bigg\{\mathbf{P}_{\blue{(3+2m^{\text{in}})},\nu}^{2\omega}\eu^{+ \blue{(3+2m^{\text{in}})}\iu\varphi + 3\nu \iu\varphi}\bigg\}\eu^{- 3\iu\beta} + 
     \\\nonumber & + \bigg\{\mathbf{P}_{\blue{(-3+2m^{\text{in}})},\nu}^{2\omega}\eu^{+ \blue{(-3+2m^{\text{in}})}\iu\varphi + 3\nu \iu\varphi}\bigg\}\eu^{+ 3\iu\beta} + 
     \\\nonumber & + \bigg\{\mathbf{P}_{\blue{2m^{\text{in}}},\nu}^{2\omega}\eu^{+ \blue{2m^{\text{in}}}\iu\varphi + 3\nu \iu\varphi}\bigg\}\bigg]\bigg),
\end{align}
\begin{align}
     \label{integralbig6}
     \ket{R_m}\colon & \; \text{D}_j = \int\limits_{\text{C}_{3\text{v}}} \dd V' \mathbf{E}^{2\omega, \text{A}_1}_{j} \left(\mathbf{r}^{\prime}\right) \mathbf{P}^{2\omega} (\mathbf{r}^{\prime}) \propto
     \\\nonumber  \propto\int\limits_{\text{C}_{3\text{v}}} \dd V' &\sum_{n,\nu}\bigg(\bigg[\left[{E}^{r}_{j,e3n} \left(r,z\right)\hat{\mathbf{e}}_r + {E}^z_{j,e3n} \left(r,z\right)\hat{\mathbf{e}}_z\right]\cos{3\varphi} +
     {E}^{\varphi}_{j,e3n} \left(r,z\right)\hat{\mathbf{e}}_\varphi\sin{3\varphi} + \dots\bigg]\cdot
     \\\nonumber &\cdot\bigg[ \bigg\{\mathbf{P}_{\blue{(3+2m^{\text{in}})},\nu}^{2\omega}\eu^{- \blue{(3+2m^{\text{in}})}\iu\varphi + 3\nu \iu\varphi}\bigg\}\eu^{+ 3\iu\beta} + 
     \\\nonumber & + \bigg\{
     \mathbf{P}_{\blue{(-3+2m^{\text{in}})},\nu}^{2\omega}\eu^{- \blue{(-3+2m^{\text{in}})}\iu\varphi + 3\nu \iu\varphi}\bigg\}\eu^{- 3\iu\beta} + 
     \\\nonumber & + \bigg\{\mathbf{P}_{\blue{2m^{\text{in}}},\nu}^{2\omega}\eu^{- \blue{2m^{\text{in}}}\iu\varphi + 3\nu \iu\varphi}\bigg\}\bigg]\bigg),
\end{align}
{and there are two other integrals for the eigenmode $\mathbf{E}^{2\omega, \text{A}_2}_{j} \left(\mathbf{r}^{\prime}\right)$ excited by left- and right-handed circularly polarized beams:}
\begin{align}
     \label{integralbig7}
     \ket{L_m}\colon & \; \text{D}_j = \int\limits_{\text{C}_{3\text{v}}} \dd V' \mathbf{E}^{2\omega, \text{A}_2}_{j} \left(\mathbf{r}^{\prime}\right) \mathbf{P}^{2\omega} (\mathbf{r}^{\prime}) \propto
     \\\nonumber \propto \int\limits_{\text{C}_{3\text{v}}} \dd V' &\sum_{n,\nu}\bigg(\bigg[\left[{E}^{r}_{j,o3n} \left(r,z\right)\hat{\mathbf{e}}_r + {E}^z_{j,o3n} \left(r,z\right)\hat{\mathbf{e}}_z\right]\sin{3\varphi} + {E}^{\varphi}_{j,o3n} \left(r,z\right)\hat{\mathbf{e}}_\varphi\cos{3\varphi} + \dots\bigg]\cdot\\\nonumber&\cdot\bigg[ \bigg\{\mathbf{P}_{\blue{(3+2m^{\text{in}})},\nu}^{2\omega}\eu^{+ \blue{(3+2m^{\text{in}})}\iu\varphi + 3\nu \iu\varphi}\bigg\}\eu^{- 3\iu\beta} + 
     \\\nonumber & + \bigg\{\mathbf{P}_{\blue{(-3+2m^{\text{in}})},\nu}^{2\omega}\eu^{+ \blue{(-3+2m^{\text{in}})}\iu\varphi + 3\nu \iu\varphi}\bigg\}\eu^{+ 3\iu\beta} + 
     \\\nonumber & + \bigg\{\mathbf{P}_{\blue{2m^{\text{in}}},\nu}^{2\omega}\eu^{+ \blue{2m^{\text{in}}}\iu\varphi + 3\nu \iu\varphi}\bigg\}\bigg]\bigg),
\end{align}
\begin{align}
      \label{integralbig8}
      \ket{R_m}\colon& \; \text{D}_j = \int\limits_{\text{C}_{3\text{v}}}\!\dd V' \mathbf{E}^{2\omega, \text{A}_2}_{j} \left(\mathbf{r}^{\prime}\right) \mathbf{P}^{2\omega} (\mathbf{r}^{\prime}) \propto
      \\\nonumber \propto \int\limits_{\text{C}_{3\text{v}}}\!  \dd V' &\sum_{n,\nu}\! \bigg(\! \bigg[\! \left[{E}^{r}_{j,o3n}\!  \left(r,z\right)\hat{\mathbf{e}}_r + {E}^z_{j,o3n}\!  \left(r,z\right)\hat{\mathbf{e}}_z\right]\sin{3\varphi} +
      {E}^{\varphi}_{j,o3n} \left(r,z\right)\hat{\mathbf{e}}_\varphi\cos{3\varphi} + \dots\bigg]\cdot
      \\\nonumber&\cdot\bigg[ \bigg\{\mathbf{P}_{\blue{(3+2m^{\text{in}})},\nu}^{2\omega}\eu^{- \blue{(3+2m^{\text{in}})}\iu\varphi + 3\nu \iu\varphi}\bigg\}\eu^{+ 3\iu\beta} + 
      \\\nonumber & + \bigg\{
      \mathbf{P}_{\blue{(-3+2m^{\text{in}})},\nu}^{2\omega}\eu^{- \blue{(-3+2m^{\text{in}})}\iu\varphi + 3\nu \iu\varphi}\bigg\}\eu^{- 3\iu\beta} + 
      \\\nonumber & + \bigg\{\mathbf{P}_{\blue{2m^{\text{in}}},\nu}^{2\omega}\eu^{- \blue{2m^{\text{in}}}\iu\varphi + 3\nu \iu\varphi}\bigg\}\bigg]\bigg).
\end{align}

\subsubsection{Excitation of the modes by different angular momentum}
{Let us find out, which of the expressions 
\eqref{integralbig1}, \eqref{integralbig2}, \eqref{integralbig3}, \eqref{integralbig4},~\eqref{integralbig5},~\eqref{integralbig6},~\eqref{integralbig7}, and~\eqref{integralbig8} for coupling integrals $\text{D}_j$ give nontrivial answer}. 
For the integral to be non-zero, the integrand should be transformed under the trivial irreducible representation, i.e. $\text{A}_1$ irrep of $\text{C}_{3\text{v}}$ group~\cite{Landau}. 
In the group $\text{C}_{3\text{v}}$, if the integrand contains exponents $e^{\pm 3\iu s \varphi}$, where $ s \in \mathbb {Z}_+$, then the integral expression can be non-zero. Given the above, {there will be three different cases dependent on the value of the index $m^{\text{in}}$ of incident vortex:}
\begin{enumerate}
\item {$m^{\text{in}}=3s$, where $ s \in \mathbb {Z}_+$}
    \paragraph{Eigenmodes, which are transformed under $\text{E}$ irrep} 
    {Coupling integrals~\eqref{integralbig1} and~\eqref{integralbig2} for the eigenmode $\mathbf{E}^{2\omega, x+\iu y}_{j} \left(\mathbf{r}^{\prime}\right)$, as well as~\eqref{integralbig3} and~\eqref{integralbig4} for the eigenmode $\mathbf{E}^{2\omega, x-\iu y}_{j} \left(\mathbf{r}^{\prime}\right)$ can only give zero. Thus, modes transformed under $\text{E}$ irrep aren't excited.}
    \paragraph{{Eigenmodes, which are transformed under $\text{A}_1$ and $\text{A}_2$ irrep}}
    {Coupling integrals~\eqref{integralbig5},~\eqref{integralbig6} for the eigenmode $\mathbf{E}^{2\omega, \text{A}_1}_{j} \left(\mathbf{r}^{\prime}\right)$ give nonzero results:}
    \begin{align}
        \label{integralA1}
        & \ket{L_m}\colon \; \text{D}_j \propto c_j^{\text{A}_1} \eu^{-3\iu\beta} + c_j^{\prime \text{A}_1} \eu^{+3\iu\beta} + c_j^{\prime\prime \text{A}_1},
        \\\nonumber
        & \ket{R_m}\colon\; \text{D}_j \propto c_j^{\text{A}_1} \eu^{+3\iu\beta} + c_j^{\prime \text{A}_1} \eu^{-3\iu\beta} + c_j^{\prime\prime \text{A}_1},
    \end{align}
    {in turn, coupling integrals~\eqref{integralbig7} and~\eqref{integralbig8} for the eigenmode $\mathbf{E}^{2\omega, \text{A}_2}_{j} \left(\mathbf{r}^{\prime}\right)$ also give nonzero results:}
    \begin{align}
        \label{integralA2}
        &\ket{L_m}\colon  \; \text{D}_j \propto c_j^{\text{A}_2} \eu^{-3\iu\beta} + c_j^{\prime \text{A}_2} \eu^{+3\iu\beta} + c_j^{\prime\prime \text{A}_2},
        \\\nonumber
        &\ket{R_m}\colon \; \text{D}_j \propto c_j^{\text{A}_2} \eu^{+3\iu\beta} + c_j^{\prime \text{A}_2} \eu^{-3\iu\beta} + c_j^{\prime\prime \text{A}_2},
    \end{align}
    {where coefficients $c_j^{\text{A}_1}, c_j^{\prime \text{A}_1}, c_j^{\prime\prime \text{A}_1}$, and $c_j^{\text{A}_2}, c_j^{\prime \text{A}_2}, c_j^{\prime\prime \text{A}_2}$ were introduced, and, importantly, they are equal for different polarizations.}
    {Substituting the expression~\eqref{integralA1} for the coupling integral $\text{D}_j$ with the eigenmode $\mathbf{E}^{2\omega, \text{A}_1}_{j} \left(\mathbf{r}^{\prime}\right)$ into the decomposition of the second harmonic electric field~\eqref{efieldsh_anyharmonic}: }
    \begin{align}
        \label{integralA1Efield}
        &\ket{L_m}\colon \; \mathbf{E}^{2\omega}({\mathbf{r}}) \propto 
        \sum_{j} \mathbf{E}^{2\omega, \text{A}_1}_{j}(\mathbf{r}) \left(c_j^{\text{A}_1} \eu^{-3\iu\beta} + c_j^{\prime\text{A}_1} \eu^{+3\iu\beta} + c_j^{\prime\prime\text{A}_1}\right),
        \\\nonumber
        &\ket{R_m}\colon \;\mathbf{E}^{2\omega}({\mathbf{r}}) \propto 
         \sum_{j} \mathbf{E}^{2\omega, \text{A}_1}_{j}(\mathbf{r}) \left(c_j^{\text{A}_1} \eu^{+3\iu\beta} + c_j^{\prime\text{A}_1} \eu^{-3\iu\beta} + c_j^{\prime\prime\text{A}_1}\right),
    \end{align}
    {and the expression~\eqref{integralA2} for the coupling integral $\text{D}_j$ with the eigenmode $\mathbf{E}^{2\omega, \text{A}_2}_{j} \left(\mathbf{r}^{\prime}\right)$:}
    \begin{align}
        \label{integralA2Efield}
        &\ket{L_m}\colon  \; \mathbf{E}^{2\omega}({\mathbf{r}}) \propto 
         \sum_{j} \mathbf{E}^{2\omega, \text{A}_2}_{j}(\mathbf{r}) \left(c_j^{\text{A}_2} \eu^{-3\iu\beta} + c_j^{\prime\text{A}_2} \eu^{+3\iu\beta} + c_j^{\prime\prime\text{A}_2}\right),
        \\\nonumber
        &\ket{R_m}\colon \;\mathbf{E}^{2\omega}({\mathbf{r}}) \propto 
        \sum_{j} \mathbf{E}^{2\omega, \text{A}_2}_{j}(\mathbf{r}) \left(c_j^{\text{A}_2} \eu^{+3\iu\beta} + c_j^{\prime\text{A}_2} \eu^{-3\iu\beta} + c_j^{\prime\prime\text{A}_2}\right).
    \end{align}
    {Thus, modes transformed under $\text{A}_1$ and $\text{A}_2$ irrep are excited and the fields are of the form~\eqref{integralA1Efield} and~\eqref{integralA2Efield}.} {Note that total electric field $\mathbf{E}^{2\omega}({\mathbf{r}})$ is the sum of all excited eigenmodes transformed under different irreducible representations but in equations above~\eqref{integralA1Efield} and~\eqref{integralA2Efield}, as well as below we write only sum of eigenmodes transformed under the particular irreducible representation that we consider at the moment. We can do it because only modes with the same symmetry can interfere with each other, thus at one moment we can pay attention only to eigenmodes transformed under the same irrep in order to avoid unnecessarily long formulae.}
\item {$m^{\text{in}}=3s-2$, where $ s \in \mathbb {Z}_+$}   
    \paragraph{Eigenmodes, which are transformed under $\text{E}$ irrep} 
    {Coupling integrals~\eqref{integralbig1} and~\eqref{integralbig2} with the eigenmode $\mathbf{E}^{2\omega, x+\iu y}_{j} \left(\mathbf{r}^{\prime}\right)$ give the values:}
    \begin{align}
        \label{integral1}
        &\ket{L_m}\colon \; \text{D}_j \propto c_j^{+} \eu^{-3\iu\beta} + c_j^{\prime+} \eu^{+3\iu\beta} + c_j^{\prime\prime+},
        \\\nonumber
        &\ket{R_m}\colon\; \text{D}_j = 0,
    \end{align}
    {where coefficients $c_j^{+}, c_j^{\prime+}, c_j^{\prime\prime+}$ were introduced. 
    Analogously, expressions~\eqref{integralbig3} and~\eqref{integralbig4} with the eigenmode $\mathbf{E}^{2\omega, x-\iu y}_{j} \left(\mathbf{r}^{\prime}\right)$ take the form:}
    \begin{align}
        \label{integral2}
        &\ket{L_m}\colon \; \text{D}_j = 0,
        \\\nonumber
        &\ket{R_m}\colon \;\text{D}_j \propto c_j^{-} \eu^{+3\iu\beta} + c_j^{\prime-} \eu^{-3\iu\beta} + c_j^{\prime\prime-},
    \end{align}
    {where new coefficients $c_j^{-}, c_j^{\prime-}, c_j^{\prime\prime-}$ equal {up to a sign} 
    to the old coefficients $c_j^{+}, c_j^{\prime+}, c_j^{\prime\prime+}$ from the expression~\eqref{integral1}, respectively. 
    It can be proved by changing the $\varphi \to -\varphi$ in the expression~\eqref{integralbig4}. Next, we substitute the obtained expression~\eqref{integral1} with the eigenmode $\mathbf{E}^{2\omega, x+\iu y}_{j} \left(\mathbf{r}^{\prime}\right)$ into the decomposition of the second harmonic electric field~\eqref{efieldsh_anyharmonic}:}
    \begin{align}
        \label{integral1field}
        &\ket{L_m}\colon  \; \mathbf{E}^{2\omega}({\mathbf{r}}) \propto 
         \sum_{j} \mathbf{E}^{2\omega, x+\iu y}_{j}(\mathbf{r}) \left(c_j^{+} \eu^{-3\iu\beta} + c_j^{\prime+} \eu^{+3\iu\beta} + c_j^{\prime\prime+}\right),
        \\\nonumber
        &\ket{R_m}\colon \;\mathbf{E}^{2\omega}({\mathbf{r}}) = 0,
    \end{align}
    {as well as expression~\eqref{integral2} for the coupling integral $\text{D}_j$ with the eigenmode $\mathbf{E}^{2\omega, x-\iu y}_{j} \left(\mathbf{r}^{\prime}\right)$:}
    \begin{align}
        \label{integral2field}
        &\ket{L_m}\colon  \; \mathbf{E}^{2\omega}({\mathbf{r}})  = 0,
        \\\nonumber
        &\ket{R_m}\colon \;\mathbf{E}^{2\omega}({\mathbf{r}}) \propto 
         \sum_{j} \mathbf{E}^{2\omega, x-\iu y}_{j}(\mathbf{r}) \left(c_j^{+} \eu^{+3\iu\beta} + c_j^{\prime+} \eu^{-3\iu\beta} + c_j^{\prime\prime+}\right).
    \end{align}
    {Thus, modes transformed under $\text{E}$ irrep are excited and their fields have the forms~\eqref{integral1field} and~\eqref{integral2field}.}
    \paragraph{{Eigenmodes, which are transformed under $\text{A}_1$ and $\text{A}_2$ irrep}}
    {Coupling integrals~\eqref{integralbig5} and~\eqref{integralbig6} for the eigenmode $\mathbf{E}^{2\omega, \text{A}_1}_{j} \left(\mathbf{r}^{\prime}\right)$, and~\eqref{integralbig7} and~\eqref{integralbig8} for the eigenmode $\mathbf{E}^{2\omega, \text{A}_2}_{j} \left(\mathbf{r}^{\prime}\right)$ give only zero answers. Thus, modes transformed under these irreps aren't excited.}
\item {$m^{\text{in}}=3s-1$, where $ s \in \mathbb {Z}_+$}
    \paragraph{Eigenmodes, which are transformed under $\text{E}$ irrep}
    {Coupling integrals~\eqref{integralbig1} and~\eqref{integralbig2} for the eigenmode $\mathbf{E}^{2\omega, x+\iu y}_{j} \left(\mathbf{r}^{\prime}\right)$ give nonzero values:}
    \begin{align}
        \label{integral3}
        &\ket{L_m}\colon \; \text{D}_j = 0,
        \\\nonumber
        &\ket{R_m}\colon\; \text{D}_j \propto \tilde{c}_j^{+} \eu^{+3\iu\beta} + \tilde{c}_j^{\prime+} \eu^{-3\iu\beta} + \tilde{c}_j^{\prime\prime+}.
    \end{align}
    {Analogously, expressions~\eqref{integralbig3} and~\eqref{integralbig4} for the eigenmode $\mathbf{E}^{2\omega, x-\iu y}_{j} \left(\mathbf{r}^{\prime}\right)$ take the form:}
    \begin{align}
        \label{integral4}
        &\ket{L_m}\colon  \; \text{D}_j \propto \tilde{c}_j^{-} \eu^{-3\iu\beta} + \tilde{c}_j^{\prime-} \eu^{+3\iu\beta} + \tilde{c}_j^{\prime\prime-},
        \\\nonumber
        &\ket{R_m}\colon\;\text{D}_j = 0,
    \end{align}
    {where new coefficients $\tilde{c}_j^{-}, \tilde{c}_j^{\prime-}, \tilde{c}_j^{\prime\prime-}$ equal up to a sign to the old coefficients $\tilde{c}_j^{+}, \tilde{c}_j^{\prime+}, \tilde{c}_j^{\prime\prime+}$ from the expression~\eqref{integral3}, {due to the same reasons as for coefficients $c_{j}^{\pm}$}. 
    Next, we substitute the obtained expression~\eqref{integral3} for the coupling integral $\text{D}_j$ with the eigenmode $\mathbf{E}^{2\omega, x+\iu y}_{j} \left(\mathbf{r}^{\prime}\right)$ into~\eqref{efieldsh_anyharmonic}:}
    \begin{align}
        \label{integral4field} 
        &\ket{L_m}\colon \; \mathbf{E}^{2\omega}({\mathbf{r}})  = 0,\\\nonumber
        &\ket{R_m}\colon \;\mathbf{E}^{2\omega}({\mathbf{r}}) \propto 
         \sum_{j} \mathbf{E}^{2\omega, x+\iu y}_{j}(\mathbf{r}) \left(\tilde{c}_j^{+} \eu^{+3\iu\beta} + \tilde{c}_j^{\prime+} \eu^{-3\iu\beta} + \tilde{c}_j^{\prime\prime+}\right).
    \end{align}
    {and for the coupling integral $\text{D}_j$~\eqref{integral4} with the eigenmode $\mathbf{E}^{2\omega, x-\iu y}_{j} \left(\mathbf{r}^{\prime}\right)$:}
    \begin{align}
        \label{integral3field}
        &\ket{L_m}\colon  \; \mathbf{E}^{2\omega}({\mathbf{r}}) \propto 
         \sum_{j} \mathbf{E}^{2\omega, x-\iu y}_{j}(\mathbf{r}) \left(\tilde{c}_j^{+} \eu^{-3\iu\beta} + \tilde{c}_j^{\prime+} \eu^{+3\iu\beta} + \tilde{c}_j^{\prime\prime+}\right),
        \\\nonumber
        &\ket{R_m}\colon \;\mathbf{E}^{2\omega}({\mathbf{r}}) = 0,
    \end{align}
    {Thus, modes transformed under $\text{E}$ have the form~\eqref{integral4field} and~\eqref{integral3field}.}
    \paragraph{{Eigenmodes, which are transformed under $\text{A}_1$ and $\text{A}_2$ irrep}}
    {Coupling integrals~\eqref{integralbig5} and~\eqref{integralbig6} with the eigenmode $\mathbf{E}^{2\omega, \text{A}_1}_{j} \left(\mathbf{r}^{\prime}\right)$, and~\eqref{integralbig7} and~\eqref{integralbig8} with the eigenmode $\mathbf{E}^{2\omega, \text{A}_2}_{j} \left(\mathbf{r}^{\prime}\right)$ give zero. Thus, modes transformed under these irreps aren't excited.}
\end{enumerate}

{To summarize, dependent on the value of the projection of the angular momentum $m^{\text{in}}$ of the incident vortex we have three cases, but despite this we can consider only one of them because electric field 
$\mathbf{E}^{2\omega}({\mathbf{r}})$ 
always looks the same, i.e., contains three items proportional to the exponents 
$\eu^{\pm 3\iu \beta}$, $\eu^{\mp 3\iu \beta}$, or just constant, where the sign depends on the sign of the polarization of the incident vortex. 
We can relate these terms to the terms in polarization~\eqref{GaAs[111]polnew}. 
\subsubsection{Total integral second harmonic intensity}
\label{subsec:intensity}
We calculate the total integral second harmonic intensity $I^{2\omega}$ in a two modes approximation for the second case, where $m^{\text{in}}=3s-2$, and $s \in \mathbb {Z}_+$. {This means that we use decompositions of the fields~\eqref{integral1field} and~\eqref{integral2field} and take a sum only of two modes with indexes $i$ and $j$.}
{Thus, according to the equations, a left-handed beam excites the eigenmodes $ \mathbf{E}^{2\omega,x+\iu y}_{j}({\mathbf{r}})$, while a right-handed beam excites the eigenmodes $ \mathbf{E}^{2\omega,x-\iu y}_{j}({\mathbf{r}})$, and the total integral second harmonic intensity $I^{2\omega}$ for the left- and right-handed beams can be written in the following form:}
\begin{align}
    \label{intensityLCP}
    &I_{{\ket{L_m}},{\ket{R_m}}}^{2\omega} \propto
      \int_{\text{sph}} \dd V \bigg|\mathbf{E}^{2\omega, x\pm\iu y}_{j}(\mathbf{r})\left(a_j^{+} \eu^{\mp 3\iu\beta} + a_j^{\prime+} \eu^{\pm 3\iu\beta} + a_j^{\prime\prime+}\right) + 
    \\\nonumber &+ \eu^{\iu \alpha}\mathbf{E}^{2\omega, x\pm\iu y}_{i}(\mathbf{r})\left(a_i^{+} \eu^{-3\iu\beta} + a_i^{\prime+} \eu^{+3\iu\beta} + a_i^{\prime\prime+}\right)\bigg|^2, 
\end{align}  
{where the integral is taken over the big sphere in the far-field, containing the nanostructure.}
Frequency-dependent phase $\alpha$ between the eigenmodes $\mathbf{E}^{2\omega,x\pm\iu y}_{i,j}({\mathbf{r}})$ which is related to the resonant excitation of the modes is isolated from complex coefficients $c_{j,i}^{+}, c_{j,i}^{\prime+}, c_{j,i}^{\prime\prime+}$,
and new coefficients 
$a_{j,i}^{+}, a_{j,i}^{\prime+}, a_{j,i}^{\prime\prime+}$ are introduced.

{To avoid cluttering the main text with lengthy technical derivations, we have transferred them to Appendix~\ref{sec:app:intensity} and present here the final expression for the interference terms $I_{{\ket{L_m}},{\ket{R_m}}}^{2\omega,\text{interf.}}$ of the total intensity, in which we are primarily interested:
\begin{align}
    \label{intensityLCPRCPinterfoutappendix} &I_{{\ket{L_m}},{\ket{R_m}}}^{2\omega,\text{interf.}} \propto
    \int_{\text{sph}} \dd V \eu^{-\iu\alpha}\mathbf{E}^{2\omega, x\pm\iu y}_{j}(\mathbf{r})\left(\mathbf{E}^{2\omega, x\pm\iu y}_{i}(\mathbf{r})\right)^* \cdot
    \\\nonumber& \cdot\left(A_{ji} + B_{ji}\eu^{\pm6\iu\beta} + C_{ji}\eu^{\mp6\iu\beta} + D_{ji}\eu^{\pm3\iu\beta} + E_{ji}\eu^{\mp3\iu\beta}\right) + \text{c.c.},
\end{align}
where, for simplicity, constants $A_{ji}, B_{ji}, C_{ji}, D_{ji}, E_{ji}$ were introduced.} {We should pay attention to the following:
\begin{enumerate}
    \item The eigenmode is excited simultaneously by several terms of the polarization. It is possible, because $\Delta m_\chi$ coincides with $\mathfrak n\nu$ for some $\nu \in \mathbb{Z}$.
    \item The multiplier $e^{-\iu\alpha}$ does not depend on the polarization.
    \item The multipliers $e^{\iu \Delta m_\chi\beta}$ appear and change their sign by changing the polarization of the beam.
\end{enumerate}
This behavior is common and appears every time the dichroism is observed.}
\subsubsection{Second-harmonic circular dichroism}
\paragraph{Result}
{The last step of our proof is to examine the dependence of the total intensity 
$I_{{\ket{L_m}},{\ket{R_m}}}^{2\omega}$ on the sign of the polarization in detail. 
{The expression for the interference terms 
$I_{{\ket{L_m}},{\ket{R_m}}}^{2\omega,\text{interf.}}$~\eqref{intensityLCPRCPinterfoutappendix} can be simplified by} taking into account that \\
$\mathbf{E}^{2\omega, x+\iu y}_{j}(\mathbf{r})\left(\mathbf{E}^{2\omega, x+\iu y}_{i}(\mathbf{r})\right)^* = \mathbf{E}^{2\omega, x-\iu y}_{j}(\mathbf{r})\left(\mathbf{E}^{2\omega, x-\iu y}_{i}(\mathbf{r})\right)^*$:}
\begin{align}
    \label{intensityLCPRCP}
    &I_{{\ket{L_m}},{\ket{R_m}}}^{2\omega,\text{interf.}}  \propto \eu^{-\iu\alpha}\text{const}_{ji}\cdot
    \left(A_{ji} + B_{ji}\eu^{\pm 6\iu\beta}  + C_{ji}\eu^{\mp 6\iu\beta} + D_{ji}\eu^{\pm 3\iu\beta} + E_{ji}\eu^{\mp 3\iu\beta}\right) + \text{c.c.},
\end{align}
where $\pm$ corresponds to the sign of polarization of the incident beam and $\text{const}_{ji}$ does not depend on polarization. 
Next, we rewrite this expression in more detail, using the fact that complex coefficients $A_{ji},B_{ji},C_{ji},D_{ji}$, and $E_{ji}$ can be expanded as $A_{ji} = A^\prime_{ji} + \iu A^{\prime\prime}_{ji}$ through real ${A^\prime_{ji}}$ and the imaginary part $A^{\prime\prime}_{ji}$:
\begin{align}
    \label{intensityLCPRCPinmoredetail}
    &I_{{\ket{L_m}},{\ket{R_m}}}^{2\omega,\text{interf.}}  \propto \nonumber 2\text{const}_{ji}\bigg(A^\prime_{ji}\cos(\alpha) + A^{\prime\prime}_{ji}\sin(\alpha) + 
     B^\prime_{ji}\cos(\alpha\mp 6\beta ) + B^{\prime\prime}_{ji}\sin(\alpha\mp 6\beta) + 
    \\\nonumber & + C^\prime_{ji}\cos(\alpha\pm 6\beta ) + C^{\prime\prime}_{ji}\sin(\alpha\pm 6\beta) + 
     D^\prime_{ji}\cos(\alpha\mp 3\beta ) + D^{\prime\prime}_{ji}\sin(\alpha\mp 3\beta) + 
    \\ & + E^\prime_{ji}\cos(\alpha\pm 3\beta ) + E^{\prime\prime}_{ji}\sin(\alpha\pm 3\beta)\bigg).
\end{align}
Note that interference contribution~\eqref{intensityLCPRCP},~\eqref{intensityLCPRCPinmoredetail} belonging to the expression for the total integral second harmonic intensity $I_{{\ket{L_m}},{\ket{R_m}}}^{2\omega}$ contains terms like $\cos(\alpha\pm 3\beta)$. 
These terms are different for different signs of the polarization of the incident beam for the most values of $\alpha$ and $\beta$. 
In other words, the total integral second harmonic intensity $I^{2\omega}$ contains the interference contribution $I^{2\omega,\text{interf.}}$ 
that depends on the polarization of the incident beam, if angles $\beta$ are not equal to $\pi\nu/3$, where $\nu \in \mathbb {Z}$. 
From the above the conclusion follows: SH-CD can appear appear in nanostructure GaAs[111]$\parallel$z with symmetry $\text{C}_{3\text{v}}$, only if angles $\beta \neq \pi\nu/3$, where $\nu \in \mathbb {Z}$. 
\paragraph{Comments}
{Let us add some additional comments for better understanding the result.}
\begin{enumerate}
    \item {Occurrence of the SH-CD in the structure doesn't depend on the absolute value of the angular momentum projection of the incident vortex $|m^{\text{in}}|$. 
    Even though we in details analyzed the case $m^{\text{in}}=3s-2$, where $ s \in \mathbb {Z}_+$, nothing will change for the two other cases, when $m^{\text{in}}=3s$ or $m^{\text{in}}=3s-1$ because the dependence of the second harmonic electric field and, consequently, of the total integral second harmonic intensity on the sign of the polarization of the incident field are the same for all cases, as it can be seen in expressions~\eqref{integralA1Efield},~\eqref{integralA2Efield},~\eqref{integral1field},~\eqref{integral2field},~\eqref{integral4field} and~\eqref{integral3field}. 
    {The main thing is the interplay between $\Delta m_\chi$ and $\mathfrak n$, which contributes in the same way as it was discussed in~\cite{Nikitina2023-Nonlinearcirculardi}.}} 
    \item {We calculated total integral intensity only for two modes transformed under the same irreducible representation because only between them interference is possible.}
    \item {Coefficients $A_{ji},B_{ji},C_{ji},D_{ji}$, and $E_{ji}$ can be either real or complex coefficients depending upon whether coefficients $a_{j}^{+},a_{j}^{\prime+},a_{j}^{\prime\prime+}$ from the decomposition of the intensity~\eqref{intensityLCP} are all in-phase or not. 
    In the case of the real coefficients we need to use at least two-mode approximation to obtain the condition for the appearance of the SH-CD because noninterference terms from the equation~\eqref{longlongintegral} will be just equal to each other for different beam polarizations. 
    However, if coefficients are complex, it's enough to use only single-mode approximation because not only interference but also noninterference terms contribute to the SH-CD and it leads to the same result. 
    The reason why we made all calculations for the two-mode approximation is that interference contribution works in any case regardless of whether coefficients are complex or real. } 
\end{enumerate}

\subsection{Second harmonic in GaAs[111]\texorpdfstring{$\parallel$}{Lg}z \texorpdfstring{$\text{C}_{4\text{v}}$}{Lg} nanostructure (no dichroism)}
\label{sec:GaAsC4v}
{In this section, we will show that in the nanostructure with the same crystal lattice GaAs[111]$\parallel$z with nanoparticle symmetry $\text{C}_{4\text{v}}$ second-harmonic circular dichroism cannot ever be obtained, using the same theoretical approach, as in the previous subsection~\ref{sec:GaAsC3v}. 
The decomposition of the nonlinear polarization $\mathbf{P}^{2\omega}(\mathbf{r})$ will be defined as in~\eqref{GaAs[111]polnew} due to the same crystal lattice but with $\mathfrak{n}=4$. 
Firstly, as in the previous example, we prove that the excitement of eigenmodes transformed under irreducible representations of the $\text{C}_{4\text{v}}$, i.e. 
$\text{E},\text{A}_1,\text{A}_2,\text{B}_1$, and $\text{B}_2$ depends on the specific value of the projection $m_{\text{in}}$. 
Specifically, modes transformed under $\text{E}$ can always be excited, modes transformed under 
$\text{A}_1$ and $\text{A}_2$ can be excited only if $m^{\text{in}}=4s$ and $m^{\text{in}}=4s-2$, and modes transformed under 
$\text{B}_1$ and $\text{B}_2$ can be excited only if $m^{\text{in}}=4s-3$ and $m^{\text{in}}=4s-1$, where $ s \in \mathbb {Z}_+$.}
\subsubsection{Eigenmode presentations}
Let us consider decompositions of eigenmodes transformed under 
$\text{E}$, and $\text{A}_1$, $\text{A}_2$, as well as $\text{B}_1$ and $\text{B}_2$ in magnetic and electric vector spherical harmonics $\mathbf{M}_{^e_omn}, \mathbf{N}_{^e_omn}$ one by one.
\paragraph{Eigenmodes, which are transformed under E irrep}
Eigenmodes transformed under $\text{E}$ can be decomposed into series of vector spherical harmonics $\mathbf{M}_{^e_omn}, \mathbf{N}_{^e_omn}$ with $m=2s-1$, where $ s \in \mathbb {Z}_+$. 
Recalling that representation $\text{E}$ is a 2-dimensional irreducible representation, we can define two orthogonal eigenmodes transformed under this representation in the following form:
\begin{align}
    &\mathbf{E}^{2\omega, x}_{j} \left(\mathbf{r}^{\prime}\right) = a_{e11}\mathbf{N}_{e11} + a_{e12}\mathbf{N}_{e12} + b_{o11}\mathbf{M}_{o11} +\dots, \label{modesеc4vx} 
    \\ &\mathbf{E}^{2\omega, y}_{j} \left(\mathbf{r}^{\prime}\right) = a_{o11}\mathbf{N}_{o11} + a_{o12}\mathbf{N}_{o12} + b_{e11}\mathbf{M}_{e11} +\dots.
    \label{modesc4vy} 
\end{align} 
Using the same speculations as in the Appendix~\eqref{sec:app:4:1}, we reach the convenient expressions for orthogonal eigenmodes $\mathbf{E}^{2\omega, x\pm \iu y}_{j}$ transformed under $\text{E}$:
\begin{align}
    \label{modesеc4vnewfinal} 
    &\mathbf{E}^{2\omega, x\pm \iu y}_{j} \left(\mathbf{r}^{\prime}\right) =
    \\\nonumber &= \sum_{n}\bigg\{ \mathbf{E}_{j,1n}(r,z)\eu^{\pm 1\iu\varphi} +  \mathbf{E}_{j,3n}(r,z)\eu^{\mp 3\iu\varphi}+ \mathbf{E}_{j,5n}(r,z)\eu^{\pm 5\iu\varphi} + \mathbf{E}_{j,7n}(r,z)\eu^{\mp 7\iu\varphi} +\dots\bigg\},
\end{align}
where $\mathbf{E}_{j,mn}(r,z)$ contains linear combinations $\mathbf{N}(\mathbf{M})_{mn}(r,z)$ with complex coefficients $a_{mn}, b_{mn}$. 
Note that signs here are related to the specific symmetry behavior of these exponential terms under C$_{4\text{v}}$ transformations, i.e. $e^{1\iu\varphi}$ behaves as $e^{-3\iu\varphi}$ but not $e^{3\iu\varphi}$.
\paragraph{Eigenmodes, which are transformed under $\text{A}_1$ and $\text{A}_2$ irrep}
Eigenmodes transformed under irreducible representations $\text{A}_1$ and $\text{A}_2$ contain vector spherical harmonics $\mathbf{M}_{^e_o(4s)n}$ and $\mathbf{N}_{^e_o(4s)n}$, where $ s \in \mathbb {Z}_+$. 
The eigenmodes transformed under $\text{A}_1$ are invariant under reflection in the plane XZ, and eigenmodes transformed under $\text{A}_2$ are not. {Then, their expansions can be written as:}
\begin{align}
    \label{modesc4vA1A2} 
    &{\mathbf{E}^{2\omega,\text{A}_1}_{j} \left(\mathbf{r}^{\prime}\right) = a_{e01}\mathbf{N}_{e01} + a_{e44}\mathbf{N}_{e44} + b_{o44}\mathbf{M}_{o44} + \dots,}
    \\
    \label{modesc4vA1A2two}
    &{\mathbf{E}^{2\omega,\text{A}_2}_{j} \left(\mathbf{r}^{\prime}\right) = b_{e01}\mathbf{M}_{e01} + b_{e44}\mathbf{M}_{e44} + a_{o44}\mathbf{N}_{o44} +\dots.}
\end{align}
Using the explicit form of the vector spherical harmonics (Appendix~\ref{sec:app:1}) and new notations, we rewrite the expressions above~\eqref{modesc4vA1A2} and~\eqref{modesc4vA1A2two} in the following form:
\begin{align}
    \label{modesc4vA1A2finalA1} 
    &{\mathbf{E}^{2\omega, \text{A}_1}_{j} \left(\mathbf{r}^{\prime}\right)} = 
    \\\nonumber&= \sum_{n} \bigg\{\left[{E}^{r}_{j,e4n} \left(r,z\right)\hat{\mathbf{e}}_r + {E}^z_{j,e4n} \left(r,z\right)\hat{\mathbf{e}}_z\right]\cos{4\varphi} {+ {E}^{\varphi}_{j,e4n} \left(r,z\right)\hat{\mathbf{e}}_\varphi\sin{4\varphi} + \dots\bigg\},}
\end{align}
\begin{align}
    \label{modesc4vA1A2finalA2}
    &{\mathbf{E}^{2\omega, \text{A}_2}_{j} \left(\mathbf{r}^{\prime}\right)} =
    \\\nonumber &{= \sum_{n} \bigg\{\left[{E}^{r}_{j,o4n} \left(r,z\right)\hat{\mathbf{e}}_r + {E}^z_{j,o4n} \left(r,z\right)\hat{\mathbf{e}}_z\right]\sin{4\varphi} }{+ {E}^{\varphi}_{j,o4n} \left(r,z\right)\hat{\mathbf{e}}_\varphi\cos{4\varphi} + \dots\bigg\},}
\end{align}
where coefficients ${E}^{r}_{j,^e_o4n}\left(r,z\right), {E}^{z}_{j,^e_o4n}\left(r,z\right), {E}^{\varphi}_{j,^e_o4n}\left(r,z\right)$ contain all functions and constants independent on $\varphi$.
\paragraph{Eigenmodes, which are transformed under $\text{B}_1$ and $\text{B}_2$ irrep}
Eigenmodes transformed under irreducible representations $\text{B}_1$ and $\text{B}_2$ contain vector spherical harmonics $\mathbf{M}_{^e_o(4s-2)n}$ and $\mathbf{N}_{^e_o(4s-2)n}$, where $ s \in \mathbb {Z}_+$:
\begin{align}
    \label{modesc4vb1b2}
    &{\mathbf{E}^{2\omega,\text{B}_1}_{j} \left(\mathbf{r}^{\prime}\right) = a_{e22}\mathbf{N}_{e22} + a_{e26}\mathbf{N}_{e26} + b_{o22}\mathbf{M}_{o22} + \dots,}
    \\
    \label{modesc4vb1b2two}
    &{\mathbf{E}^{2\omega,\text{B}_2}_{j} \left(\mathbf{r}^{\prime}\right) = b_{e22}\mathbf{M}_{e22} + a_{o22}\mathbf{N}_{o22} + a_{o26}\mathbf{N}_{o26} +\dots.}
\end{align}
which can be rewritten as
\begin{align}
    \label{modesc4vb1b2finalb1} 
    &{\mathbf{E}^{2\omega, \text{B}_1}_{j} \left(\mathbf{r}^{\prime}\right)}= 
    \\\nonumber&= \sum_{n} \bigg\{\left[{E}^{r}_{j,e2n} \left(r,z\right)\hat{\mathbf{e}}_r + {E}^z_{j,e2n} \left(r,z\right)\hat{\mathbf{e}}_z\right]\cos{2\varphi} 
    {+ {E}^{\varphi}_{j,e2n} \left(r,z\right)\hat{\mathbf{e}}_\varphi\sin{2\varphi} + \dots\bigg\},}
\end{align}
\begin{align}
    \label{modesc4vb1b2finalb2}
    &{\mathbf{E}^{2\omega, \text{B}_2}_{j} \left(\mathbf{r}^{\prime}\right)}=
    \\\nonumber &{= \sum_{n} \bigg\{\left[{E}^{r}_{j,o2n} \left(r,z\right)\hat{\mathbf{e}}_r + {E}^z_{j,o2n} \left(r,z\right)\hat{\mathbf{e}}_z\right]\sin{2\varphi} +}
    { {E}^{\varphi}_{j,o2n} \left(r,z\right)\hat{\mathbf{e}}_\varphi\cos{2\varphi} + \dots\bigg\},}
\end{align}
where coefficients ${E}^{r}_{j,^e_o2n}\left(r,z\right), {E}^{z}_{j,^e_o2n}\left(r,z\right), {E}^{\varphi}_{j,^e_o2n}\left(r,z\right)$ contain all functions and constants independent on coordinate~$\varphi$.
\subsubsection{Coupling integrals}
\paragraph{{Eigenmodes, which are transformed under $\text{E}$ irrep}}
Using the expressions~\eqref{modesеc4vnewfinal} for eigenmodes $\mathbf{E}^{2\omega, x\pm\iu y}_{j} \left(\mathbf{r}^{\prime}\right)$ and decomposition of nonlinear polarization $\mathbf{P}^{2\omega}(\mathbf{r})$~\eqref{GaAs[111]polnew} we obtain four  coupling integrals $\text{D}_j = \int\limits_{\text{C}_{4\text{v}}} \dd V' \mathbf{E}^{2\omega, x\pm\iu y}_{j} \left(\mathbf{r}^{\prime}\right) \mathbf{P}^{2\omega} (\mathbf{r}')$. 
In particular, there are two integrals for eigenmode $\mathbf{E}^{2\omega, x+\iu y}_{j} \left(\mathbf{r}^{\prime}\right)$ excited by left- and right-handed circularly polarized beam:
\begin{align}
    \label{integralbig1c4v}
    \ket{L_m}\colon \; &\text{D}_j = \int\limits_{\text{C}_{4\text{v}}} \dd V' \mathbf{E}^{2\omega, x+\iu y}_{j} \left(\mathbf{r}^{\prime}\right) \mathbf{P}^{2\omega} (\mathbf{r}^{\prime}) \propto 
    \\\nonumber & \propto \int\limits_{\text{C}_{4\text{v}}} \dd V' \sum_{n,\nu}\bigg(\bigg[\mathbf{E}_{j,1n}\eu^{+1\iu\varphi} + \mathbf{E}_{j,3n}\eu^{- 3\iu\varphi} + 
     \mathbf{E}_{j,5n}\eu^{+5\iu\varphi} + \mathbf{E}_{j,7n}\eu^{- 7\iu\varphi} +\dots\bigg]\cdot
    \\\nonumber & \cdot\bigg[ \bigg\{\mathbf{P}_{\blue{(3+2m^{\text{in}})},\nu}^{2\omega}\eu^{+ \blue{(3+2m^{\text{in}})}\iu\varphi + 4\nu \iu\varphi}\bigg\}\eu^{- 3\iu\beta} + 
    \\\nonumber & + \bigg\{
    \mathbf{P}_{\blue{(-3+2m^{\text{in}})},\nu}^{2\omega}\eu^{+ \blue{(-3+2m^{\text{in}})}\iu\varphi + 4\nu \iu\varphi}\bigg\}\eu^{+ 3\iu\beta} + 
    \\\nonumber & + \bigg\{\mathbf{P}_{\blue{2m^{\text{in}}},\nu}^{2\omega}\eu^{+ \blue{2m^{\text{in}}}\iu\varphi + 4\nu \iu\varphi}\bigg\}\bigg]\bigg),
\end{align}
\begin{align}
    \label{integralbig2c4v}
    \ket{R_m}\colon \; &\text{D}_j = \int\limits_{\text{C}_{4\text{v}}} \dd V' \mathbf{E}^{2\omega, x+\iu y}_{j} \left(\mathbf{r}^{\prime}\right) \mathbf{P}^{2\omega} (\mathbf{r}^{\prime}) \propto 
    \\\nonumber & \propto \int\limits_{\text{C}_{4\text{v}}} \dd V' \sum_{n,\nu}\bigg(\bigg[\mathbf{E}_{j,1n}\eu^{+1\iu\varphi} + \mathbf{E}_{j,3n}\eu^{- 3\iu\varphi} + 
    \mathbf{E}_{j,5n}\eu^{+5\iu\varphi} + \mathbf{E}_{j,7n}\eu^{- 7\iu\varphi} +\dots\bigg]\cdot
    \\\nonumber& \cdot\bigg[ \bigg\{\mathbf{P}_{\blue{(3+2m^{\text{in}})},\nu}^{2\omega}\eu^{- \blue{(3+2m^{\text{in}})}\iu\varphi + 4\nu \iu\varphi}\bigg\}\eu^{+ 3\iu\beta} + 
    \\\nonumber & + \bigg\{
    \mathbf{P}_{\blue{(-3+2m^{\text{in}})},\nu}^{2\omega}\eu^{- \blue{(-3+2m^{\text{in}})}\iu\varphi + 4\nu \iu\varphi}\bigg\}\eu^{- 3\iu\beta} + 
    \\\nonumber & + \bigg\{\mathbf{P}_{\blue{2m^{\text{in}}},\nu}^{2\omega}\eu^{- \blue{2m^{\text{in}}}\iu\varphi + 4\nu \iu\varphi}\bigg\}\bigg]\bigg),
\end{align}
and there are two other integrals for eigenmode $\mathbf{E}^{2\omega, x-\iu y}_{j} \left(\mathbf{r}^{\prime}\right)$ excited by left- and right-handed circularly polarized beam:
\begin{align}
    \label{integralbig3c4v}
    \ket{L_m}\colon \; &\text{D}_j = \int\limits_{\text{C}_{4\text{v}}} \dd V' \mathbf{E}^{2\omega, x-\iu y}_{j} \left(\mathbf{r}^{\prime}\right) \mathbf{P}^{2\omega} (\mathbf{r}^{\prime}) \propto 
    \\\nonumber & \propto \int\limits_{\text{C}_{4\text{v}}} \dd V' \sum_{n,\nu}\bigg(\bigg[\mathbf{E}_{j,1n}\eu^{-1\iu\varphi} + \mathbf{E}_{j,3n}\eu^{+ 3\iu\varphi} + 
     \mathbf{E}_{j,5n}\eu^{-5\iu\varphi} + \mathbf{E}_{j,7n}\eu^{+ 7\iu\varphi} +\dots\bigg]\cdot
    \\\nonumber&\cdot\bigg[ \bigg\{\mathbf{P}_{\blue{(3+2m^{\text{in}})},\nu}^{2\omega}\eu^{+ \blue{(3+2m^{\text{in}})}\iu\varphi + 4\nu \iu\varphi}\bigg\}\eu^{- 3\iu\beta} + 
    \\\nonumber & + \bigg\{
    \mathbf{P}_{\blue{(-3+2m^{\text{in}})},\nu}^{2\omega}\eu^{+ \blue{(-3+2m^{\text{in}})}\iu\varphi + 4\nu \iu\varphi}\bigg\}\eu^{+ 3\iu\beta} + 
    \\\nonumber & + \bigg\{\mathbf{P}_{\blue{2m^{\text{in}}},\nu}^{2\omega}\eu^{+ \blue{2m^{\text{in}}}\iu\varphi + 4\nu \iu\varphi}\bigg\}\bigg]\bigg),
\end{align}
\begin{align}
    \label{integralbig4c4v}
    \ket{R_m}\colon \; &\text{D}_j = \int\limits_{\text{C}_{4\text{v}}} \dd V' \mathbf{E}^{2\omega, x-\iu y}_{j} \left(\mathbf{r}^{\prime}\right) \mathbf{P}^{2\omega} (\mathbf{r}^{\prime}) \propto 
    \\\nonumber & \propto \int\limits_{\text{C}_{4\text{v}}} \dd V' \sum_{n,\nu}\bigg(\bigg[\mathbf{E}_{j,1n}\eu^{-1\iu\varphi} + \mathbf{E}_{j,3n}\eu^{+ 3\iu\varphi} + 
     \mathbf{E}_{j,5n}\eu^{-5\iu\varphi} + \mathbf{E}_{j,7n}\eu^{+ 7\iu\varphi} +\dots\bigg]\cdot
    \\\nonumber &\cdot\bigg[ \bigg\{\mathbf{P}_{\blue{(3+2m^{\text{in}})},\nu}^{2\omega}\eu^{- \blue{(3+2m^{\text{in}})}\iu\varphi + 4\nu \iu\varphi}\bigg\}\eu^{+ 3\iu\beta} + 
    \\\nonumber & + \bigg\{
    \mathbf{P}_{\blue{(-3+2m^{\text{in}})},\nu}^{2\omega}\eu^{- \blue{(-3+2m^{\text{in}})}\iu\varphi + 4\nu \iu\varphi}\bigg\}\eu^{- 3\iu\beta} + 
    \\\nonumber & + \bigg\{\mathbf{P}_{\blue{2m^{\text{in}}},\nu}^{2\omega}\eu^{- \blue{2m^{\text{in}}}\iu\varphi + 4\nu \iu\varphi}\bigg\}\bigg]\bigg).
\end{align}
\paragraph{Eigenmodes, which are transformed under $\text{A}_1$ and $\text{A}_2$ irrep}
Next, let us use expressions~\eqref{modesc4vA1A2finalA1} and~\eqref{modesc4vA1A2finalA2} for eigenmodes $\mathbf{E}^{2\omega, \text{A}_1}_{j} \left(\mathbf{r}^{\prime}\right)$ and $\mathbf{E}^{2\omega, \text{A}_2}_{j} \left(\mathbf{r}^{\prime}\right)$ and again the decomposition of nonlinear polarization $\mathbf{P}^{2\omega}(\mathbf{r})$~\eqref{GaAs[111]polnew} to obtain four coupling integrals $\text{D}_j$. 
Two integrals for eigenmode $\mathbf{E}^{2\omega, \text{A}_1}_{j} \left(\mathbf{r}^{\prime}\right)$ excited by left- and right-handed circularly polarized beam:
\begin{align}
    \label{integralbig5c4v}
    \ket{L_m}\colon \; &\text{D}_j = \int\limits_{\text{C}_{4\text{v}}} \dd V' \mathbf{E}^{2\omega, \text{A}_1}_{j} \left(\mathbf{r}^{\prime}\right) \mathbf{P}^{2\omega} (\mathbf{r}^{\prime}) \propto
    \\\nonumber   \propto \int\limits_{\text{C}_{4\text{v}}} \dd V' &\sum_{n,\nu}\bigg(\bigg[\left[{E}^{r}_{j,e4n} \left(r,z\right)\hat{\mathbf{e}}_r + {E}^z_{j,e4n} \left(r,z\right)\hat{\mathbf{e}}_z\right] \cos{4\varphi} +
    {E}^{\varphi}_{j,e4n} \left(r,z\right)\hat{\mathbf{e}}_\varphi\sin{4\varphi} + \dots\bigg]\cdot
    \\\nonumber &\cdot\bigg[ \bigg\{\mathbf{P}_{\blue{(3+2m^{\text{in}})},\nu}^{2\omega}\eu^{+ \blue{(3+2m^{\text{in}})}\iu\varphi + 4\nu \iu\varphi}\bigg\}\eu^{- 3\iu\beta} + 
    \\\nonumber & + \bigg\{
    \mathbf{P}_{\blue{(-3+2m^{\text{in}})},\nu}^{2\omega}\eu^{+ \blue{(-3+2m^{\text{in}})}\iu\varphi + 4\nu \iu\varphi}\bigg\}\eu^{+ 3\iu\beta} + 
    \\\nonumber & + \bigg\{\mathbf{P}_{\blue{2m^{\text{in}}},\nu}^{2\omega}\eu^{+ \blue{2m^{\text{in}}}\iu\varphi + 4\nu \iu\varphi}\bigg\}\bigg]\bigg),
\end{align}
\begin{align}
    \label{integralbig6c4v}
    \ket{R_m}\colon \; &\text{D}_j = \int\limits_{\text{C}_{4\text{v}}} \dd V' \mathbf{E}^{2\omega, \text{A}_1}_{j} \left(\mathbf{r}^{\prime}\right) \mathbf{P}^{2\omega} (\mathbf{r}^{\prime}) \propto
    \\\nonumber \propto  \int\limits_{\text{C}_{4\text{v}}} \dd V' &\sum_{n,\nu}\bigg(\bigg[\left[{E}^{r}_{j,e4n} \left(r,z\right)\hat{\mathbf{e}}_r + {E}^z_{j,e4n} \left(r,z\right)\hat{\mathbf{e}}_z\right]\cdot \cos{4\varphi} 
    + {E}^{\varphi}_{j,e4n} \left(r,z\right)\hat{\mathbf{e}}_\varphi\sin{4\varphi} + \dots\bigg]\cdot
    \\\nonumber &\cdot\bigg[ \bigg\{\mathbf{P}_{\blue{(3+2m^{\text{in}})},\nu}^{2\omega}\eu^{- \blue{(3+2m^{\text{in}})}\iu\varphi + 4\nu \iu\varphi}\bigg\}\eu^{+ 3\iu\beta} + 
    \\\nonumber & + \bigg\{
    \mathbf{P}_{\blue{(-3+2m^{\text{in}})},\nu}^{2\omega}\eu^{- \blue{(-3+2m^{\text{in}})}\iu\varphi + 4\nu \iu\varphi}\bigg\}\eu^{- 3\iu\beta} + 
    \\\nonumber & + \bigg\{\mathbf{P}_{\blue{2m^{\text{in}}},\nu}^{2\omega}\eu^{- \blue{2m^{\text{in}}}\iu\varphi + 4\nu \iu\varphi}\bigg\}\bigg]\bigg),
\end{align}
{and for the eigenmode $\mathbf{E}^{2\omega, \text{A}_2}_{j} \left(\mathbf{r}^{\prime}\right)$:}
\begin{align}
    \label{integralbig7c4v}
    \ket{L_m}\colon \; &\text{D}_j = \int\limits_{\text{C}_{4\text{v}}} \dd V' \mathbf{E}^{2\omega, \text{A}_2}_{j} \left(\mathbf{r}^{\prime}\right) \mathbf{P}^{2\omega} (\mathbf{r}^{\prime}) \propto
    \\\nonumber \propto  \int\limits_{\text{C}_{4\text{v}}} \dd V' &\sum_{n,\nu}\bigg(\bigg[\left[{E}^{r}_{j,o4n} \left(r,z\right)\hat{\mathbf{e}}_r + {E}^z_{j,o4n} \left(r,z\right)\hat{\mathbf{e}}_z\right] \sin{4\varphi} + {E}^{\varphi}_{j,o4n} \left(r,z\right)\hat{\mathbf{e}}_\varphi\cos{4\varphi} + \dots\bigg]\cdot
    \\\nonumber & \cdot\bigg[ \bigg\{\mathbf{P}_{\blue{(3+2m^{\text{in}})},\nu}^{2\omega}\eu^{+ \blue{(3+2m^{\text{in}})}\iu\varphi + 4\nu \iu\varphi}\bigg\}\eu^{- 3\iu\beta} + 
    \\\nonumber & + \bigg\{
    \mathbf{P}_{\blue{(-3+2m^{\text{in}})},\nu}^{2\omega}\eu^{+ \blue{(-3+2m^{\text{in}})}\iu\varphi + 4\nu \iu\varphi}\bigg\}\eu^{+ 3\iu\beta} + 
    \\\nonumber & + \bigg\{\mathbf{P}_{\blue{2m^{\text{in}}},\nu}^{2\omega}\eu^{+ \blue{2m^{\text{in}}}\iu\varphi + 4\nu \iu\varphi}\bigg\}\bigg]\bigg),
\end{align}
\begin{align}
    \label{integralbig8c4v}
    \ket{R_m}\colon \; &\text{D}_j = \int\limits_{\text{C}_{4\text{v}}} \dd V' \mathbf{E}^{2\omega, \text{A}_2}_{j} \left(\mathbf{r}^{\prime}\right) \mathbf{P}^{2\omega} (\mathbf{r}^{\prime}) \propto
    \\\nonumber \propto \int\limits_{\text{C}_{4\text{v}}} \dd V' & \sum_{n,\nu}\bigg(\bigg[\left[{E}^{r}_{j,o4n} \left(r,z\right)\hat{\mathbf{e}}_r + {E}^z_{j,o4n} \left(r,z\right)\hat{\mathbf{e}}_z\right]\sin{4\varphi} +
    {E}^{\varphi}_{j,o4n} \left(r,z\right)\hat{\mathbf{e}}_\varphi\cos{4\varphi} + \dots\bigg]\cdot
    \\\nonumber & \cdot\bigg[ \bigg\{\mathbf{P}_{\blue{(3+2m^{\text{in}})},\nu}^{2\omega}\eu^{- \blue{(3+2m^{\text{in}})}\iu\varphi + 4\nu \iu\varphi}\bigg\}\eu^{+ 3\iu\beta} + 
    \\\nonumber & + \bigg\{
    \mathbf{P}_{\blue{(-3+2m^{\text{in}})},\nu}^{2\omega}\eu^{- \blue{(-3+2m^{\text{in}})}\iu\varphi + 4\nu \iu\varphi}\bigg\}\eu^{- 3\iu\beta} + 
    \\\nonumber & + \bigg\{\mathbf{P}_{\blue{2m^{\text{in}}},\nu}^{2\omega}\eu^{- \blue{2m^{\text{in}}}\iu\varphi + 4\nu \iu\varphi}\bigg\}\bigg]\bigg).
\end{align}
\paragraph{Eigenmodes, which are transformed under $\text{B}_1$ and $\text{B}_2$}
Finally, in similar manner for eigenmodes 
$\mathbf{E}^{2\omega, \text{B}_1}_{j} \left(\mathbf{r}^{\prime}\right)$ and $\mathbf{E}^{2\omega, \text{B}_2}_{j} \left(\mathbf{r}^{\prime}\right)$, using ~\eqref{modesc4vb1b2finalb1} and~\eqref{modesc4vb1b2finalb2}, and again the decomposition of nonlinear polarization $\mathbf{P}^{2\omega}(\mathbf{r})$~\eqref{GaAs[111]polnew}, we obtain two integrals for eigenmode $\mathbf{E}^{2\omega, \text{B}_1}_{j} \left(\mathbf{r}^{\prime}\right)$:
\begin{align}
    \label{integralbig9c4v}
    \ket{L_m}\colon \; &\text{D}_j = \int\limits_{\text{C}_{4\text{v}}} \dd V' \mathbf{E}^{2\omega, \text{B}_1}_{j} \left(\mathbf{r}^{\prime}\right) \mathbf{P}^{2\omega} (\mathbf{r}^{\prime}) \propto
    \\\nonumber \propto  \int\limits_{\text{C}_{4\text{v}}} \dd V' &\sum_{n,\nu}\bigg(\bigg[\left[{E}^{r}_{j,e2n} \left(r,z\right)\hat{\mathbf{e}}_r + {E}^z_{j,e2n} \left(r,z\right)\hat{\mathbf{e}}_z\right]\cos{2\varphi} +
     {E}^{\varphi}_{j,e2n} \left(r,z\right)\hat{\mathbf{e}}_\varphi\sin{2\varphi} + \dots\bigg]\cdot
    \\\nonumber &\cdot\bigg[ \bigg\{\mathbf{P}_{\blue{(3+2m^{\text{in}})},\nu}^{2\omega}\eu^{+ \blue{(3+2m^{\text{in}})}\iu\varphi + 4\nu \iu\varphi}\bigg\}\eu^{- 3\iu\beta} + 
    \\\nonumber & + \bigg\{
    \mathbf{P}_{\blue{(-3+2m^{\text{in}})},\nu}^{2\omega}\eu^{+ \blue{(-3+2m^{\text{in}})}\iu\varphi + 4\nu \iu\varphi}\bigg\}\eu^{+ 3\iu\beta} + 
    \\\nonumber & + \bigg\{\mathbf{P}_{\blue{2m^{\text{in}}},\nu}^{2\omega}\eu^{+ \blue{2m^{\text{in}}}\iu\varphi + 4\nu \iu\varphi}\bigg\}\bigg]\bigg),
\end{align}
\begin{align}
    \label{integralbig10c4v}
    \ket{R_m}\colon \; &\text{D}_j = \int\limits_{\text{C}_{4\text{v}}} \dd V' \mathbf{E}^{2\omega, \text{B}_1}_{j} \left(\mathbf{r}^{\prime}\right) \mathbf{P}^{2\omega} (\mathbf{r}^{\prime}) \propto
    \\\nonumber  \propto \int\limits_{\text{C}_{4\text{v}}} \dd V' &\sum_{n,\nu}\bigg(\bigg[\left[{E}^{r}_{j,e2n} \left(r,z\right)\hat{\mathbf{e}}_r + {E}^z_{j,e2n} \left(r,z\right)\hat{\mathbf{e}}_z\right]\cos{2\varphi} +
    {E}^{\varphi}_{j,e2n} \left(r,z\right)\hat{\mathbf{e}}_\varphi\sin{2\varphi} + \dots\bigg]\cdot
    \\\nonumber & \cdot\bigg[ \bigg\{\mathbf{P}_{\blue{(3+2m^{\text{in}})},\nu}^{2\omega}\eu^{- \blue{(3+2m^{\text{in}})}\iu\varphi + 4\nu \iu\varphi}\bigg\}\eu^{+ 3\iu\beta} + 
    \\\nonumber & + \bigg\{
    \mathbf{P}_{\blue{(-3+2m^{\text{in}})},\nu}^{2\omega}\eu^{- \blue{(-3+2m^{\text{in}})}\iu\varphi + 4\nu \iu\varphi}\bigg\}\eu^{- 3\iu\beta} + 
    \\\nonumber & + \bigg\{\mathbf{P}_{\blue{2m^{\text{in}}},\nu}^{2\omega}\eu^{- \blue{2m^{\text{in}}}\iu\varphi + 4\nu \iu\varphi}\bigg\}\bigg]\bigg),
\end{align}
{and two other integrals for eigenmode $\mathbf{E}^{2\omega, \text{B}_2}_{j} \left(\mathbf{r}^{\prime}\right)$:}
\begin{align}
    \label{integralbig11c4v}
    \ket{L_m}\colon \; &\text{D}_j = \int\limits_{\text{C}_{4\text{v}}} \dd V' \mathbf{E}^{2\omega, \text{B}_2}_{j} \left(\mathbf{r}^{\prime}\right) \mathbf{P}^{2\omega} (\mathbf{r}^{\prime}) \propto
    \\\nonumber \propto \int\limits_{\text{C}_{4\text{v}}} \dd V' &\sum_{n,\nu}\bigg(\bigg[\left[{E}^{r}_{j,o2n} \left(r,z\right)\hat{\mathbf{e}}_r + {E}^z_{j,o2n} \left(r,z\right)\hat{\mathbf{e}}_z\right] \sin{2\varphi} + {E}^{\varphi}_{j,o2n} \left(r,z\right)\hat{\mathbf{e}}_\varphi\cos{2\varphi} + \dots\bigg]\cdot
    \\\nonumber & \cdot\bigg[ \bigg\{\mathbf{P}_{\blue{(3+2m^{\text{in}})},\nu}^{2\omega}\eu^{+ \blue{(3+2m^{\text{in}})}\iu\varphi + 4\nu \iu\varphi}\bigg\}\eu^{- 3\iu\beta} + 
    \\\nonumber & + \bigg\{
    \mathbf{P}_{\blue{(-3+2m^{\text{in}})},\nu}^{2\omega}\eu^{+ \blue{(-3+2m^{\text{in}})}\iu\varphi + 4\nu \iu\varphi}\bigg\}\eu^{+ 3\iu\beta} + 
    \\\nonumber & + \bigg\{\mathbf{P}_{\blue{2m^{\text{in}}},\nu}^{2\omega}\eu^{+ \blue{2m^{\text{in}}}\iu\varphi + 4\nu \iu\varphi}\bigg\}\bigg]\bigg),
\end{align}
\begin{align}
    \label{integralbig12c4v}
    \ket{R_m}\colon \; &\text{D}_j = \int\limits_{\text{C}_{4\text{v}}} \dd V' \mathbf{E}^{2\omega, \text{B}_2}_{j} \left(\mathbf{r}^{\prime}\right) \mathbf{P}^{2\omega} (\mathbf{r}^{\prime}) \propto
    \\\nonumber  \propto\int\limits_{\text{C}_{4\text{v}}} \dd V' &\sum_{n,\nu}\bigg(\bigg[\left[{E}^{r}_{j,o2n} \left(r,z\right)\hat{\mathbf{e}}_r + {E}^z_{j,o2n} \left(r,z\right)\hat{\mathbf{e}}_z\right]\sin{2\varphi} +
    {E}^{\varphi}_{j,o2n} \left(r,z\right)\hat{\mathbf{e}}_\varphi\cos{2\varphi} + \dots\bigg]\cdot
    \\\nonumber & \cdot\bigg[ \bigg\{\mathbf{P}_{\blue{(3+2m^{\text{in}})},\nu}^{2\omega}\eu^{- \blue{(3+2m^{\text{in}})}\iu\varphi + 4\nu \iu\varphi}\bigg\}\eu^{+ 3\iu\beta} + 
    \\\nonumber & + \bigg\{
    \mathbf{P}_{\blue{(-3+2m^{\text{in}})},\nu}^{2\omega}\eu^{- \blue{(-3+2m^{\text{in}})}\iu\varphi + 4\nu \iu\varphi}\bigg\}\eu^{- 3\iu\beta} + 
    \\\nonumber & + \bigg\{\mathbf{P}_{\blue{2m^{\text{in}}},\nu}^{2\omega}\eu^{- \blue{2m^{\text{in}}}\iu\varphi + 4\nu \iu\varphi}\bigg\}\bigg]\bigg).
\end{align}
\subsubsection{Excitation of the modes by different angular momentum}
{Let us consider, which of the expressions 
\eqref{integralbig1c4v}, \eqref{integralbig2c4v}, \eqref{integralbig3c4v}, \eqref{integralbig4c4v},~\eqref{integralbig5c4v}, \eqref{integralbig6c4v}, as well as ~\eqref{integralbig7c4v},~\eqref{integralbig8c4v},~\eqref{integralbig9c4v},~\eqref{integralbig10c4v},~\eqref{integralbig11c4v}, and~\eqref{integralbig12c4v} for coupling integrals $\text{D}_j$ give a nontrivial answer}. 
For that, the integrand should be transformed under the trivial irreducible representation, i.e. $\text{A}_1$ of group symmetry $\text{C}_{4\text{v}}$. 
Therefore, if the integrand contains exponents $e^{\pm 4 \iu s \varphi}$, where $ s \in \mathbb {Z}_+$}, then the integral can be nonzero. 
In view of the above, {there will be four different cases dependent on the value of the index $m^{\text{in}}$ of the incident vortex:}
\begin{enumerate}
    \item {$m^{\text{in}}=4s$, where $ s \in \mathbb {Z}_+$}:   
    \paragraph{Eigenmodes, which are transformed under $\text{E}$ irrep}
    {Coupling integrals~\eqref{integralbig1c4v},~\eqref{integralbig2c4v} for the eigenmode $\mathbf{E}^{2\omega, x+\iu y}_{j} \left(\mathbf{r}^{\prime}\right)$ give nonzero results:}
\begin{align}
    \label{integral1mequal4s}
    &\ket{L_m}\colon  \; \text{D}_j \propto c_j \eu^{-3\iu\beta},
    \\\nonumber
    &\ket{R_m}\colon \; \text{D}_j \propto c_j^{\prime} \eu^{-3\iu\beta}.
\end{align}
    {Coupling integrals~\eqref{integralbig3c4v},~\eqref{integralbig4c4v} for the eigenmode $\mathbf{E}^{2\omega, x-\iu y}_{j} \left(\mathbf{r}^{\prime}\right)$ give nonzero results:}
\begin{align}
    \label{integral2mequal4s}
    &\ket{L_m}\colon  \; \text{D}_j \propto c^{\prime}_j \eu^{+3\iu\beta},
    \\\nonumber
    &\ket{R_m}\colon \; \text{D}_j \propto c_j \eu^{+3\iu\beta}.
\end{align}
    \paragraph{Eigenmodes, which are transformed under $\text{A}_1$ and $\text{A}_2$ irrep}
    {Coupling integrals~\eqref{integralbig5c4v},~\eqref{integralbig6c4v} for the eigenmode $\mathbf{E}^{2\omega,\text{A}_1}_{j} \left(\mathbf{r}^{\prime}\right)$ give nonzero results:}
\begin{align}
    \label{integral3mequal4s}
    &\ket{L_m}\colon  \; \text{D}_j \propto c^{\text{A}_1}_j,
    \\\nonumber
    &\ket{R_m}\colon \; \text{D}_j \propto c^{\text{A}_1}_j.
\end{align}
    {Coupling integrals~\eqref{integralbig7c4v},~\eqref{integralbig8c4v} for the eigenmode $\mathbf{E}^{2\omega,\text{A}_2}_{j} \left(\mathbf{r}^{\prime}\right)$ give nonzero results:}
\begin{align}
    \label{integral4mequal4s}
    &\ket{L_m}\colon  \; \text{D}_j \propto c^{\text{A}_2}_j, 
    \\\nonumber
    &\ket{R_m}\colon \; \text{D}_j \propto c^{\text{A}_2}_j.
\end{align}
    \paragraph{Eigenmodes, which are transformed under $\text{B}_1$ and $\text{B}_2$ irrep}
    {Coupling integrals~\eqref{integralbig9c4v} and~\eqref{integralbig10c4v} for the eigenmode $\mathbf{E}^{2\omega, \text{B}_1}_{j} \left(\mathbf{r}^{\prime}\right)$, as well as~\eqref{integralbig11c4v} and~\eqref{integralbig12c4v} for the eigenmode $\mathbf{E}^{2\omega, \text{B}_2}_{j} \left(\mathbf{r}^{\prime}\right)$ give only trivial answers. Thus, modes transformed under these irrep aren't excited.}

    {Comment: we omit all discussions related to the coefficients because that all was mentioned in the previous section~\ref{sec:GaAsC3v}.}
    \item {$m^{\text{in}}=4s-2$, where $ s \in \mathbb {Z}_+$}:
    \paragraph{Eigenmodes, which are transformed under $\text{E}$ irrep}
      {Coupling integrals~\eqref{integralbig1c4v},~\eqref{integralbig2c4v} for the eigenmode $\mathbf{E}^{2\omega, x+\iu y}_{j} \left(\mathbf{r}^{\prime}\right)$:}
\begin{align}
     \label{integral1mequal4s-2}
     &\ket{L_m}\colon \; \text{D}_j \propto \tilde{c}_j \eu^{-3\iu\beta},
     \\\nonumber
     &\ket{R_m}\colon \; \text{D}_j \propto \tilde{c}_j^{\prime} \eu^{-3\iu\beta}.
\end{align}
    {Coupling integrals~\eqref{integralbig3c4v},~\eqref{integralbig4c4v} for the eigenmode $\mathbf{E}^{2\omega, x-\iu y}_{j} \left(\mathbf{r}^{\prime}\right)$:}
\begin{align}
    \label{integral2mequal4s-2}
    &\ket{L_m}\colon  \; \text{D}_j \propto \tilde{c}^{\prime}_j \eu^{+3\iu\beta},
    \\\nonumber
    &\ket{R_m}\colon \; \text{D}_j \propto \tilde{c}_j \eu^{+3\iu\beta}.
\end{align}
    \paragraph{Eigenmodes, which are transformed under $\text{A}_1$ and $\text{A}_2$ irrep}
    {Coupling integrals~\eqref{integralbig5c4v},~\eqref{integralbig6c4v} for the eigenmode $\mathbf{E}^{2\omega,\text{A}_1}_{j} \left(\mathbf{r}^{\prime}\right)$:}
\begin{align}
    \label{integral3mequal4s-2}
    &\ket{L_m}\colon \; \text{D}_j \propto \tilde{c}^{\text{A}_1}_j, 
    \\\nonumber
    &\ket{R_m}\colon \; \text{D}_j \propto \tilde{c}^{\text{A}_1}_j.
\end{align}
    {Coupling integrals~\eqref{integralbig7c4v},~\eqref{integralbig8c4v} for the eigenmode $\mathbf{E}^{2\omega,\text{A}_2}_{j} \left(\mathbf{r}^{\prime}\right)$:}
\begin{align}
    \label{integral4mequal4s-2}
    &\ket{L_m}\colon  \; \text{D}_j \propto \tilde{c}^{\text{A}_2}_j,
    \\\nonumber
    &\ket{R_m}\colon \; \text{D}_j \propto \tilde{c}^{\text{A}_2}_j.
\end{align}
    \paragraph{Eigenmodes, which are transformed under $\text{B}_1$ and $\text{B}_2$ irrep}
    {Coupling integrals~\eqref{integralbig9c4v} and~\eqref{integralbig10c4v} with the eigenmode $\mathbf{E}^{2\omega, \text{B}_1}_{j} \left(\mathbf{r}^{\prime}\right)$, and~\eqref{integralbig11c4v} and~\eqref{integralbig12c4v} for $\mathbf{E}^{2\omega, \text{B}_2}_{j} \left(\mathbf{r}^{\prime}\right)$ give zeros, and these modes aren't excited.}
    \item {$m^{\text{in}}=4s-3$, where $ s \in \mathbb {Z}_+$}:
    \paragraph{Eigenmodes, which are transformed under $\text{E}$ irrep}
    {Integrals~\eqref{integralbig1c4v},~\eqref{integralbig2c4v} with $\mathbf{E}^{2\omega, x+\iu y}_{j} \left(\mathbf{r}^{\prime}\right)$ are:}
\begin{align}
    \label{integral1mequal4s-3}
    &\ket{L_m}\colon \; \text{D}_j \propto C_j \eu^{+3\iu\beta},
    \\\nonumber
    &\ket{R_m}\colon\; \text{D}_j \propto C_j^{\prime} \eu^{+3\iu\beta}.
\end{align}
    {Coupling integrals~\eqref{integralbig3c4v},~\eqref{integralbig4c4v} for the eigenmode $\mathbf{E}^{2\omega, x-\iu y}_{j} \left(\mathbf{r}^{\prime}\right)$ give nonzero results:}
\begin{align}
    \label{integral2mequal4s-3}
    &\ket{L_m}\colon \; \text{D}_j \propto C^{\prime}_j \eu^{-3\iu\beta},
    \\\nonumber
    &\ket{R_m}\colon\; \text{D}_j \propto C_j \eu^{-3\iu\beta}.
\end{align}
    \paragraph{Eigenmodes, which are transformed under $\text{A}_1$ and $\text{A}_2$ irrep}
    {Integrals~\eqref{integralbig5c4v} and~\eqref{integralbig6c4v} for $\mathbf{E}^{2\omega, \text{A}_1}_{j} \left(\mathbf{r}^{\prime}\right)$, and~\eqref{integralbig7c4v},~\eqref{integralbig8c4v} for $\mathbf{E}^{2\omega, \text{A}_2}_{j} \left(\mathbf{r}^{\prime}\right)$ give zero.}
    \paragraph{Eigenmodes, which are transformed under $\text{B}_1$ and $\text{B}_2$ irrep}
    {Coupling integrals~\eqref{integralbig9c4v},~\eqref{integralbig10c4v} for the eigenmode $\mathbf{E}^{2\omega,\text{B}_1}_{j} \left(\mathbf{r}^{\prime}\right)$:}
\begin{align}
    \label{integral3mequal4s-3}
    &\ket{L_m}\colon \; \text{D}_j \propto c^{\text{B}_1}_j, 
    \\\nonumber
    &\ket{R_m}\colon\; \text{D}_j \propto c^{\text{B}_1}_j.
\end{align}
    {Coupling integrals~\eqref{integralbig11c4v},~\eqref{integralbig12c4v} for the eigenmode $\mathbf{E}^{2\omega,\text{B}_2}_{j} \left(\mathbf{r}^{\prime}\right)$:}
\begin{align}
    \label{integral4mequal4s-3}
    &\ket{L_m}\colon \; \text{D}_j \propto c^{\text{B}_2}_j, 
    \\\nonumber
    &\ket{R_m}\colon\; \text{D}_j \propto c^{\text{B}_2}_j.
\end{align}
    \item {$m^{\text{in}}=4s-1$, where $ s \in \mathbb {Z}_+$}:
    \paragraph{Eigenmodes, which are transformed under $\text{E}$ irrep}
        {Coupling integrals~\eqref{integralbig1c4v},~\eqref{integralbig2c4v} for the eigenmode $\mathbf{E}^{2\omega, x+\iu y}_{j} \left(\mathbf{r}^{\prime}\right)$:}
\begin{align}
    \label{integral1mequal4s-1}
    &\ket{L_m}\colon \; \text{D}_j \propto \tilde{C}_j \eu^{+3\iu\beta},
    \\\nonumber
    &\ket{R_m}\colon\; \text{D}_j \propto \tilde{C}_j^{\prime} \eu^{+3\iu\beta}.
\end{align}
    {Coupling integrals~\eqref{integralbig3c4v},~\eqref{integralbig4c4v} for the eigenmode $\mathbf{E}^{2\omega, x-\iu y}_{j} \left(\mathbf{r}^{\prime}\right)$:}
\begin{align}
    \label{integral2mequal4s-1}
    &\ket{L_m}\colon \; \text{D}_j \propto \tilde{C}^{\prime}_j \eu^{-3\iu\beta},
    \\\nonumber
    &\ket{R_m}\colon\; \text{D}_j \propto \tilde{C}_j \eu^{-3\iu\beta}.
\end{align}
    \paragraph{Eigenmodes, which are transformed under $\text{A}_1$ and $\text{A}_2$ irrep}
    {Integrals~\eqref{integralbig5c4v},~\eqref{integralbig6c4v} for  $\mathbf{E}^{2\omega, \text{A}_1}_{j} \left(\mathbf{r}^{\prime}\right)$, as well as~\eqref{integralbig7c4v},~\eqref{integralbig8c4v} for $\mathbf{E}^{2\omega, \text{A}_2}_{j} \left(\mathbf{r}^{\prime}\right)$ give zero.}
        \paragraph{Eigenmodes, which are transformed under $\text{B}_1$ and $\text{B}_2$ irrep}
    {Integrals~\eqref{integralbig9c4v},~\eqref{integralbig10c4v} for the eigenmode $\mathbf{E}^{2\omega,\text{B}_1}_{j} \left(\mathbf{r}^{\prime}\right)$:}
\begin{align}
    \label{integral3mequal4s-1}
    &\ket{L_m}\colon \; \text{D}_j \propto \tilde{c}^{\text{B}_1}_j, 
    \\\nonumber
    &\ket{R_m}\colon\; \text{D}_j \propto \tilde{c}^{\text{B}_1}_j.
\end{align}
    {Integrals~\eqref{integralbig11c4v},~\eqref{integralbig12c4v} for the eigenmode $\mathbf{E}^{2\omega,\text{B}_2}_{j} \left(\mathbf{r}^{\prime}\right)$:}
\begin{align}
    \label{integral4mequal4s-1}
    &\ket{L_m}\colon \; \text{D}_j \propto \tilde{c}^{\text{B}_2}_j,
    \\\nonumber
    &\ket{R_m}\colon\; \text{D}_j \propto \tilde{c}^{\text{B}_2}_j.
\end{align}
\end{enumerate}
To summarize, depending on the value of the projection of the angular momentum \(m^{\text{in}}\) of the incident vortex, we have four cases. 
However, as in the previous example (Sec.~\ref{sec:GaAsC3v}), due to the somewhat similar expressions for the coupling integrals which are proportional to either the exponents \(\eu^{+ 3\iu \beta}\), \(\eu^{- 3\iu \beta}\), or just constants, we can consider only one of them.
Let us proceed with further calculations, assuming \(m^{\text{in}}=4s-3\), where \( s \in \mathbb{Z}_+\).
\subsubsection{Second-harmonic electric field}
Let us obtain expressions for the second-harmonic electric field $\mathbf{E}^{2\omega}$, assuming $m^{\text{in}}=4s-3$, where $ s \in \mathbb {Z}_+$. {This means that we use coupling integrals~\eqref{integral1mequal4s-3},~\eqref{integral2mequal4s-3},~\eqref{integral3mequal4s-3}, and~\eqref{integral4mequal4s-3}}
\paragraph{Eigenmodes, which are transformed under $\text{E}$}
We calculate second-harmonic electric field for the eigenmode $\mathbf{E}^{2\omega, x+\iu y}_{j} \left(\mathbf{r}^{\prime}\right)$~\eqref{integral1mequal4s-3}:
\begin{align}
    \label{integral1fieldfour}
    &\ket{L_m}\colon  \; \mathbf{E}^{2\omega} \propto \sum_{j } \mathbf{E}^{2\omega, x+\iu y}_{j}(\mathbf{r})C_j\eu^{+3\iu\beta},
    \\\nonumber
    &\ket{R_m}\colon \;\mathbf{E}^{2\omega} \propto \sum_{j } \mathbf{E}^{2\omega, x+\iu y}_{j}(\mathbf{r}) C_j^{\prime} \eu^{+3\iu\beta},
\end{align}
for the eigenmode $\mathbf{E}^{2\omega, x-\iu y}_{j} \left(\mathbf{r}^{\prime}\right)$~\eqref{integral2mequal4s-3}:
\begin{align}
    \label{integral2fieldfour}
    &\ket{L_m}\colon  \; \mathbf{E}^{2\omega} \propto \sum_{j }\mathbf{E}^{2\omega, x-\iu y}_{j}(\mathbf{r})C_j^{\prime} \eu^{-3\iu\beta},
    \\\nonumber
    &\ket{R_m}\colon \;\mathbf{E}^{2\omega} \propto \sum_{j } \mathbf{E}^{2\omega, x-\iu y}_{j}(\mathbf{r})C_j \eu^{-3\iu\beta},
\end{align}
for the eigenmode $\mathbf{E}^{\text{B}_1}_{j}\left(\mathbf{r}^{\prime}\right)~\eqref{integral3mequal4s-3}$:
\begin{align}
    \label{integral1fieldfourB1}
    &\ket{L_m}\colon  \; \mathbf{E}^{2\omega} \propto \sum_{j } \mathbf{E}^{\text{B}_1}_{j}(\mathbf{r})c_j^{\text{B}_1},
    \\\nonumber
    &\ket{R_m}\colon \;\mathbf{E}^{2\omega} \propto \sum_{j } \mathbf{E}^{\text{B}_1}_{j}(\mathbf{r})c_j^{\text{B}_1},
\end{align}
and, finally, for the eigenmode $\mathbf{E}^{\text{B}_2}_{j}\left(\mathbf{r}^{\prime}\right)$~\eqref{integral4mequal4s-3}:
\begin{align}
    \label{integral1fieldfourB2}
    &\ket{L_m}\colon  \; \mathbf{E}^{2\omega} \propto \sum_{j } \mathbf{E}^{\text{B}_2}_{j}(\mathbf{r})c_j^{\text{B}_2},
    \\\nonumber
    &\ket{R_m}\colon \;\mathbf{E}^{2\omega} \propto \sum_{j }  \mathbf{E}^{\text{B}_2}_{j}(\mathbf{r})c_j^{\text{B}_2}.
\end{align}
\subsubsection{Total integral second harmonic intensity}
We calculate the total integral second harmonic intensity $I^{2\omega}$ in a two modes approximation, using the decomposistions of the field above $\mathbf{E}^{2\omega}$~\eqref{integral1fieldfour},~\eqref{integral2fieldfour},~\eqref{integral1fieldfourB1},~\eqref{integral1fieldfourB2}. 
{We take a sum only of two modes with indexes $i$ and $j$. 
Recall that we consider eigenmodes transformed under $\text{E}$ irrep separately because modes transformed under different irreps don't interfere with each other.} 
A left-handed and right-handed beam excites the eigenmodes $ \mathbf{E}^{2\omega,x+\iu y}_{j}({\mathbf{r}})$, and the second-harmonic intensity $I_{{\ket{L_m}}}^{2\omega}$ and $I_{{\ket{R_m}}}^{2\omega}$ can be written as follows:
\begin{align}
    \label{intensityLCPRCPfour}
    &I_{{\ket{L_m}}}^{2\omega} \propto    \int_{\text{sph}} \dd V \bigg|\mathbf{E}^{2\omega, x+\iu y}_{j}(\mathbf{r}) \widetilde{C}_j \eu^{+3\iu\beta} + 
     \eu^{\iu \alpha}\mathbf{E}^{2\omega, x+\iu y}_{i}(\mathbf{r}) \widetilde{C}_i \eu^{+3\iu\beta}\bigg|^2,
    \\
    \label{intensityLCPRCPfourzwei}
    &I_{{\ket{R_m}}}^{2\omega} \propto \int_{\text{sph}} \dd V \bigg|\mathbf{E}^{2\omega, x+\iu y}_{j}(\mathbf{r})\widetilde{C}_j^{\prime}\eu^{+3\iu\beta} +
     \eu^{\iu \alpha}\mathbf{E}^{2\omega, x+\iu y}_{i}(\mathbf{r})\widetilde{C}_i^{\prime} \eu^{+3\iu\beta}\bigg|^2,
\end{align}
where in the explicit form the phase $\alpha$ between eigenmodes $\mathbf{E}^{2\omega,x+\iu y}_{i,j}({\mathbf{r}})$ was extracted from coefficients $C_{j,i}, C_{j,i}^{\prime}$, and, as a consequence the new coefficients $\widetilde{C}_{j,i}, \widetilde{C}_{j,i}^{\prime}$ were introduced. 
Interference terms are proportional to:
\begin{align}
    &I_{{\ket{L_m}}}^{2\omega,\text{interf.}} \propto   \int_{\text{sph}} \dd V \eu^{-\iu\alpha}\mathbf{E}^{2\omega, x+\iu y}_{j}(\mathbf{r})\left(\mathbf{E}^{2\omega, x+\iu y}_{i}(\mathbf{r})\right)^* A_{ji} 
    \label{intensityLCPRCPfourinterf} + \text{c.c.},
    \\
    &I_{{\ket{R_m}}}^{2\omega,\text{interf.}} \propto \int_{\text{sph}} \dd V  \eu^{-\iu\alpha}\mathbf{E}^{2\omega, x+\iu y}_{j}(\mathbf{r})\left(\mathbf{E}^{2\omega, x+\iu y}_{i}(\mathbf{r})\right)^* A^\prime_{ji} 
    \label{intensityLCPRCPfourzweiinterf} + \text{c.c.},
\end{align}
where $I_{{\ket{L_m}},{\ket{R_m}}}^{2\omega,\text{interf.}}$ are related to the interference terms from the expressions~\eqref{intensityLCPRCPfour} and~\eqref{intensityLCPRCPfourzwei} for the total integral second harmonic intensity $I_{{\ket{L_m}},{\ket{R_m}}}^{2\omega}$, and new coefficients $A_{ji}, A^\prime_{ji}$ contain all combinations of the old coefficients $\widetilde{C}_{j,i}, \widetilde{C}_{j,i}^{\prime}$.
Analogously, we obtain the total integral second harmonic intensity $I_{{\ket{L_m}},{\ket{R_m}}}^{2\omega}$ for two eigenmodes $\mathbf{E}^{2\omega,x-\iu y}_{j}({\mathbf{r}})$ excited by a left-handed and right-handed beam:
\begin{align}
    \label{intensityLCPRCPfour2} 
    I_{{\ket{L_m}}}^{2\omega} \propto  \int_{\text{sph}} \dd V \bigg|\mathbf{E}^{2\omega, x-\iu y}_{j}(\mathbf{r})\widetilde{C}_j^{\prime} \eu^{-3\iu\beta} 
    + \eu^{\iu \alpha}\mathbf{E}^{2\omega, x-\iu y}_{i}(\mathbf{r})\widetilde{C}_i^{\prime} \eu^{-3\iu\beta}\bigg|^2,
    \\
    \label{intensityLCPRCPfour2zwei}
    I_{{\ket{R_m}}}^{2\omega}  \propto  \int_{\text{sph}} \dd V \bigg|\mathbf{E}^{2\omega, x-\iu y}_{j}(\mathbf{r})\widetilde{C}_j \eu^{-3\iu\beta} 
    + \eu^{\iu \alpha}\mathbf{E}^{2\omega, x-\iu y}_{i}(\mathbf{r})\widetilde{C}_i \eu^{-3\iu\beta}\bigg|^2.
\end{align}
Interference terms are proportional to:
\begin{align}
    \label{intensityLCPRCPfour2interf}
    &I_{{\ket{L_m}}}^{2\omega,\text{interf.}} \!\propto \! \int_{\text{sph}}\! \dd V \eu^{-\iu\alpha}\mathbf{E}^{2\omega, x-\iu y}_{j}(\mathbf{r})\left(\mathbf{E}^{2\omega, x-\iu y}_{i}(\mathbf{r})\right)^*\! A^\prime_{ji} + \text{c.c.},
    \\
    \label{intensityLCPRCPfour2zweiinterf}
     &I_{{\ket{R_m}}}^{2\omega,\text{interf.}}  \!\propto \!\int_{\text{sph}}\! \dd V \eu^{-\iu\alpha}\mathbf{E}^{2\omega, x-\iu y}_{j}(\mathbf{r})\left(\mathbf{E}^{2\omega, x-\iu y}_{i}(\mathbf{r})\right)^*\! A_{ji}  + \text{c.c.},
\end{align}
where we should also note that the terms of the form $\mathbf{E}^{2\omega, x+\iu y}_{j}(\mathbf{r})\!\left(\mathbf{E}^{2\omega, x+\iu y}_{i}(\mathbf{r})\right)^* \!=\mathbf{E}^{2\omega, x-\iu y}_{j}(\mathbf{r})\!\left(\mathbf{E}^{2\omega, x-\iu y}_{i}(\mathbf{r})\right)^*$ are equal to each other due to the achiral symmetry of the nanoparticle.

{We should pay attention to the following:
\begin{enumerate}
    \item The eigenmode is excited 
    by only one term of the polarization, since $\Delta m_\chi$ is never equal to $\mathfrak n \nu$, where $\nu \in \mathbb{Z}$. 
    \item The multiplier $e^{\iu \Delta m_\chi\beta}$ finally disappears.
    \item A lot of coefficients are equal to each other.
\end{enumerate}}
\subsubsection{Second-harmonic circular dichroism}
Therefore, using this equality and comparing expressions above~\eqref{intensityLCPRCPfourinterf}, and~\eqref{intensityLCPRCPfour2zweiinterf}, as well as~\eqref{intensityLCPRCPfourzweiinterf}, and~\eqref{intensityLCPRCPfour2interf}, we can conclude that interference terms $I^{2\omega,\text{interf.}}$, and, as a consequence, the total integral second harmonic intensity $I^{2\omega}$ look absolutely the same for a left- and right-handed beams. 

Comment: the same independence of intensity on the incident vortex polarization can be obtained for interference terms $I^{2\omega,\text{interf.}}$ for interference of two excited eigenmodes transformed under irreducible representation $\text{B}_1$, as well as under irreducible representation $\text{B}_2$ with help of expressions~\eqref{integral1fieldfourB1}, and~\eqref{integral1fieldfourB2}. 
Consequently, the total integral second harmonic intensity $I^{2\omega}$ doesn't depend on the sign of the polarisation of the incident vortex. 
Let us remind that despite the fact that all expressions for intensity were obtained for the specific case, when $m^{\text{in}}=4s-3$, $s \in \mathbb {Z}_+$, it's not necessary to calculate all other cases because the expressions and the final result will be always the same due to the same expressions for coupling integrals. 
Overall, the final conclusion can be introduced: circular dichroism in the second-harmonic signal doesn't appear in a nanostructure GaAs[111]$\parallel$z with symmetry $\text{C}_{4\text{v}}$ under any circumstances.

\section{Derivation of the main condition for a general case}
\label{universalmaincondition}
{We've presented two examples, the choice of which is linked, firstly, to the popularity of the materials considered, and secondly, to the nontrivial nature of the examples themselves. 
In this section, we will identify the common patterns observed, ultimately leading to a simple general formula.
\begin{enumerate}
    \item {An incident vortex beam excites intrinsic modes, consisting of the same multipoles as the beam itself (with the same $m^{\text{in}}$), as well as those that {are transformed under the} same irreducible representation. 
    By symmetry considerations (for instance, a function describing the structure's shape expands into series by $\cos(\mathfrak{n}\varphi))$, these will be multipoles with $\pm m^{\text{in}}+\mathfrak{n}\nu$, $\nu\in \mathbb Z$. 
    It is important to note that at this stage, a mirror beam excites mirror modes, regardless of whether they were originally degenerate (transformed according to representation E) or not (transformed according to other representations).}
    \item {Subsequently, it is necessary to derive an expression for the nonlinear polarization, into which the field inside the particles enters $q$ times. 
    The factor responsible for rotational symmetry, $\sum_\nu [...] e^{\iu(\pm m^{\text{in}}+\mathfrak{n}\nu)\varphi}$, $\nu\in \mathbb Z$, multiplied by itself $q$ times, yields $\sum_\nu [...]e^{\iu(\pm qm^{\text{in}}+\mathfrak{n}\nu)\varphi}$, $\nu\in \mathbb Z$.}
    \item {At the previous stage, the symmetry of the particle was already considered. 
    We regard the particle as fixed relative to the coordinate system, and the relative rotation of the lattice is represented by the rotation of the tensor describing this lattice.
    Thus, one of the most important points is that tensor contains terms of the form $\sum_{m_\chi}[...] e^{\iu m_\chi (\varphi-\beta)}$.
    This does not depend on the handedness, since it characterizes the lattice.  
    Note that phase differences between such terms are proportional to $\Delta m_\chi \beta$.}
    \item{Another crucial observation is that nonlinear susceptibility is real (contains rather $\cos(m_\chi\varphi)$ or $\sin(m_\chi\varphi)$) terms than exponential). 
    Thus polarizability can always be written in a ``mirror'' form for other handedness (and also terms with $m_\chi=0$).
    Generally, for some pair with $m_\chi$ and $-m_\chi$: 
    \begin{align}
    \label{generalpol}
     \mathbf{P}^{q\omega}(r,z,\varphi)\propto
    & \bigg\{\sum\limits_{\nu} \mathbf{P}_{(m_\chi + qm^{\text{in}} + \mathfrak{n}\nu)}^{q\omega}\eu^{\pm (m_\chi + qm^{\text{in}} + \mathfrak{n}\nu) \iu\varphi}\bigg\}\eu^{\mp m_\chi\iu\beta}+
    \\\nonumber
    + &\bigg\{
    \!\sum\limits_{\nu} \mathbf{P}_{(-m_\chi+qm^{\text{in}}+ \mathfrak{n}\nu)}^{q\omega}\eu^{\pm (-m_\chi+qm^{\text{in}}+ \mathfrak{n}\nu) \iu\varphi}\!\bigg\}\eu^{\pm m_\chi\iu\beta} 
    \end{align}
    At this stage it is not clear, how the dichroism could appear. 
    Note that in general case, first and second terms exciting eigenmodes of a different symmetry (with different total m).
    For example, if $m_\chi=2$, $\mathfrak n = 3$, for the first term we have $qm^{\text{in}}+2+3\nu$ and $qm^{\text{in}}-2+3\nu$, which never coincide for $\nu \in \mathbb Z$ or any $qm^{\text{in}}$. }
    \item{Let us consider the case, when these terms can coincide, exciting the eigenmodes of the same symmetry, e.g.  $m_\chi=2$, $\mathfrak n = \Delta m_{\chi}= 4$. 
    This is the stage, where the dichroism appears. Imagine the first term of the polarization excites one mode, and the second one excites another mode of the same symmetry. 
    One of these two modes can be in the vicinity of resonance, thus assume that the mode, excited by the second term has a relative phase $\alpha$. 
    Note that $\alpha$ is the same (with the same sign) for both beam handednesses, because these two cases mirror each other. 
    On the other hand, multipliers with $\beta$ have opposite signs for left and right beams.
    Thus, in one case, the relative phase will be $\alpha+\Delta m_\chi \beta$, and in the other $\alpha-\Delta m_\chi \beta$, which provides constructive and destructive interference.
    Each mode will be excited by both terms of the polarization, which makes the calculations cumbersome, but does not change the general idea.}
    \item{The main formula immediately follows from the fact that to obtain the dichroism, different terms of nonlinear polarization should excite eigenmodes of the same symmetry, and also phases should be different, which does not fulfill for $\beta=\pi\nu/s$.}
\end{enumerate}
The two described cases provide the consideration of all these processes in detail, and it can be seen that regardless of the symmetry of the excited eigenmodes on each stage, this general consideration is still applicable. 
{We also note that not the symmetry of the excited eigenmodes that matters, but the difference between the rotational symmetry of different modes excited.
Due to the modified law of total angular momentum projection conservation, each mode consists of the multipoles, whose $m$ differs by $\mathfrak n \nu$. 
Different terms of the tensor refer to the excitation of the modes, for which $m$ differs by $\Delta m_\chi$.
If the multipolar content of the modes, ``excited by different terms of the tensor'' coincide, the modes can interact.
Mirror reflection of the incident beam changes the sign of additional phase between these modes due to the lattice rotation. 
Hence, the considerations eventually do not depend on the order of the harmonic or absolute value of the incident angular momentum projection.}}
{\section{Numerical Calculations}
\label{sec:app:plots}}
{In Section~\eqref{universalmaincondition}, we introduced a universal recipe for defining the possibility of circular dichroism (CD) appearance in perturbative harmonic generation, based on the illustrative examples previously discussed. 
Let us apply it to the second-harmonic generation and the nanostructure BaTiO$_3$[001]$\|x$ with symmetry $\text{D}_{3\text{h}}$ irradiated by an incident vortex beam with the total angular momentum projection on the $z$-axis $m^{\text{in}} = 3$.
}

{According to the step 1, vortex beam excited intrinsic modes, consisting of the multipoles with $m=\pm 3 + 3\nu, \; \nu \in \mathbb{Z}$. Let us refer to our previous work~\cite{Nikitina2023-Nonlinearcirculardi} and find the values $m_{\chi}$ for the crystalline lattice BaTiO$_3$[001]$\|x$: $m_{\chi} \in \{ \pm 1, \; \pm 3 \}$. 
Using it and relying on the general form of the nonlinear vector polarization $\mathbf{P}^{q\omega}(r,z,\varphi)$~\eqref{generalpol}, we can determine which eigenmodes can be excited at the doubled frequency. 
In fact, eigenmodes with $\pm m_{\chi} \pm 2m^{\text{in}} +3\nu =\pm m_{\chi} \pm 6 +3\nu, \; \nu \in \mathbb{Z}$, with the additional phase $\mp m_{\chi}\beta$, where the sign $\pm$ denotes the handedness of the incident beam, are expected to be excited. 
Particularly, there are eigenmodes with $\pm 7+ 3\nu$, $\pm 5+ 3\nu$, $\pm 9+ 3\nu$, and $\pm 3+ 3\nu$, and the main thing is which of them can interfere with each other. In this particular case only eigenmodes with $\pm 9+ 3\nu$ and $\pm 3+ 3\nu$ have the same symmetry and interfere. The phase's difference between them is equal to $\mp 6\beta$. Therefore, for the possibility of SH-CD in considered system it's required to rotate the nanostructure at $\beta \neq \pi/6$ with respect to the crystalline lattice. }

{For the validation of our theoretical results, we present numerical calculations of the SH intensity in COMSOL Multiphysics\textsuperscript{\textregistered} for the described system above with the following parameters: a trimer that has the $\text{D}_{3\text{h}}$ symmetry consisting of three equidistant discs each of which has the diameter equal to 500 nm and height equal to 450 nm. 
The distance between the discs was equal to 55 nm and refractive index  of the material was equal to 3.5 which is close to the typical values of refractive index for dielectric materials.
Incident wave was not a particular vortex beam, but just a circularly polarized plane wave, multiplied by $e^{\pm2 \iu \varphi}/r$, where $r=\sqrt{x^2+y^2}$. 
Generally, this choice is arbitrary, and one can take any incident field, which obey the symmetry requirements.
Second harmonic computation was conducted, using standard approach, described, e.g. in~\cite{Saerens2020-EngineeringoftheSe, Gladyshev2024-FastSimulationofLi}.
In Fig.~\ref{numericalpieceofart}, in the range of 1580--1630 nm, we illustrate the SH intensity in arbitrary units for right- and left-handed beams for angle $\beta = 15^\circ$. 
As expected, SH-CD can be obtained because SH intensity differs for different handednesses.
In Fig.~\ref{numericalpieceofart}  (in the insert), absolute value of SH-CD is shown for different values of angle $\beta$ that we numerically calculated using the eq.~\eqref{SHCDsecond}: SH-CD appears for all $\beta \neq \pi/6$ which is consistent with the speculations described above.\\
Relatively small values of dichroism are explained by arbitrarily chosen wavelengths, structure, and sizes.
Higher values could be achieved if the proper engineering is conducted. 
Particularly, eigenmodes, which contribute to the dichroism, should be close to their resonances on the SH wavelength.
\begin{figure}[ht!]
    \begin{minipage}[h]{1\linewidth}
        \center{\includegraphics[width=0.7\linewidth]{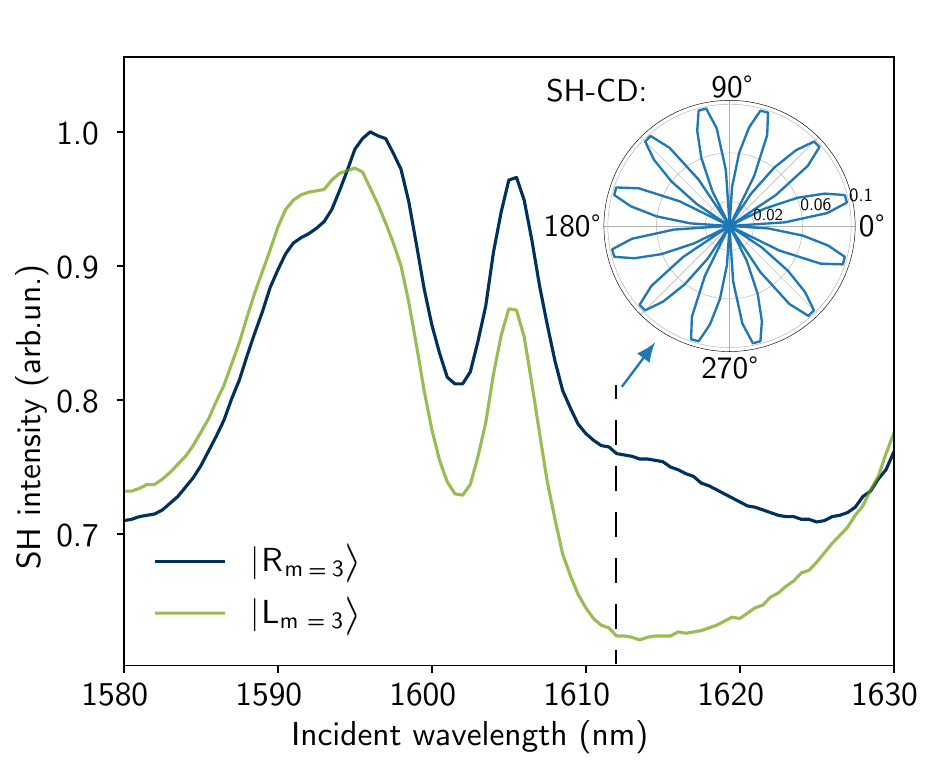}}
        \label{SHintensity}
    \end{minipage}
    \caption{{Numerically calculated values of SH intensity in arbitrary units for the nanostructure BaTiO$_3$[001]$\|x$ with symmetry $\text{D}_{3\text{h}}$ for right- and left-handed incident beams with $m=3$. 
    The angle $\beta$ between the structure's symmetry plane and the crystalline lattice was chosen $15^\circ$. 
    On the insert: Numerically calculated absolute values of SH-CD depending on angle $\beta$. 
    SH-CD is possible for all $\beta \neq \pi/6$.}}
    \label{numericalpieceofart}
\end{figure}
}
\section{Discussions}
\label{sec:discussions}
\subsection{Cascaded generation}
{Cascading is also possible to achieve high-harmonic generation~\cite{Meredith1981-Cascadinginoptical, Zalogina2023-High-harmonicgenerat, Celebrano2019-EvidenceofCascaded} via lower-order nonlinearities. 
{Let us} consider the cascading process {in stages}. 
In the first step, the condition for the dichroism will be just the same as for the nonlinearity of the considered order, e.g., the second harmonic.
If there is a dichroism, there is no need to go further.
However, if in the first step, the dichroism does not appear, we should consider the next process with a different incident field. 
Here, the consideration is more tricky than the usual one. \\
The simplest example is the third harmonic cascaded process via $\hat\chi^{(2)}$. 
The second step is a sum-frequency process $2\omega+\omega$.
One of the fields ($2\omega$) will contain several terms of different symmetry, with $m$, differ{ing} by $\Delta m_\chi$ since they are being generated in the first step ``by different components of the tensor''.
They also have corresponding phases, depending on $\beta$.
Let us again consider~\eqref{generalpol} (and only one handedness). 
After the first step, the first term will generate the SH field with the momentum projections $m_\chi + 2m^{\text{in}} + \mathfrak{n}\nu$ and phase $-m_\chi\beta$, and the second with $-m_\chi + 2m^{\text{in}} + \mathfrak{n}\nu$ and phase $m_\chi\beta$.\\
Then these terms participate in the sum-frequency process as a part of the incident field. 
Let us consider the angular momentum projection, generated by the first term on the second step. 
It will contain two terms, with $m_\chi+m_\chi + 2m^{\text{in}}+m^{\text{in}} + \mathfrak{n}\nu$ and phase $-2m_\chi\iu\beta$ and $-m_\chi+m_\chi + 2m^{\text{in}}+m^{\text{in}} + \mathfrak{n}\nu$ and phase 0.
The second term in the second step will produce
$-m_\chi+m_\chi + 2m^{\text{in}}+m^{\text{in}} + \mathfrak{n}\nu$ and phase $0$ and $-m_\chi-m_\chi + 2m^{\text{in}}+m^{\text{in}} + \mathfrak{n}\nu$ and phase $2m_\chi\iu\beta$. 
Even though two of these four terms will have the same $m$, and the same symmetry and could be the candidates to produce dichroism, their phase does not depend on $\beta$. 
This will happen to all such terms in other processes, because each $m_\chi$ carries its corresponding phase, and terms with the same $m$ will thus possess the same phase, which cannot lead to dichroism.
The first and fourth terms in general, can have the same symmetry in the second step (2$\Delta m_\chi = \mathfrak n \nu$), even if $\Delta m_\chi \neq \mathfrak n \nu$ (since we restrict ourselves to the case when we do not have dichroism in the first step).
However, this does not happen often. For example, if one can imagine the lattice which provides only $\Delta m_\chi=2$, but $\mathfrak n = 4$, this will be the case.
Another tricky case is when we have $\mathfrak n > 6$.
For the second harmonic, dichroism is not possible due to~\eqref{highestn}.
But here, for the first and fourth terms we have effectively $2m_\chi$ for each one, thus $\Delta m_\chi^\text{eff} \leq 12$.
This difference, though insignificant, does not allow us to apply the same formula for cascade processes directly. 
However, the selection rules will be still similar, and the method in general is still applicable.}

\subsection{Angular incidence and substrate}
{All these considerations are applicable to a normal incidence of a beam, aligned with the $z$-axis, which is also the symmetry axis of a nanostructure.
The inclination of the beam would break the symmetry. 
In this case, we suggest considering a perfectly aligned beam, but a rotated nanostructure, which effectively will have a lower symmetry and the crystalline lattice should be rotated (the Matlab code for the tensor rotation could be found in~\cite{Nikitina2023-Nonlinearcirculardi}).
We also note that the presence of substrate does not alter the results, because it does not change the symmetry which is of interest to us.
The dichroism will appear both in transmission and reflection.}

\subsection{{Other subtleties}}
{Throughout the paper, we avoided considering the case $m^{\text{in}}=0$, while the Poincare sphere also exists for this case, and multipolar building blocks of the corresponding beams are  $\vb N_{e01}\pm \vb M_{e01}$.
Initially, it might appear that the nonlinear polarization~\eqref{generalpol} terms are totally symmetric,  and polarization coincides for both handednesses so the circular dichroism can not be obtained. 
However, this is the case where particular form of $\vb P^{q\omega}_{(\pm m_\chi+\mathfrak{n} \nu)} (r,z)$ plays a role. 
Such terms are not symmetric in general, reflecting the interplay of the lattice with the field, carrying particular helicity~\cite{Klimmer2021-All-opticalpolarizat, deCeglia2024-Nonlinearspin-orbit}, thus our considerations are still valid and main formula applicable.}
Additionally, we wish to clarify that our analysis and results are not analogous to  considering nanoparticles as ``big molecules'' with their hyperpolarizability~\cite{Stone2013-TheTheoryofIntermo}. 
We assume that the nanostructure is large enough so its material is described by nonlinear susceptibility tensor, and surface effects are weak~\cite{Timbrell2018-Acomparativeanalysi}. 
Thus, even if a ``big molecule'' can possess no symmetry at all we can still have zero circular dichroism in this approximation.

\subsection{Other definitions of dichroism}
\label{sec:other}
{Circular dichroism under vortex beam illumination can be introduced in several different ways~\cite{Ye2019-ProbingMolecularChi}.
Besides mirror-reflected vortex beams with opposite signs of $m$ one can consider changing $\ell$ while keeping $\lambda$ the same (see Appendix~\ref{app:mult}) or changing $\lambda$ while keeping $\ell$ the same. 
However, for both cases, in nanostuctures of considered symmetries a different response will be always observed even in the linear regime. 
This happens due to the fact that incident beams have different $|m|$ values, but eigenmodes with different $|m|$ belong to different irreps and have different resonance frequencies~\cite{Gladyshev_Frizyuk_Bogdanov_2020, Xiong_Xiong_Yang_Yang_Chen_Wang_Xu_Xu_Xu_Liu_2020}.
Another possible case is considering incident beams not with single $m$, but some linear combination, e.g. $m_1$ and $m_2$. 
They can be written in the form $\ket{R_{m_1}}+e^{\iu\gamma}\ket{R_{m_2}}$ or $\ket{L_{m_1}}+e^{-\iu\gamma}\ket{L_{m_2}}$.
This case is tricky, because we will have many different terms in~\eqref{generalpol}. 
One can not immediately tell, if the nonlinear dichroism is observable, because in the case, e.g. $m_1-m_2 = \Delta m_\chi$ some terms of the polarization can interfere, even though we didn't have the dichroism for a single $m$. 
Such cases should be considered separately even in the linear regime, where the dichroism could also appear if $m_1-m_2 = \mathfrak n \nu$ (this is similar to the results obtained in~\cite{Surzhykov2015-Interactionoftwiste}), depending on the phase $\gamma$.
Indeed, in this case two terms of the incident field will both excite eigenmodes with the same symmetry, and with a relative phase $\alpha$, which is the same for $\ket {R}$ and $\ket {L}$ (while the $\gamma$ has a different sign). 
Thus, the overall intensity will differ for similar reasons as in the previously discussed nonlinear case, where $\gamma$ plays similar role as $\beta$.
}

\section{Dichroism under vector beams illumination}
\label{vectorbeam}
{Another interesting case occurs with vector beam illumination (see~\eqref{vb_e}) and~\eqref{vb_o}).
This situation is analogous to linear dichroism in the linear regime. 
Excitation by radial and azimuthal beams is widely studied and has many applications~\cite{Das2015-Beamengineeringfor, Melik-Gaykazyan2018-SelectiveThird-Harmo, Bautista2012-Second-HarmonicGener, Kroychuk2020-EnhancedNonlinearLi}. 
Here, we discuss the general case of such beams.
In linear regime, we can have different response to $x$- and $y$-polarized plane waves in rectangular particle (C$_{2\text{v}}$ or D$_{2\text{h}}$), while we do not have a circular dichroism in this case  (when we consider the total intensity of the scattered wave).
Due to the similar reasons, one can obtain different response to $\ket{\text{B}_{me}}$ and $\ket{\text{B}_{mo}}$ in the linear regime:
\begin{equation}
    \text{e/o-D}^{\text{sca}}_{m} = \frac{(I_{\text{B}_{me}}^{\text{sca}}-I_{\text{B}_{mo}}^{\text{sca}})}{(I_{\text{B}_{me}}^{\text{sca}}+I_{\text{B}_{mo}}^{\text{sca}})},
\end{equation}
where $I_{\text{B}_{me/o}}^{\text{sca}}$ is total scattering intensity, integrated by a sphere in the far-field, as previously, under illumination by the corresponding vector beam.
The main condition to obtain this type of dichroism: 
\begin{equation}
    \exists\nu \in \mathbb{Z}, \ \ 2m = \mathfrak{n}\nu.
    \label{lincond}
\end{equation}
One can prove it in two different ways. 
From the representation theory~\cite{Gladyshev_Frizyuk_Bogdanov_2020}, one can see that if the modes excited by $\ket{\text{B}_{me}}$ and $\ket{\text{B}_{mo}}$ have different symmetries (are transformed according to different irreps), their resonance frequencies will be different (avoiding specific cases of accidental degeneration). 
Thus they will be excited with different phases and amplitudes, leading to different responses. 
When $ 2m \neq \mathfrak{n}\nu$, both modes are degenerate, and transformed according to the two-dimensional irrep E.
Each beam will excite one of two orthogonal degenerate eigenmodes.
The second way is to just look at the scattered field symmetry, using the general approach and the total angular momentum projection conservation, modified by a nanostructure.
Scattered fields by the $(|L_m\rangle \pm |R_m\rangle)$ will have different intensity in case when the fields produced by the first and the second term interfere. This happens, when they have the same symmetry, i.e. $\exists \nu, \nu'\colon \ m^{\text{in}}+\mathfrak n \nu = -m^{\text{in}}+\mathfrak n \nu'$. The condition~\eqref{lincond} immediately follows.
The relative rotation of the incident beams and nanoparticle will alter the results. 
At certain angles, the two incident beams are mirror-symmetric with respect to the symmetry plane of a nanostructure, and dichroism will not be observed.
We leave a detailed discussion of relative angles beyond the scope of this work.

Let us now assume that $ 2m \neq \mathfrak{n}\nu$ is satisfied, and look at the nonlinear dichroism, which is defined analogously:
\begin{equation}
    \text{e/o-D}_{m}^{q\omega} = \frac{(I_{\text{B}_{me}}^{q\omega}-I_{\text{B}_{mo}}^{q\omega})}{(I_{\text{B}_{me}}^{q\omega}+I_{\text{B}_{mo}}^{q\omega})}.
\end{equation}
Let us present the incident wave, which contributes $q$ times, as a tensor product:
\begin{equation}
    \underbrace{(|L_m\rangle \pm |R_m\rangle)\otimes (|L_m\rangle \pm |R_m\rangle) \dots \otimes(|L_m\rangle \pm |R_m\rangle)}_{q \text{ times}}.
\end{equation}
Expanding the brackets, we obtain several terms, for which we can consider the process separately.
Importantly, signs before some of them depend on the incident beam. 
If the ``beam-dependent'' terms have the same symmetry as independent ones, they will interfere with each other, leading to the different responses. 
Let us first consider second harmonic, for the convenience.
The terms $|L_m\rangle \otimes |L_m\rangle$ or $|R_m\rangle \otimes |R_m\rangle$ will have the same symmetry as $|L_m\rangle \otimes |R_m\rangle$ or $|R_m\rangle \otimes |L_m\rangle$ if the following condition is satisfied:
$\exists \nu,\nu'\colon \ m_\chi + 2 \cdot m^{\text{in}}+\mathfrak n \nu = m_\chi'+ m^{\text{in}} - m^{\text{in}} +\mathfrak n \nu' $, which converts into
\begin{equation}
    \exists\nu \in \mathbb{Z}, \ \ \Delta m_\chi+2m^{\text{in}}=\mathfrak n \nu.     
\end{equation}
If this condition is satisfied, we may obtain $\text{e/o-D}^{2\omega}_m$ for some relative angles.
For example, for GaAs C$_{3\text{v}}$ trimer with [111]$||z$ we have $\Delta m_\chi \in \{\pm 3, \pm 6\}$, and we will not obtain the SH dichroism for  $m^{\text{in}} = 2$, because $3(6)+4 \neq 3\nu $.
However, for the same trimer but crystalline lattice is oriented as [001]$||z$, we have $\Delta m_\chi = \pm 4$, and $-4+4=0$ is satisfied, leading to appearance of the SH dichroism for most relative angles. 
For high-harmonic generation, the condition is modified as follows:
\begin{align}
     \label{vectorcondq}
     \exists \nu, \nu'\colon \ \Delta m_\chi+2\nu'm^{\text{in}}=\mathfrak n\nu, \ \   \nu \in \mathbb Z, \\
     \nu' \in 2\mathbb Z+1, \ \ \nonumber \nu'\in [-q...q].  
\end{align}
To determine the specific relative angles between the beams, structure, and lattice at which dichroism may be prohibited, it is necessary to accurately account for the relative phases at each step. We leave this discussion beyond the scope of our paper, providing a foundation for further calculations.}

{Interestingly, this type of dichroism can be even obtained in cylindrically symmetric nanostructure if $\Delta m_\chi+2m^{\text{in}}=0$ (for the second harmonic). This is obvious, for example, for tensor which only has $\chi_{yyy}$ or $\chi_{xxx}$ component and linearly polarized plane waves. We've conducted the modelling for this case, and provide the results in Fig.~\ref{Fig:disc}.
Diameter of the cylinder is 500 nm, height is 450 nm, $m^{\text{in}}$ of the incident beams equals 3, and $\Delta m_\chi \in \{\pm 3, \pm 6\}$ for GaAs, [111]$||z$. $y=0$ symmetry plane is common for the even beam and a lattice. 
If $\Delta m_\chi+2m^{\text{in}} \neq 0 \ \  \forall \Delta m_\chi$, for example, for $m^{\text{in}} = 2$, the dichroism is not observed in the cylindrically-symmetric nanostructure.
\begin{figure}[ht!]
        \center{\includegraphics[width=0.6\linewidth]{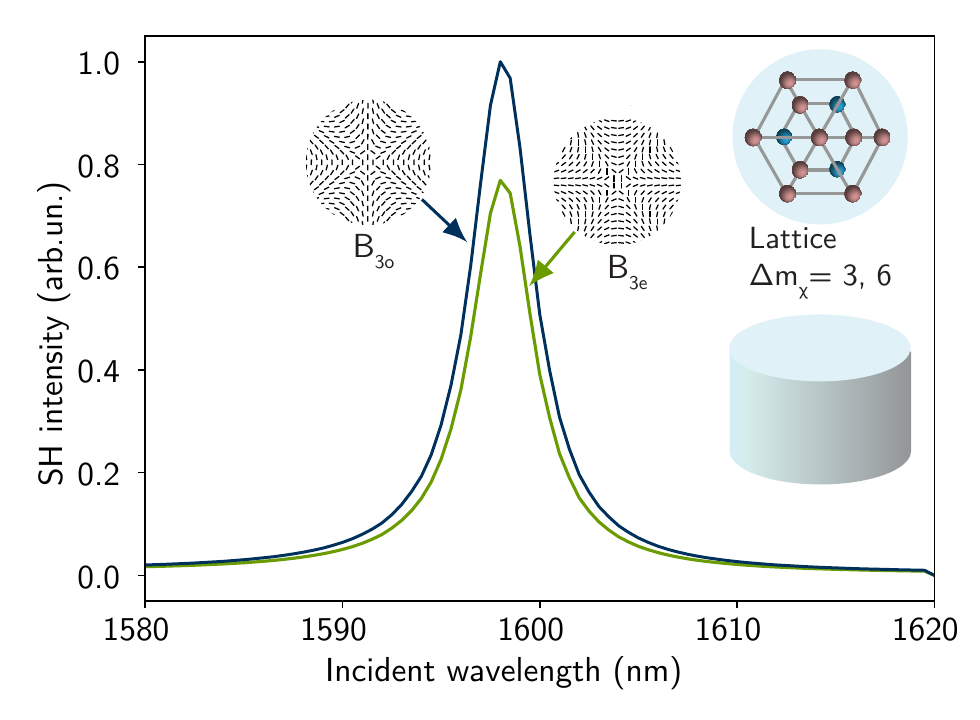}}
        \caption{Numerically calculated values of SH intensity in arbitrary units for the GaAs[111]$\|z$ for even and odd incident beams with $m = 3$. }
        \label{Fig:disc}
\end{figure}

}

\section{Conclusions}

\label{sec:conclusions}

To conclude, we performed a thoughtful theoretical analysis of nonlinear circular dichroism and provided the complete answer for all possible perturbative harmonic orders, incident vortex beams, and a broad range of materials and achiral nanostructures, which would help to accelerate the further analysis of such structures. 
Using these findings as a basis, we've discussed the possibility of linear and nonlinear dichroism under vector beams excitation.
These results can play a particularly important role for chiral sensing with nanostructures~\cite{Mohammadi2018-NanophotonicPlatform, Besteiro2017-AluminumNanoparticle, Garcia-Guirado2020-EnhancedChiralSensi, Czajkowski2022-Localversusbulkcir}, since determining the chiral response of the nanoparticle itself is crucial, as well as in nanoparticle fabrication~\cite{Toyoda2013-TransferofLightHel},
beams generation~\cite{Camacho-Morales2016-NonlinearGeneration},
optical trapping~\cite{Ng2010-TheoryofOpticalTra, Kozawa2010-Opticaltrappingofm, Moine2005-Opticalforcecalcula, 
Zhang2018-Nonlinearity-Induced, Gomez-Viloria2024-On-AxisOpticalTrapp} and manipulation~\cite{Mao2024-Switchableopticaltr, Bishop2003-Opticalapplicationa}, where the subtleties of the interaction between structured light and nano-objects are critical.

}

\section{Acknowledgements}

We express our gratitude to our readers for their engagement and to ChatGPT for its assistance with linguistic queries. 
Special thanks are due to Yuri Kivshar, Ivan Fernandez-Corbaton, Ivan Toftul, and Kirill Koshelev for fruitful discussions.
Numerical simulations of nonlinear dichroism were supported by the Russian Science Foundation (Project No. 22-12-00204). 
K.F. would like to acknowledge the financial support from the European Community through the ``METAFAST'' Project (H2020-FETOPEN-2018-2020, Grant No. 899673) and the Ministero Italiano dell’Istruzione (MIUR) through the ``METEOR'' Project (No. PRIN-2020, 2020EYLJT\_002).
\normalem
\bibliographystyle{unsrtnat}  

\bibliography{sample} 
\newpage
\newpage
\appendix
\section{Appendix}
{\subsection{On non-perturbative harmonic generation}\label{sec:app:pert}
Even though high harmonic generation is widely studied and observed in non-perturbative regime~\cite{Goulielmakis2022-Highharmonicgenerat, Vampa2014-TheoreticalAnalysis}, our study can be applied to the systems that are still under conditions the perturbative regime, or have a perturbative contribution~\cite{Zograf2022-High-HarmonicGenerat, Shcherbakov2021-Generationofevenan, Zalogina2023-High-harmonicgenerat}. 
The symmetry dependencies in non-perturbative regime should be studied separately~\cite{You2017-Anisotropichigh-harm, Jimenez-Galan2023-Orbitalperspectiveo}.
\subsection{$\hat\chi^{(3)}$ tensor in cylindrical coordinates}\label{sec:app:tens}
By the link~\cite{kuyzirf2024} we provide a Matlab code to compute the nonlinear susceptibility tensor in cylindrical coordinates for cubic symmetries. 
It could be easily rewritten for other tensors. One can imagine the procedure as writing tensor in the form of tensor product of basis vectors:
\begin{equation}
    \hat\chi^{(3)}=\chi^{(3)}_{ijkl} \vb e_i \otimes \vb e_j \otimes \vb e_k \otimes \vb e_l,
\end{equation}
where $\vb e_i$ are $ \hat{\vb x}, \  \hat{\vb y},$ or $   \hat{\vb z}$ and then rewriting each vector $\vb e_i$ in cylindrical coordinates according to
\begin{align}
    \hat{\vb x} = \hat{\bm \rho} \cos \varphi - \hat{\bm \varphi} \sin \varphi \\ \nonumber
    \hat{\vb y} = \hat{\bm \rho}  \sin \varphi + \hat{\bm \varphi} \cos \varphi 
\end{align}
we get the answer. 
For example, for two of the most common materials, Si ($m \overline 3 m$,	$O_h$ group), and GaAs  ($\overline 43m$, $T_d$ group) we have four independent components~\cite{Boyd2003}:
\begin{align}
    \chi^{(3)}_{xxxx}=\chi^{(3)}_{yyyy}=\chi^{(3)}_{zzzz} \\
    \chi^{(3)}_{xxyy}=\chi^{(3)}_{yyzz}=\chi^{(3)}_{zzxx}=\chi^{(3)}_{zzyy}=\chi^{(3)}_{yyzz}=\chi^{(3)}_{xxzz}\\ 
    \chi^{(3)}_{xzxz}=\chi^{(3)}_{xyxy}=\chi^{(3)}_{yzyz}=\chi^{(3)}_{yxyx}=\chi^{(3)}_{zyzy}=\chi^{(3)}_{zxzx}\\
    \chi^{(3)}_{xyyx}=\chi^{(3)}_{xzzx}=\chi^{(3)}_{yxxy}=\chi^{(3)}_{yzzy}=\chi^{(3)}_{zyyz}=\chi^{(3)}_{zxxz}.
\end{align}
For example, if we provide them with values $ 8, \ 4,\ 4, \ 4$ respectively, in cylindrical coordinates we obtain
\begin{align}
    \chi^{(3)}_{zzzz} & = 8 \\ 
    \chi^{(3)}_{\rho \rho \rho \rho} = \chi^{(3)}_{\varphi \varphi \varphi \varphi} & = 9 - \cos 4\varphi \\
     - \chi^{(3)}_{\rho \varphi \varphi \varphi} = - \chi^{(3)}_{ \varphi \rho \varphi \varphi} = - \chi^{(3)}_{ \varphi \varphi \rho \varphi} = - \chi^{(3)}_{ \varphi \varphi \varphi \rho} = \nonumber \\
     = \chi^{(3)}_{ \varphi \rho \rho \rho} = \chi^{(3)}_{\rho \varphi \rho \rho} = \chi^{(3)}_{\rho  \rho \varphi \rho} =\chi^{(3)}_{\rho  \rho \rho \varphi} & =   \sin 4\varphi \\
     \chi^{(3)}_{\rho \rho \varphi \varphi} = \chi^{(3)}_{\rho \varphi  \rho \varphi} = \chi^{(3)}_{\rho \varphi \varphi \rho } =
     \chi^{(3)}_{\varphi \rho \rho  \varphi} = \chi^{(3)}_{\varphi \varphi \rho \rho } = \chi^{(3)}_{\varphi \rho \varphi \rho} &  = \cos(4\varphi)+3 \\
     \chi^{(3)}_{\varphi z \varphi z} = \chi^{(3)}_{\varphi z z \varphi } = \chi^{(3)}_{z \varphi z \varphi } = \chi^{(3)}_{z \varphi  \varphi z} 
     = \chi^{(3)}_{\varphi \varphi z z} = \chi^{(3)}_{z z \varphi  \varphi } = \nonumber
     \\ =\chi^{(3)}_{z \rho z \rho} = \chi^{(3)}_{ \rho z z \rho} 
     =  \chi^{(3)}_{\rho z \rho z} = \chi^{(3)}_{z z \rho  \rho} = \chi^{(3)}_{z \rho  \rho z} = \chi^{(3)}_{ \rho  \rho z z }&  = 4.
\end{align}
From above, we see that $m_\chi\in \{0, 4, -4\}$, so $\Delta m_\chi \in \{ \pm 4, \pm 8 \}$. Note that the second value $\pm 8$ could not be obtained for $\hat\chi^{(2)}$ or any third rank tensor.}
\subsection{Vector spherical harmonics}
\label{sec:app:1}
Magnetic and electric vector spherical harmonics $\mathbf{M}_{^e_omn}, \mathbf{N}_{^e_omn}$ are generated by the scalar functions~\cite{Bohren1998Mar}:
\begin{align}
&\psi_{e m n}=\cos m \varphi P_{n}^{m}(\cos \vartheta) z_{n}({k} r),\label{scalarfunction} \\ \nonumber &\psi_{o m n}=\sin m \varphi P_{n}^{m}(\cos \vartheta) z_{n}({k} r),
\end{align}
where $P_{n}^{m}(\cos \theta)$ are associated Legendre polynomials, and $z_{n}({k} r)$ are any of the spherical Bessel functions: $j_n, y_n, h_n^{(1)}, h_n^{(2)}$.
They can be defined as:
\begin{align}
&\mathbf{M}_{^e_o m n}=\nabla \times\left(\mathbf{r} \psi_{^e_o m n}\right),\\ \nonumber 
&\mathbf{N}_{^e_o m n}=\frac{\nabla \times \mathbf{M}_{^e_o m n}}{\mathbf{k}}.
 \end{align}
In the component form they can be written as follows:
\begin{align}
\mathbf{M}_{e m n}(k, \mathbf{r})&=\frac{-m}{\sin\theta} \sin m \varphi P_{n}^{m}(\cos\theta) z_{n}(\rho)\hat{\mathbf{e}}_{\theta}- \\ \nonumber &-\cos m \varphi \dv{P_{n}^{m}(\cos\theta)}{\theta} z_{n}(\rho) \hat{\mathbf{e}}_{\varphi},
\end{align}
\begin{align}
\mathbf{M}_{o m n}(k, \mathbf{r})&={\frac{m}{\sin\theta} \cos m \varphi P_{n}^{m}(\cos\theta) }z_{n}(\rho) \hat{\mathbf{e}}_{\theta}- \\ \nonumber &-\sin m \varphi \dv{P_{n}^{m}(\cos\theta)}{\theta} z_{n}(\rho) \hat{\mathbf{e}}_{\varphi},
\end{align}
\begin{align}
\mathbf{N}_{e m n}(k, \mathbf{r})&=\frac{z_{n}(\rho)}{\rho} \cos m \varphi \, n(n+1) P_{n}^{m}(\cos\theta) \hat{\mathbf{e}}_{r}+ \\ \nonumber &+\cos m \varphi \dv{P_{n}^{m}(\cos\theta)}{\theta} \frac{1}{\rho} \dv{\rho}\left[\rho z_{n}(\rho)\right] \hat{\mathbf{e}}_{\theta}- \\ \nonumber &-m \sin m \varphi \frac{P_{n}^{m}(\cos\theta)}{\sin\theta}\frac{1}{\rho} \dv{\rho}\left[\rho z_{n}(\rho)\right] \hat{\mathbf{e}}_{\varphi},
\end{align}
\begin{align} 
\mathbf{N}_{o m n}(k, \mathbf{r})&=\frac{z_{n}(\rho)}{\rho} \sin m \varphi \, n(n+1) P_{n}^{m}(\cos\theta) \hat{\mathbf{e}}_{r}+\\\nonumber &+\sin m \varphi \dv{P_{n}^{m}(\cos\theta)}{\theta} \frac{1}{\rho} \dv{\rho}\left[\rho z_{n}(\rho)\right] \hat{\mathbf{e}}_{\theta}+\\\nonumber &+{m \cos m \varphi \frac{P_{n}^{m}(\cos\theta)}{\sin\theta}} \frac{1}{\rho} \dv{\rho}\left[\rho z_{n}(\rho)\right] \hat{\mathbf{e}}_{\varphi},
\end{align}
where dimensionless variable $\rho = k r$ was introduced.
\subsection{On specific multipolar combinations and vortex beams}
{
\begin{figure}[ht!]
    \centering
    \includegraphics[width=0.60\linewidth]{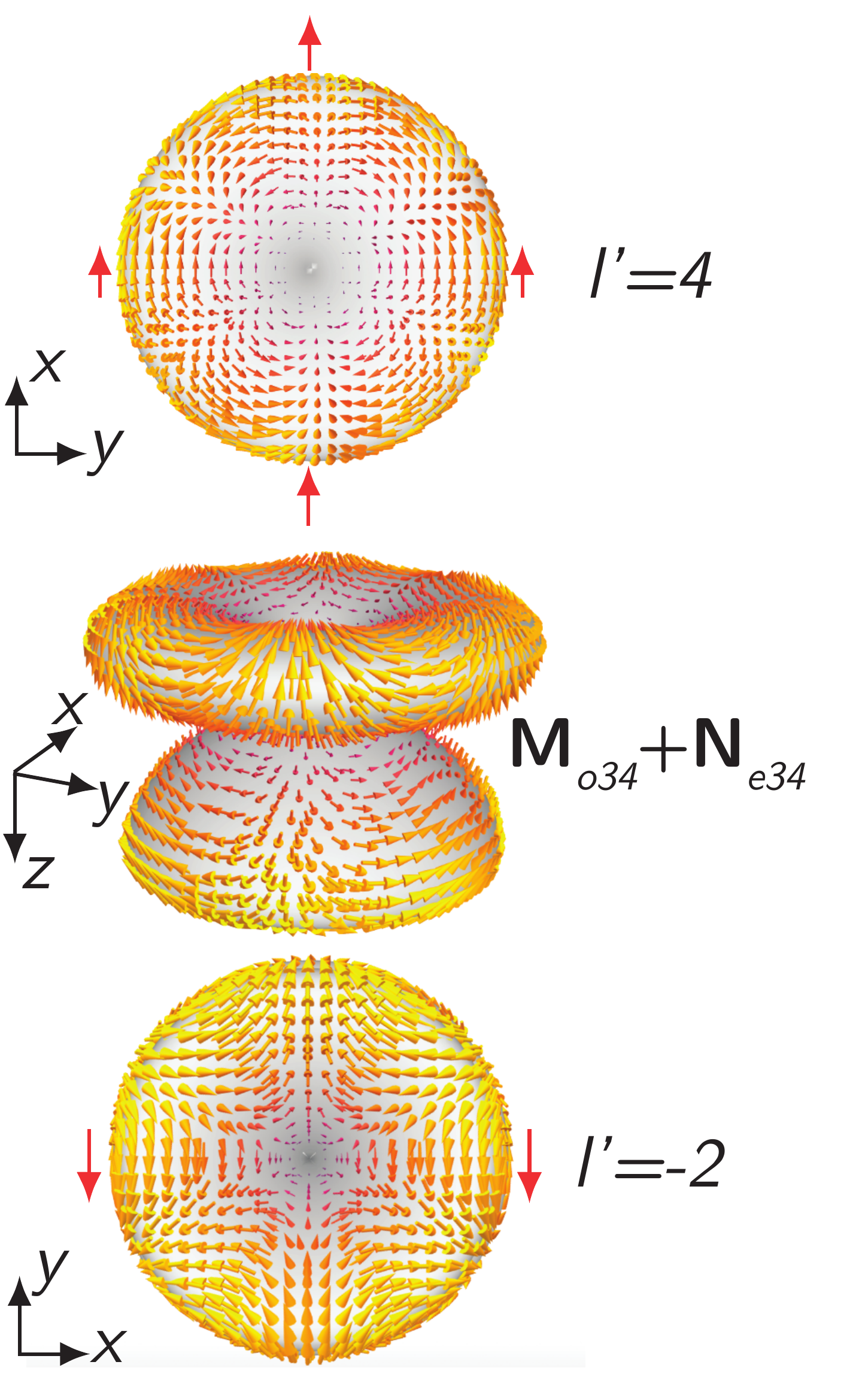}
    \caption{Visual illustration of {the far-field} real part of a multipolar combination \eqref{comb6} or \eqref{comb7}. The far-field radiation pattern is shown in the center with arrows depicted. For convenience, we provide the view from the top and bottom. Red arrows are a guide for eyes for computing the Hopf index $l'$.}
    \label{topcharge}
\end{figure}
\label{app:mult}
We provide this appendix for a better understanding of the symmetry properties of the vortex beams building blocks, which are specific multipolar combinations. Recall that in the paraxial approximation beam (\eqref{eq_rl} and~\eqref{eq_ll}) is often written in the form
\begin{equation}
    |R/L_m\rangle {\sim} \exp(\iu(m-\lambda)\varphi)(\hat{\vb x} + \lambda \iu\hat{\vb y})/\sqrt{2}
\end{equation}
where opposite signs of total angular momentum projection $m$ and helicity $\lambda$~\cite{Fernandez-Corbaton2012-Helicityandangular, Fernandez-Corbaton2014-Onthetransformation} refer to the right- and left-handed beams. Usually, the quantity $\ell=m-\lambda$ is introduced, which refers to the number of phase rotations around the singularity (see Fig.~\ref{mainscheme}) and also to the order of higher-order Poincare sphere. For a circularly polarized plane wave, $\ell=0$.
{Let us recall the expressions from~\eqref{sei} and~\eqref{sette}~\cite{Molina-Terriza2008-Determinationofthe}, which are the building blocks of vortex beams with $m$ and $\lambda$:
\begin{equation}
    (\vb N_{e|m|n}\pm \iu\vb N_{o|m|n})+(\pm\vb M_{e|m|n}+ \iu\vb M_{o|m|n}),
    \label{comb6}
\end{equation}
where upper sign is for $\lambda=1$ and $m=|m|$, and the lower sign is for $\lambda=-1$ and $m=-|m|$ and
\begin{equation}
    (\vb N_{e|m|n}\mp \iu\vb N_{o|m|n})-(\mp\vb M_{e|m|n}+ \iu\vb M_{o|m|n}),
    \label{comb7}
\end{equation}
where upper sign is for $\lambda=1$ and $m=-|m|$, and the lower sign is for $\lambda=-1$ and $m=|m|$.
These multipolar combinations are eigenvectors of helicity operator~\cite{Vavilin2024-ThepolychromaticT-m, Fernandez-Corbaton2015-HelicityandDuality, Bliokh2013-Dualelectromagnetism}, and determine the polarization of the beam. 
Moreover, their total angular momentum projection is also well-defined and equal to $m$~\cite{Akhiezer}.  
These functions are complex-valued, thus it is difficult to provide a visual presentation. 
However, if we depict {the real part of the far-field}, it provides us with some beautiful insights.
Note, that such combinations refer to the building blocks of vector beams $\ket{B_{me}}$ (the imaginary part would refer to $\ket{B_{mo}}$, respectively.) }
\begin{figure}[ht!]
    \centering
    \includegraphics[width=0.95\linewidth]{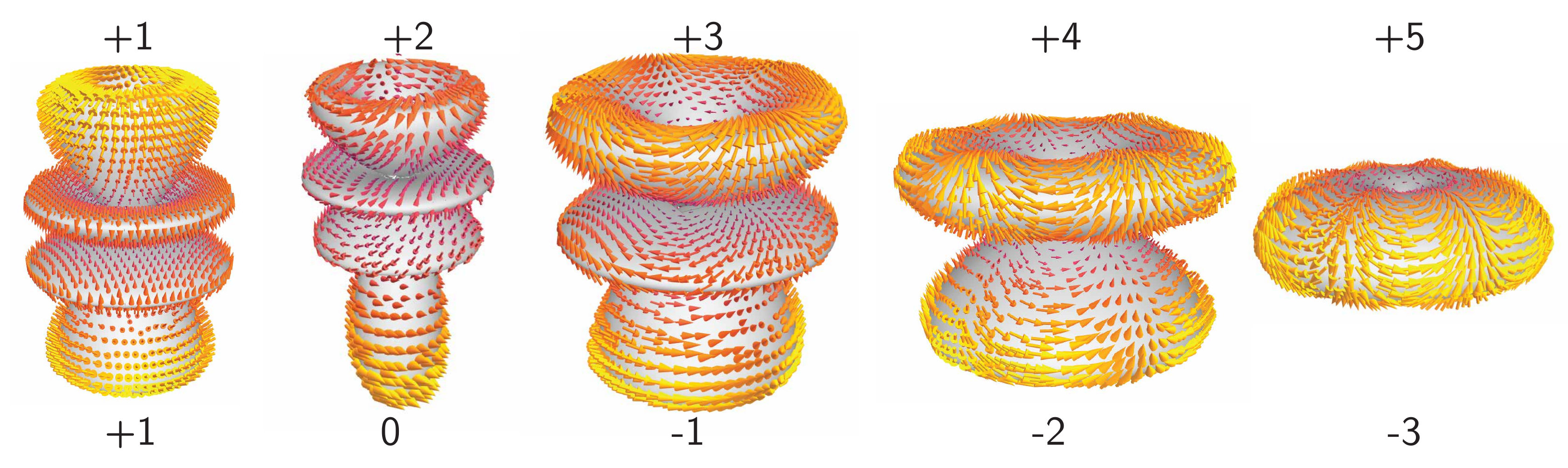}
    \caption{{Far-field of the $\vb N_{em4} + \iu \vb M_{om4}$} for different $m\in\{ 0, 1, 2, 3, 4\}$ and their Hopf indexes $l'$.}
    \label{topcharge2}
\end{figure}
For each smooth vector field on sphere $S^2$, according to Hairy ball, or Poicare-Hopf theorem~\cite{Hopf1927-Vektorfelderinn-dim, Poincare1881-Memoiresurlescourb}, the sum of Hopf indexes $l'$~\cite{Hopf-DifferentialGeometry} of the singularities is equal 2. 
Hence it is also applicable for multipoles~\cite{Chen2019-SingularitiesandPoi}.
We can calculate the Hopf index as the number of counterclockwise rotations made by the polarization vector when circling the singularity counterclockwise.
For these particular linear combinations, we have two point singularities on the south and north poles, and their Hopf indexes are always equal to either $m+1$ and $m-1$~(see Fig.~\ref{topcharge} and~\ref{topcharge2}). 
For the linear combinations with $m=1$, which contribute to Kerker effect~\cite{Zambrana-Puyalto2013-Dualitysymmetryand}, we only have one of them with index 2.
For the complex functions, the Hopf index will transform into a number of phase rotations $\ell$.
These properties could be also useful for engineering, e.g. topological charges of BICs~\cite{Kang2022-Mergingboundstates, Chen2019-SingularitiesandPoi}.}

{We should also notice that beams with the same $\ell$ can have either $m= \ell + 1$ or $m= \ell - 1$.
In some papers, the dichroism is defined as a difference between responses for the same $\ell$ but different $\lambda$. However, this would lead to the excitation of symmetrically different resonant modes (with different $m$), hence different linear response~\cite{Zambrana-Puyalto2014-Angularmomentum-indu}.
}
{\subsection{Extra technical derivations }
\label{sec:app:4}}
\subsubsection{{Multipolar content of the eigenmodes of C$_{3\text{v}}$ nanostructure}}\label{sec:app:4:1}
\paragraph{Eigenmodes, which are transformed under E irrep}
Decomposition of the eigenmodes transformed under $\text{E}$, according to the tables with a given multipolar content of the eigenmodes~\cite{Gladyshev_Frizyuk_Bogdanov_2020, Xiong_Xiong_Yang_Yang_Chen_Wang_Xu_Xu_Xu_Liu_2020} consists only of the vector spherical harmonics with the projections of the angular momentum $m=3s-1$ and $m=3s-2$, where $ s \in \mathbb {Z}_+$ with, in general, {complex coefficients.} 
However, we note that $\text{E}$ is a 2-dimensional irreducible representation, thus there are two orthogonal eigenmodes transformed under this representation. Given the above, eigenmodes can be written in the following form:
    \begin{align}
        \label{modesеtrimerx} 
        &\mathbf{E}^{2\omega, x}_{j} \left(\mathbf{r}^{\prime}\right) = a_{e11}\mathbf{N}_{e11} + a_{e12}\mathbf{N}_{e12} + b_{o11}\mathbf{M}_{o11} +\dots, 
        \\
        \label{modesеtrimery} 
        &\mathbf{E}^{2\omega, y}_{j} \left(\mathbf{r}^{\prime}\right) = a_{o11}\mathbf{N}_{o11} + a_{o12}\mathbf{N}_{o12} + b_{e11}\mathbf{M}_{e11} +\dots,
    \end{align} 
    where $\mathbf{E}^{2\omega, x}_{j} \left(\mathbf{r}^{\prime}\right)$ and $\mathbf{E}^{2\omega, y}_{j} \left(\mathbf{r}^{\prime}\right)$ are orthogonal eigenmodes transformed under irreducible representation $\text{E}$, the first one behaves as a function $f(\vb r)=x$ and the second as $f(\vb r)=y$~\cite{C3v_irreps}. {To ensure orthogonality, we've exploited the symmetry of the nanostructure under reflection in $y=0$ plane.}
    Index $j$ corresponds to the specific eigenmode {with such symmetry. For each $j$ the coefficients $a$ and $b$ are different}.
    It's worth mentioning that orthogonal eigenmodes should be transformed through each other in a certain way under transformations of group symmetry $\text{C}_{3\text{v}}$, thus coefficients $a_{emn}$ and $a_{omn}$, as well as $b_{emn}$ and $b_{omn}$ are connected with each other: $a_{e(3s-2)n} = a_{o(3s-2)n} = a_{(3s-2)n}$, $a_{e(3s-1)n} = -a_{o(3s-1)n} = a_{(3s-2)n}$, and $b_{e(3s-2)n} = -b_{o(3s-2)n} = b_{(3s-2)n}$, $b_{e(3s-1)n} = b_{o(3s-1)n} = b_{(3s-1)n}$, where $ s \in \mathbb {Z}_+$. 
    These identities can be proved by applying the symmetry transformations of the $\text{C}_{3\text{v}}$ group to each vector spherical harmonic~\cite{Zhang2008-Additiontheoremfor, Stein}. 
    Therefore, decompositions of eigenmodes~\eqref{modesеtrimerx} and~\eqref{modesеtrimery} are written as:
    \begin{align}
        &\mathbf{E}^{2\omega, x}_{j} \left(\mathbf{r}^{\prime}\right) = a_{11}\mathbf{N}_{e11} + a_{12}\mathbf{N}_{e12} + b_{11}\mathbf{M}_{o11}+\dots, \label{modesеtrimernewx} \\ 
        &\mathbf{E}^{2\omega, y}_{j} \left(\mathbf{r}^{\prime}\right) = a_{11}\mathbf{N}_{o11} + a_{12}\mathbf{N}_{o12} + (-b_{11})\mathbf{M}_{e11}+\dots.\label{modesеtrimernewy}
    \end{align} 
    Let us change the basis to a circular form, i.e. $\mathbf{E}^{2\omega, x\pm \iu y}_{j} = ({\mathbf{E}^{2\omega, x}_{j} \pm \iu\mathbf{E}^{2\omega, y}_{j}})/{2}$ and rewrite decompositions as:
    \begin{align}
        \nonumber
        \mathbf{E}^{2\omega, x+\iu y}_{j} \left(\mathbf{r}^{\prime}\right) 
         & = \frac{a_{11}}{2}(\mathbf{N}_{e11} + \iu\mathbf{N}_{o11}) + \frac{a_{22}}{2}(\mathbf{N}_{e22} - \iu\mathbf{N}_{o22}) + 
        \\
        \label{modesеtrimernew1x} 
        &+ \frac{a_{12}}{2}(\mathbf{N}_{e12}  + \iu\mathbf{N}_{o12}) + \frac{b_{11}}{2}(\mathbf{M}_{o11} - \iu\mathbf{M}_{e11}) +\dots,
    \end{align}
    \begin{align}
        \nonumber
        \mathbf{E}^{2\omega, x-\iu y}_{j} 
        \left(\mathbf{r}^{\prime}\right)  & = \frac{a_{11}}{2}(\mathbf{N}_{e11} - \iu\mathbf{N}_{o11}) + \frac{a_{22}}{2}(\mathbf{N}_{e22} + \iu\mathbf{N}_{o22}) + 
        \\
        \label{modesеtrimernew1y} 
        &+ \frac{a_{12}}{2}(\mathbf{N}_{e12} - \iu\mathbf{N}_{o12}) + \frac{b_{11}}{2}(\mathbf{M}_{o11} + \iu\mathbf{M}_{e11})+\dots,
    \end{align}
    where $\mathbf{E}^{2\omega, x\pm \iu y}_{j}$ are also orthogonal eigenmodes. 
    Using the definition of the vector spherical harmonics (See Appendix~\ref{sec:app:1}),
    we can simplify the expressions~\eqref{modesеtrimernew1x} and~\eqref{modesеtrimernew1y} above:
    \begin{align}
        \nonumber
        \mathbf{E}^{2\omega, x\pm \iu y}_{j} \left(\mathbf{r}^{\prime}\right)  & = \frac{a_{11}}{2}\mathbf{N}_{11}^{}(r,z)\eu^{\pm 1\iu\varphi} +
        \frac{a_{22}}{2}\mathbf{N}_{22}^{}(r,z)\eu^{\mp 2\iu\varphi} +
        \\
        \label{modesеtrimernew2}
        &+  \frac{a_{12}}{2}\mathbf{N}_{12}^{}(r,z)\eu^{\pm 1\iu\varphi} + \frac{b_{11}}{2}\mathbf{M}_{11}^{}(r,z)\eu^{\pm 1\iu\varphi}+\dots,
    \end{align} 
    where $\mathbf{N}(\mathbf{M})_{mn}^{}(r,z)$ do not depend on  $\varphi$. 
    Note that the functions $\mathbf{N}_{22}^{}(r,z)$ can have a different sign of $\varphi$-component for different ``polarization'' but here we omit it as well as in the polarization{, since it doesn't affect the final result} (See Suppl. Info in~\cite{Frizyuk2021-NonlinearCircularDi}).
    Finally, combining combinations all $\mathbf{N}(\mathbf{M})^{}_{mn}(r,z)$ with {complex coefficients $a_{mn}, b_{mn}$} with the same $m$ into  $\mathbf{E}^{}_{j,mn}(r,z)$ we obtain the following decompositions of orthogonal eigenmodes $\mathbf{E}^{2\omega, x\pm \iu y}_{j}$:
    \begin{align}
        \label{modesеtrimernewfinalinappendix}
        \nonumber
        \mathbf{E}^{2\omega, x\pm \iu y}_{j} \left(\mathbf{r}^{\prime}\right) = \sum_{n}\bigg\{ \mathbf{E}^{}_{j,1n}(r,z)\eu^{\pm 1\iu\varphi} &+  \mathbf{E}^{}_{j,2n}(r,z)\eu^{\mp 2\iu\varphi} + 
        \\ + \mathbf{E}^{}_{j,4n}(r,z)\eu^{\pm 4\iu\varphi}  + \mathbf{E}^{}_{j,5n}(r,z)\eu^{\mp 5\iu\varphi} & +\dots\bigg\}.
    \end{align}

\paragraph{Eigenmodes, which are transformed under $\text{A}_1$ and $\text{A}_2$ irrep}
     {Decompositions of eigenmodes transformed under irreducible representations $\text{A}_1$ and $\text{A}_2$ contain vector spherical harmonics $\mathbf{M}_{^e_o(3s)n}$ and $\mathbf{N}_{^e_o(3s)n}$, where $ s \in \mathbb {Z}_+$. Eigenmodes transformed under $\text{A}_1$ are even under reflection in the plane XZ (for a particular orientation of a trimer), while eigenmodes transformed under $\text{A}_2$ are odd.  
    Thus, decompositions can be expressed as follows~\cite{Gladyshev_Frizyuk_Bogdanov_2020}:}
    \begin{align}
        \label{modesA1A2} &{\mathbf{E}^{2\omega,\text{A}_1}_{j} \left(\mathbf{r}^{\prime}\right) = a_{e01}\mathbf{N}_{e01} + a_{e33}\mathbf{N}_{e33} + b_{o33}\mathbf{M}_{o33} + \dots,}\\
        &{\mathbf{E}^{2\omega,\text{A}_2}_{j} \left(\mathbf{r}^{\prime}\right) = b_{e01}\mathbf{M}_{e01} + b_{e33}\mathbf{M}_{e33} + a_{o33}\mathbf{N}_{o33} +\dots.}\label{modesA1A2two}
    \end{align}
    {Using the explicit form of the vector spherical harmonics (Appendix~\ref{sec:app:1}), and new notations, we can rewrite the expressions above~\eqref{modesA1A2} and~\eqref{modesA1A2two} in the following form:}
    \begin{align}
        \label{modesA1A2finalA1inappendix} 
        &{\mathbf{E}^{2\omega, \text{A}_1}_{j} \left(\mathbf{r}^{\prime}\right)}= 
        \\\nonumber&= \sum_{n} \bigg\{\left[{E}^{r}_{j,e3n} \left(r,z\right)\hat{\mathbf{e}}_r + {E}^z_{j,e3n} \left(r,z\right)\hat{\mathbf{e}}_z\right]\cos{3\varphi} {+ {E}^{\varphi}_{j,e3n} \left(r,z\right)\hat{\mathbf{e}}_\varphi\sin{3\varphi} + \dots\bigg\},}
    \end{align}
    \begin{align}
        \label{modesA1A2finalA2inappendix}
        &{\mathbf{E}^{2\omega, \text{A}_2}_{j} \left(\mathbf{r}^{\prime}\right)}=
        \\\nonumber &{= \sum_{n} \bigg\{\left[{E}^{r}_{j,o3n} \left(r,z\right)\hat{\mathbf{e}}_r + {E}^z_{j,o3n} \left(r,z\right)\hat{\mathbf{e}}_z\right]\sin{3\varphi} +}
        { {E}^{\varphi}_{j,o3n} \left(r,z\right)\hat{\mathbf{e}}_\varphi\cos{3\varphi} + \dots\bigg\},}
    \end{align}
    {where coefficients ${E}^{r}_{j,^e_o3n}\left(r,z\right)$,  ${E}^{z}_{j,^e_o3n}\left(r,z\right)$,  ${E}^{\varphi}_{j,^e_o3n}\left(r,z\right)$ contain all functions and constants independent on coordinate $\varphi$.}

\subsubsection{{Expression for the total intensity}}\label{sec:app:intensity}
{In the Section~\eqref{subsec:intensity}, we obtained the following expression for the total integral second harmonic intensity $I_{{\ket{L_m}},{\ket{R_m}}}^{2\omega}$ for the case of the nanostructure GaAs[111]$\parallel$z with symmetry $\text{C}_{3\text{v}}$~\eqref{intensityLCP}:}
\begin{align}
    \label{intensityLCPinappendix}
    &I_{{\ket{L_m}},{\ket{R_m}}}^{2\omega} \propto
      \int_{\text{sph}} \dd V \bigg|\mathbf{E}^{2\omega, x\pm\iu y}_{j}(\mathbf{r})\left(a_j^{+} \eu^{\mp 3\iu\beta} + a_j^{\prime+} \eu^{\pm 3\iu\beta} + a_j^{\prime\prime+}\right) + 
    \\\nonumber &+ \eu^{\iu \alpha}\mathbf{E}^{2\omega, x\pm\iu y}_{i}(\mathbf{r})\left(a_i^{+} \eu^{-3\iu\beta} + a_i^{\prime+} \eu^{+3\iu\beta} + a_i^{\prime\prime+}\right)\bigg|^2 =\int_{\text{sph}} \dd V \cdot J^\pm,
\end{align} 
where, for clarity, we introduce the integrand $J^\pm$.
\paragraph{A left-handed beam}
{Let us precisely calculate the integrand from the expression~\eqref{intensityLCPinappendix} for the left-handed excitation:}
\begin{align}
    \label{longlongintegral}
    J^{+} &= \bigg|\mathbf{E}^{2\omega, x+\iu y}_{j}(\mathbf{r})\left(a_j^{+} \eu^{-3\iu\beta} + a_j^{\prime+} \eu^{+3\iu\beta} + a_j^{\prime\prime+}\right) + 
    \\\nonumber &+ \eu^{\iu \alpha}\mathbf{E}^{2\omega, x+\iu y}_{i}(\mathbf{r})\left(a_i^{+} \eu^{-3\iu\beta} + a_i^{\prime+} \eu^{+3\iu\beta} + a_i^{\prime\prime+}\right)\bigg|^2 =
    \\\nonumber &= \bigg[\eu^{-\iu \alpha}\mathbf{E}^{2\omega, x+\iu y}_{j}(\mathbf{r})\left(\mathbf{E}^{2\omega, x+\iu y}_{i}(\mathbf{r})\right)^*\cdot
    \\\nonumber &\cdot\bigg(a_j^{+} \eu^{-3\iu\beta} + a_j^{\prime+} \eu^{+3\iu\beta} +
     a_j^{\prime\prime+}\bigg)\left(a_i^{+*} \eu^{+3\iu\beta} + a_i^{\prime+*} \eu^{-3\iu\beta} + a_i^{\prime\prime+*}\right)+ \text{c.c.}\bigg] +
    \\\nonumber&+ \left(\left|\mathbf{E}^{2\omega, x+\iu y}_{j}(\mathbf{r})\right|^2\cdot\left|a_j^{+} \eu^{-3\iu\beta} + a_j^{\prime+} \eu^{+3\iu\beta} + a_j^{\prime\prime+}\right|^2 + \right.
    (j\to i)\bigg)=
    \\\nonumber &= \bigg[\eu^{-\iu \alpha}\mathbf{E}^{2\omega, x+\iu y}_{j}(\mathbf{r})\left(\mathbf{E}^{2\omega, x+\iu y}_{i}(\mathbf{r})\right)^*\cdot\\\nonumber &\cdot\bigg(\bigg\{a^{+}_j a^{+ *}_i + a^{\prime +}_j a^{\prime+ *}_i +
     a^{\prime\prime+}_j a^{\prime\prime+ *}_i \bigg\} +
      a^{\prime+}_j a^{+ *}_i \eu^{+6\iu\beta} + a^{+}_j a^{\prime + *}_i \eu^{-6\iu\beta} + 
    \\\nonumber&
    + \bigg\{a^{\prime\prime+}_j a^{+ *}_i \eu^{+3\iu\beta}  +
    a^{\prime+}_j a^{\prime\prime + *}_i \eu^{+3\iu\beta}\bigg\} +
     \bigg\{a^{+}_j a^{\prime\prime + *}_i \eu^{-3\iu\beta} + a^{\prime\prime+}_j a^{\prime+ *}_i \eu^{-3\iu\beta}\bigg\}\bigg) + \text{c.c.}\bigg] + 
    \\\nonumber &+ \bigg[\left|\mathbf{E}^{2\omega, x+\iu y}_{j}(\mathbf{r})\right|^2\bigg(\bigg\{|a^{+}_j|^2 + |a^{\prime +}_j|^2 + |a^{\prime\prime+}_j|^2\bigg\} + 
      a^{\prime+}_j a^{+ *}_j \eu^{+6\iu\beta} +
     a^{+}_j a^{\prime + *}_j \eu^{-6\iu\beta}  +
     \\\nonumber&+\bigg\{a^{\prime\prime+}_j a^{+ *}_j \eu^{+3\iu\beta} + a^{\prime+}_j a^{\prime\prime + *}_j \eu^{+3\iu\beta} \bigg\} +  \bigg\{a^{+}_j a^{\prime\prime + *}_j \eu^{-3\iu\beta} + a^{\prime\prime+}_j a^{\prime+ *}_j \eu^{-3\iu\beta}\bigg\}\bigg) + (j\to i) \bigg] = 
    \\\nonumber&= \bigg[\eu^{-\iu \alpha}\mathbf{E}^{2\omega, x+\iu y}_{j}(\mathbf{r})\left(\mathbf{E}^{2\omega, x+\iu y}_{i}(\mathbf{r})\right)^*\cdot
    \\\nonumber& \cdot\bigg(A_{ji} + B_{ji}\eu^{+6\iu\beta} + C_{ji}\eu^{-6\iu\beta} +
     D_{ji}\eu^{+3\iu\beta} + E_{ji}\eu^{-3\iu\beta}\bigg) + \text{c.c.}\bigg] +
    \\\nonumber & + \bigg[\left|\mathbf{E}^{2\omega, x+\iu y}_{j}(\mathbf{r})\right|^2\bigg(A_{jj} + 
     B_{jj}\eu^{+6\iu\beta} + 
      C_{jj}\eu^{-6\iu\beta} + D_{jj}\eu^{+3\iu\beta} + E_{jj}\eu^{-3\iu\beta}\bigg) + (j\to i)\bigg],
\end{align}
where constants $A_{ji} = a^{+}_j a^{+ *}_i + a^{\prime +}_j a^{\prime+ *}_i$, $B_{ji} = a^{\prime+}_j a^{+ *}_i \eu^{+6\iu\beta}$, $C_{ji} = a^{+}_j a^{\prime + *}_i \eu^{-6\iu\beta}$, $D_{ji} = a^{\prime\prime+}_j a^{+ *}_i \eu^{+3\iu\beta}  + a^{\prime+}_j a^{\prime\prime + *}_i \eu^{+3\iu\beta}$, and $E_{ji} = a^{+}_j a^{\prime\prime + *}_i \eu^{-3\iu\beta} + a^{\prime\prime+}_j a^{\prime+ *}_i \eu^{-3\iu\beta}$ were introduced, for simplicity. 
{Substituting the expression for the integrand $J^+$~\eqref{longlongintegral} into formula for the intensity $I_{{\ket{L_m}}}^{2\omega}$~\eqref{intensityLCP}, we get:}
\begin{align}
    \label{intensityLCPinterf} &I_{{\ket{L_m}}}^{2\omega,\text{interf.}} \propto
    \int_{\text{sph}} \dd V \eu^{-\iu\alpha}\mathbf{E}^{2\omega, x+\iu y}_{j}(\mathbf{r})\left(\mathbf{E}^{2\omega, x+\iu y}_{i}(\mathbf{r})\right)^* \cdot
    \\\nonumber&\left(A_{ji} + B_{ji}\eu^{+6\iu\beta} + C_{ji}\eu^{-6\iu\beta} + D_{ji}\eu^{+3\iu\beta} + E_{ji}\eu^{-3\iu\beta}\right) + \text{c.c.},
\end{align}
{where we focus only on interference terms of the intensity.}
\paragraph{A right-handed beam}
{For the right-handed excitation, we omit the similar calculations~\eqref{longlongintegral} and present the interference terms $I_{{\ket{R_m}}}^{2\omega}$:}}
\begin{align}
    \label{intensityRCPinterf}
    I_{{\ket{R_m}}}^{2\omega,\text{interf.}}    &\propto \nonumber \int_{\text{sph}} \dd V \eu^{-\iu\alpha}\mathbf{E}^{2\omega, x-\iu y}_{j}(\mathbf{r})\left(\mathbf{E}^{2\omega, x-\iu y}_{i}(\mathbf{r})\right)^* \cdot
    \\    & \cdot\left(A_{ji} + B_{ji}\eu^{-6\iu\beta} + C_{ji}\eu^{+6\iu\beta}  + D_{ji}\eu^{-3\iu\beta} + E_{ji}\eu^{+3\iu\beta}\right) + \text{c.c.}.
\end{align}
{Finally, these two formulas for the interference terms of the total intensity~\eqref{intensityLCPinterf} and~\eqref{intensityRCPinterf} can be combined into a single expression that has already been introduced in the Section~\eqref{subsec:intensity}~\eqref{intensityLCPRCPinterfoutappendix}:}
\begin{align}
    \label{intensityLCPRCPinterfinappendix} &I_{{\ket{L_m}},{\ket{R_m}}}^{2\omega,\text{interf.}} \propto
     \int_{\text{sph}} \dd V \eu^{-\iu\alpha}\mathbf{E}^{2\omega, x\pm\iu y}_{j}(\mathbf{r})\left(\mathbf{E}^{2\omega, x\pm\iu y}_{i}(\mathbf{r})\right)^* \cdot
    \\\nonumber&\left(A_{ji} + B_{ji}\eu^{\pm6\iu\beta} + C_{ji}\eu^{\mp6\iu\beta} + D_{ji}\eu^{\pm3\iu\beta} + E_{ji}\eu^{\mp3\iu\beta}\right) + \text{c.c.}.
\end{align}

\end{document}